  \documentclass{aa} % for a referee version
  \usepackage{natbib}
  \usepackage{graphicx}
  \usepackage{longtable}

  \usepackage{txfonts}

\begin{document}

%     \subtitle{}
     
      \titlerunning{Comparing the host galaxies ages  of X-ray selected AGN in COSMOS}
      \authorrunning{I. Georgantopoulos et al.}
      
      \title{Comparing the host galaxy ages  of X-ray selected AGN in  COSMOS}
      \subtitle{Obscured AGN are associated with older galaxies }

     \author{I. Georgantopoulos \inst{1}, E. Pouliasis \inst{1}, G. Mountrichas \inst{2}, A. Van der Wel \inst{3,4}, S. Marchesi \inst{5,6}, G. Lanzuisi \inst{5}}

     \institute{Institute for Astronomy Astrophysics Space Applications and Remote Sensing (IAASARS), National Observatory of Athens, I. Metaxa \& V. Pavlou, Penteli, 15236, Greece\\
     \email{ig@noa.gr}
     \and
       Instituto de Fisica de Cantabria (CSIC-Universidad de Cantabria), Avenida de los Castros, 39005 Santander, Spain
       \and
       Sterrenkundig Observatorium, Universiteit Gent, Krijgslaan 281 S9, 9000 Gent, Belgium
       \and
       Max-Planck Institut f{\"u}r Astronomie K{\"o}nigstuhl, D-69117, Heidelberg, Germany
       \and
       INAF - Osservatorio di Astrofisica e Scienza dello Spazio di Bologna, Via Piero Gobetti, 93/3, 40129, Bologna, Italy 
       \and
       Department of Physics and Astronomy, Clemson University, Kinard Lab of Physics, Clemson, SC 29634, USA }

 %   \date{}

    \abstract{ We explore the properties of the host galaxies of X-ray selected AGN 
    in the COSMOS field using 
    the Chandra Legacy sample and the LEGA-C survey VLT optical spectra. 
    Our main goal is to compare the relative ages of the host galaxies of the obscured and 
    unobscured AGN by means of the calcium break $\rm D_n(4000)$ and the $H_\delta$ 
    Balmer line. The host galaxy ages are
    examined in conjunction with other properties such as the galaxy stellar mass, and star-formation rate 
    as well as the AGN Eddington ratio. Our sample consists of 50 unobscured or mildly obscured ($\rm N_H<10^{23}cm^{-2})$ and 23 heavily obscured AGN 
    ($\rm N_H>10^{23}cm^{-2})$ in the redshift range $z=0.6-1$. We take specific caution 
    to create control samples in order to match the exact luminosity and redshift distributions
    for the obscured and unobscured AGN. The majority of unobscured AGN 
    appear to live in young  galaxies  in contrast to the obscured AGN which appear to live in galaxies  located between the young and old galaxy populations.
    This finding may be in contrast to those  evolutionary AGN unification models which postulate that the AGN begin their life 
    in a heavy obscuration phase.   
    The host galaxies of the obscured AGN have significantly lower levels of  specific star-formation. 
    At the same time the obscured AGN have  lower Eddington ratios indicating a link between the star-formation and the black hole accretion. 
    We find that the distribution of the stellar masses of the host galaxies of obscured AGN is 
     skewed towards higher stellar masses in agreement with previous findings. 
     Our results on the relative age of obscured AGN are 
      valid when we match our obscured and unobscured AGN samples according to the stellar mass of 
      their host galaxies. 
     All the above results become less conspicuous when 
    a lower column density ($\rm \log N_H(cm^{-2})= 21.5$ or 22)  is used to separate 
    the obscured and unobscured AGN populations.  
    }


    \authorrunning{Georgantopoulos et al.}

    \maketitle
  %
  %________________________________________________________________

	\section{Introduction}
Active Galactic Nuclei (AGNs) are among the most luminous sources in the Universe.  They are powered by accretion onto 
supermassive black holes (SMBHs) in their centres \citep{Lynden1969}. Despite the difference in physical scale between the SMBH and the 
galaxy spheroid (about nine orders of magnitude), there is a tight correlation between the masses of the SMBH and the galaxy bulge
\citep{Silk1998, Magorrian1998, Ferrarese_Merritt2000, Gebhardt2000}. 
The physical mechanisms  that underlie this correlation are not 
properly understood but most theoretical models for galaxy evolution predict a regulating mechanism between the AGN power and the star-formation of the host galaxy. 
The models that explain the AGN and galaxy co-evolution on the basis of mergers 
\citep[e.g.][]{Hopkins2008a} suggest that when the AGN becomes active it passes a long period in an obscured phase.  
The obscuring  material feeds the AGN that eventually becomes powerful enough to push away the surrounding material and become unobscured \citep[e.g.][]{Ciotti1997, Hopkins2006a, Hopkins2006b, Somerville2008, Blecha2018}. 
Then the unobscured AGN passes a phase of coeval growth where  the SMBH accretes material at high Eddington rates 
and at the same time this stimulates intense  star-formation in the centre of the host galaxy 
\citep{DiMatteo2005, Hopkins2008a, Zubovas2013}. 
These are the evolutionary unification models which are differentiated 
from the standard unification models. The latter postulate that the obscuration in an AGN depends only  
on the inclination angle relative to the line of sight \citep{Antonucci1993}.
A number of observational works provide evidence in support of the above evolutionary  scenarios 
\citep[e.g.][]{Koss2018, Glikman2018, Banerji2021, Hatcher2021, Mountrichas2023}.

In the past years there has been a number of works which examined the validity 
of the AGN/galaxy co-evolution models. The most widely applied method to study the AGN-galaxy co-evolution is via examining the correlation of the SMBH activity and 
the star formation rate (SFR) of the host galaxy  
\cite[e.g.][]{rovilos2012, Rosario2012, Chen2013, Hickox2014, Stanley2015, Rodighiero2015, Aird2012, Aird2019, Lanzuisi2017, Stanley2017, Harrison2018, Brown2019}. In many of the above studies the star-formation is measured at far infrared wavelengths from the cold dust emission heated by young stars using data from the {\it Herschel} mission. The X-ray luminosity is used  as a very good proxy of the AGN power. More recent works attempt to disentangle the effect of the host galaxy on the star-formation by taking into account the position of the host galaxy on the star-formation main sequence as defined by its redshift and stellar mass. These works \citep{Mullaney2015, Bernhard2019, Masoura2018,  Masoura2021, Florez2020, Mountrichas2021b, Torbaniuk2021} find again  a correlation between the normalised SFR 
 ($\rm SFR_{NORM}$)  and the AGN X-ray luminosity; $\rm SFR_{NORM}$  is defined as the observed SFR divided by the mean SFR of normal galaxies at the same redshift and stellar mass.  However, it is not entirely  clear whether the  correlation between SFR and X-ray luminosity hides an underlying correlation with the galaxy’s stellar mass \citep{Fornasini2020, Mountrichas2022a}. 
 This would suggest that the AGN power and the SFR evolve 
 in a similar manner because they are fed by the same molecular gas depot of the host galaxy. Some of the above works  examined the SFR separately for type-1 and type-2 AGN. 
 They concluded that there is no concrete evidence that 
the SFR properties differ in the two types of AGN \citep{Masoura2021, Mountrichas2021b}. 
 Nevertheless, \cite{Chen2015} by analysing a sample of mid-IR selected AGN in the Bootes region, find that the SFR in type-2  AGN is a factor of two higher compared to type-1 AGN. 

Additional evidence in support of the evolutionary unification scenarios 
may come from the comparison of  
the Eddington ratio distribution in obscured and unobscured AGN
\citep{Ananna2022, Schulze2015, Kelly_Shen2013, Schulze_Wisotzki2010}.
\cite{Ananna2022} studied  the Eddington ratios of low redshift AGN in the BAT AGN spectroscopic survey (BASS).  They find that the Eddington ratio distribution of obscured AGN is skewed towards low Eddington ratios.
This is in favour of a radiation-driven scenario where the radiation pressure regulates the shape of the torus. This means that when the AGN is luminous the torus 
is pushed away while the obscured AGN 
are those which have low Eddington ratios. 
These findings are not in contrast with  
 the evolution unification scenarios 
 where the birth of an AGN is marked by a long phase of obscuration accompanied by 
 low Eddington ratios. 

On the other hand, 
recent results on the masses of obscured and unobscured    AGN
may be in contrast to the above evolutionary 
unification scenarios.
\cite{Zou2019} examined optically selected narrow and broad line AGN in the COSMOS field, finding that the type-2 systems 
have significantly higher stellar masses. This result has been corroborated by other optically selected samples 
\citep{Mountrichas2021b, Koutoulidis2022} although the difference in stellar mass is not prominent when the separation between type-1 and type-2 AGN is based on X-ray obscuration. 
This difference in stellar mass may suggest that type-2 AGN 
are more massive because they are associated  with older systems and had more time to increase their mass because of merging with satellite galaxies. 
However, more concrete evidence is necessary in order to pin down the age of these AGN and thus to better 
constrain the models of AGN and galaxy co-evolution. This evidence can be provided by optical spectroscopy.
In a pioneering work \cite{Kauffmann2003b} examined
the properties of the host galaxies of a large number of AGN in  redshifts below z=0.3  from the Sloan Digital Sky Survey. 
In particular they examined the galaxy ages using the strength of the calcium 4000-$\AA$ break as the primary indicator
of the age of the stellar population. 
They examined the stellar ages of narrow-line and broad-line AGN finding no significant difference between the two 
populations when their luminosity and redshift is taken into account. 

The LEGA-C survey provides the opportunity to expand these studies 
at redshifts $z\sim0.7$ corresponding to a look-back time of 7 Gyr. The LEGA-C survey has observed 4,000 galaxies in the COSMOS field using the VIMOS instrument on VLT
providing excellent quality spectroscopy and hence accurate measurements of the calcium break and other 
age indicators such as the $H_\delta$ absorption line.
In this work, we examine the properties of the {\it Chandra} X-ray selected AGN in the COSMOS field combining the LEGA-C optical spectroscopy 
with Spectral energy distribution fittings performed with the X-CIGALE code. 
The optica1 spectroscopy provides robust indicators of the age as well as black 
hole masses while the SED provide the stellar masses and SFR. Our goal is to examine how 
the properties of obscured and unobscured 
AGN evolve with stellar age. This may provide
 constraints on the galaxy/AGN co-evolution
models and evolutionary unification models.
	Throughout the paper, we 
	assume a standard $\rm \Lambda CDM$ cosmology with $\rm H_o=69.3km s^{-1}Mpc^{-1}$, $\rm \Omega_m=0.286$.

\begin{figure*}
\center
   \begin{tabular}{c c}
    \includegraphics[width=0.47\textwidth]{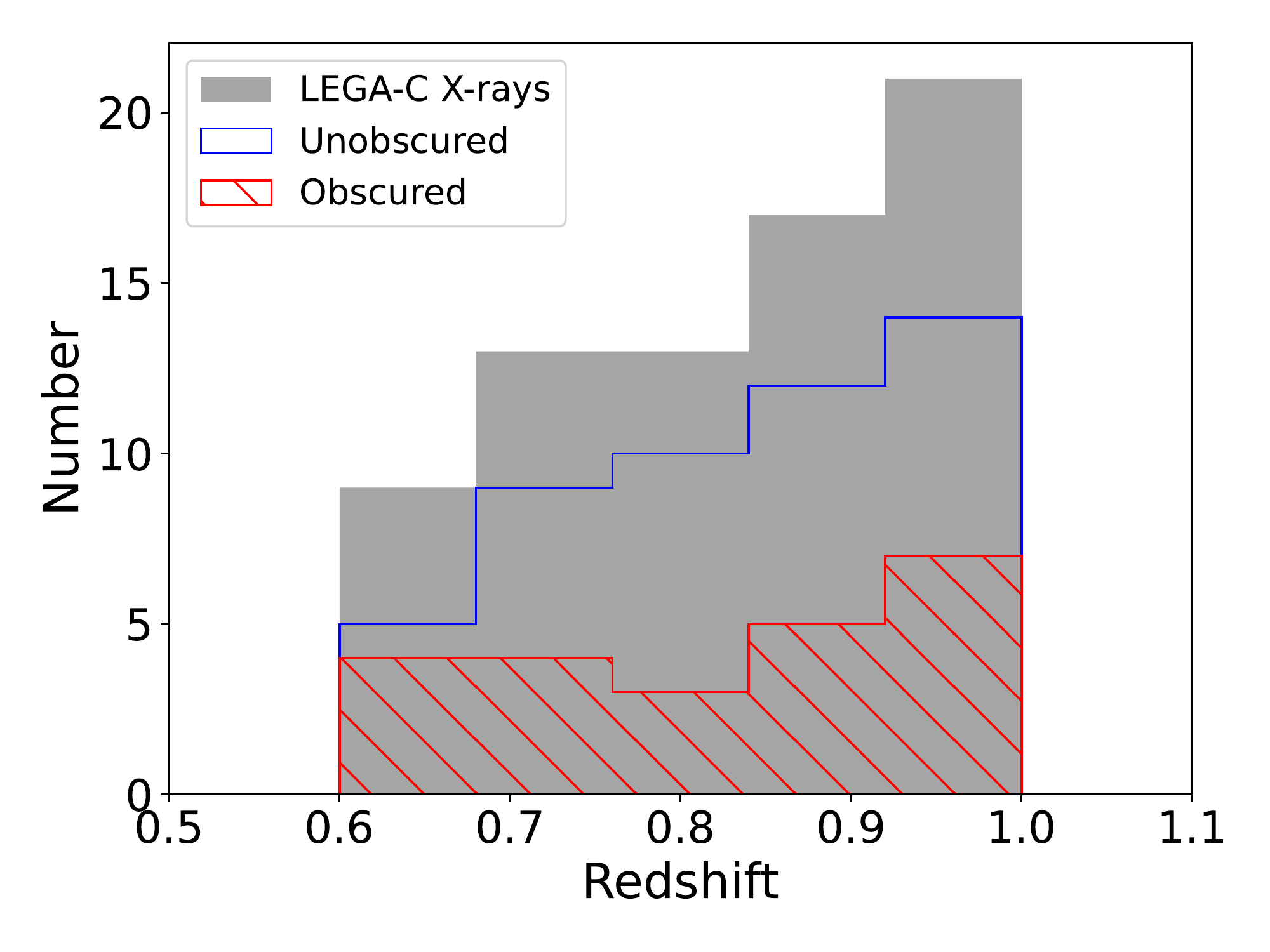} &

    \includegraphics[width=0.47\textwidth]{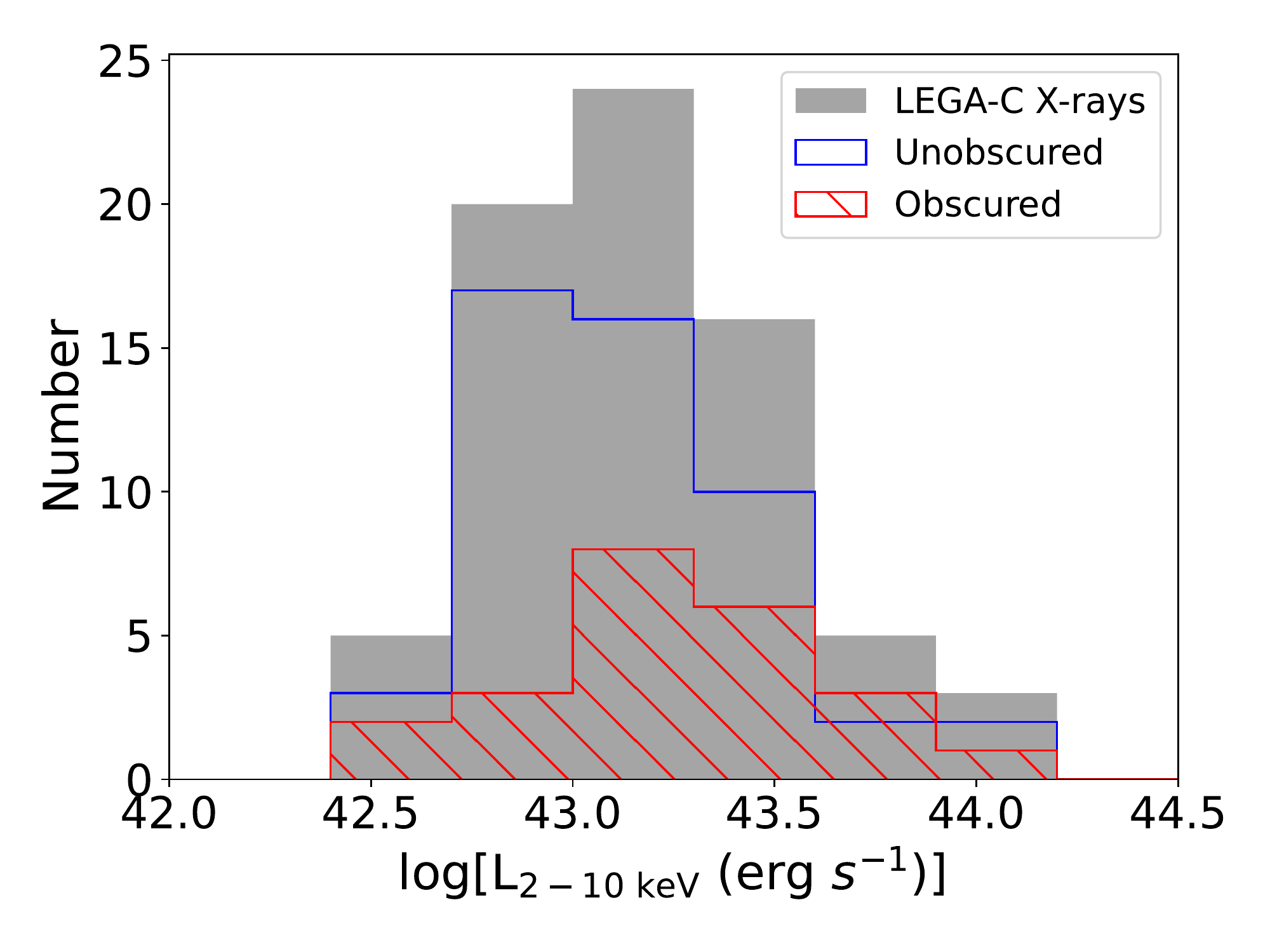}
    \end{tabular}
\caption{Distributions of redshift (left panel) and intrinsic X-ray luminosity $\rm L_X$ (right panel) for the obscured and unobscured AGN samples. The distribution of the combined  X-ray sample is over-plotted for reference.}
\label{Lzdistr}
\end{figure*}

\begin{figure}
   \begin{tabular}{c}
       \includegraphics[width=0.47\textwidth]{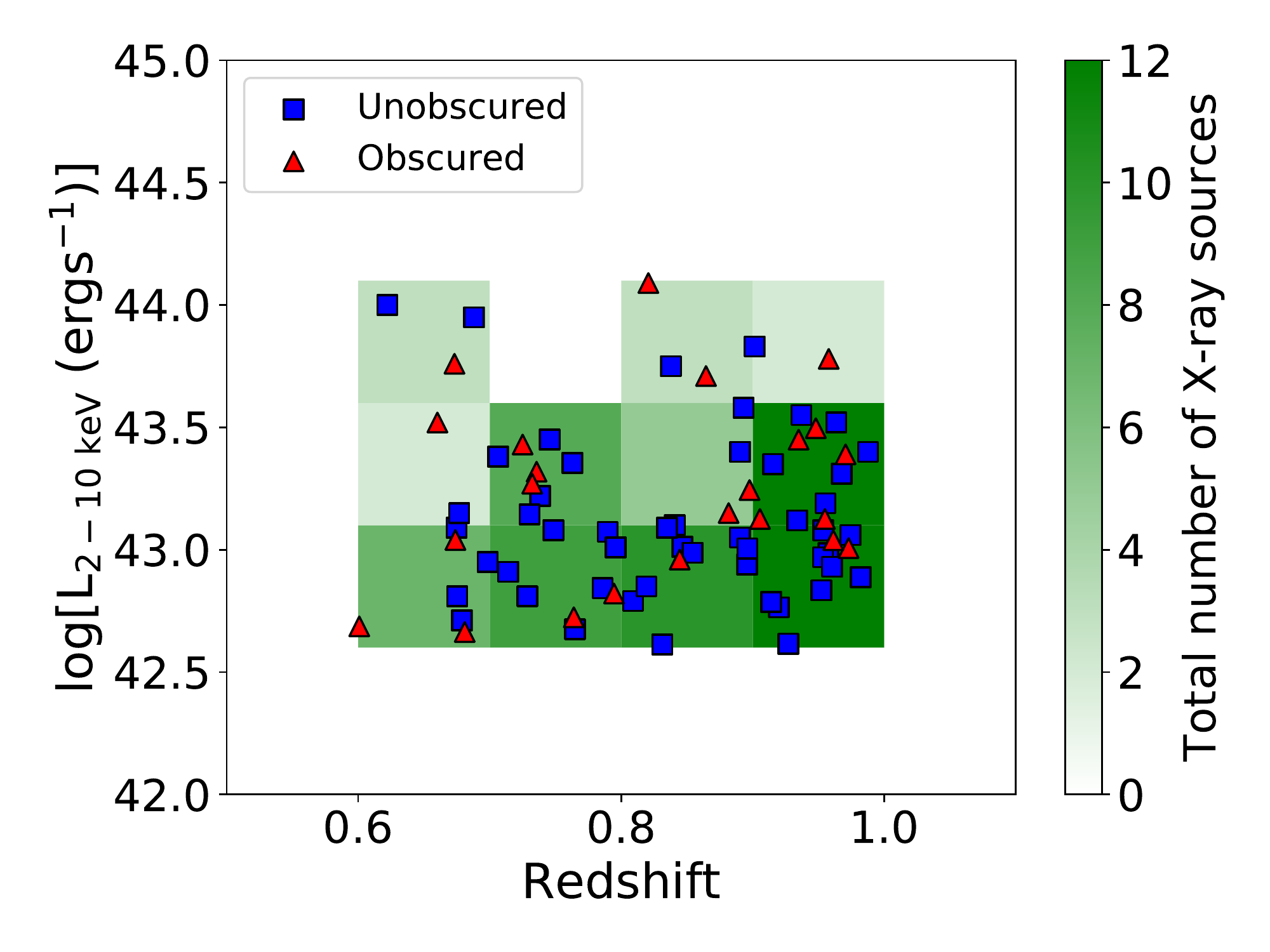}
    \end{tabular}
\caption{The distribution in the redshift and intrinsic luminosity plane for the obscured and unobscured AGN. The grid is used to assign weights 
in different $\rm z,Lx$ bins in order to take into account the different luminosity and redshift distributions of obscured and unobscured AGN (see section \ref{weights}). 
 }\label{grid}
\end{figure}

\begin{figure}
   \begin{tabular}{c}
       \includegraphics[width=0.47\textwidth]{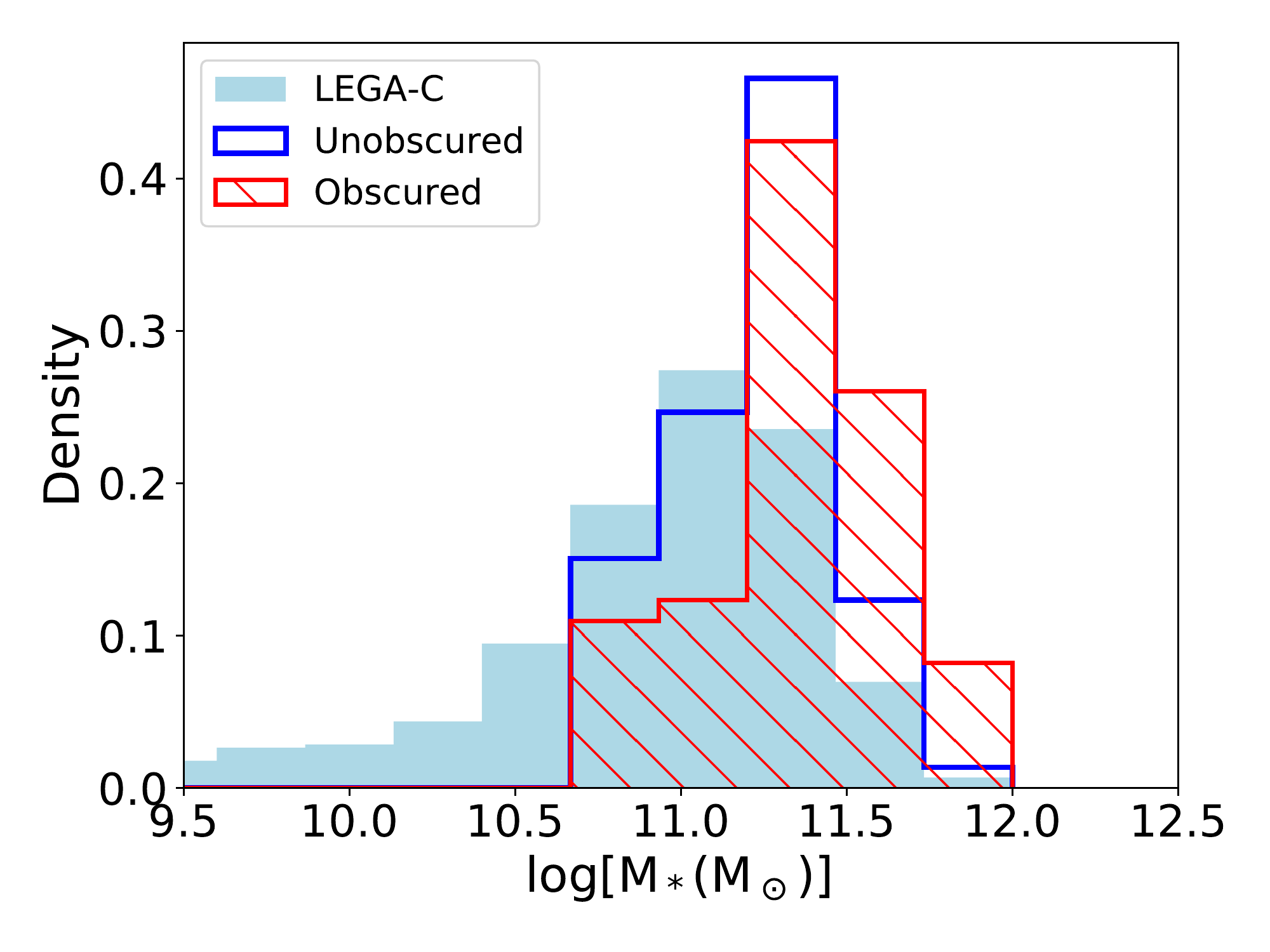} 
    \end{tabular}
\caption{The distribution of $\rm M_*$ for the obscured and unobscured AGN samples. The distribution refers to the frequency histogram so that 
the sum of all bins equals unity.
The distribution has been weighted taking into account the redshift and luminosity distributions of obscured and unobscured AGN. We over-plot the distribution of the LEGA-C galaxies for reference.}\label{Mstar}
\end{figure}

\section{Data}
\subsection{The LEGA-C survey}
The Very Large Telescope  VIMOS LEGA-C survey \citep{VanderWel2021} collects high signal-to-noise, 
high resolution spectra for thousands of galaxies in the redshift range 0.6 to 1. This allows to probe stellar populations (ages and metallicities)
and stellar kinematics (velocity dispersions) of thousands of galaxies at a look-back time of $\sim$7 Gyr. This allows to study with unprecedented 
accuracy the star-formation history of galaxies. 

The LEGA-C survey is based on the UltraVISTA 
catalogue of \cite{Muzzin2013}. This 
catalogue contains 160,070 sources 
down to K=23.4 across 1.62 sq. degrees
of the COSMOS field \citep{Scoville2007}. 
 The LEGA-C survey creates a parent sample of galaxies with (spectroscopic or photometric) redshifts in the range 0.6 < z < 1 and with 
 Ks magnitudes brighter than a redshift-dependent limit $\rm K_s= 20.7 - 7.5 \log[(1 + z)/1.8]$.
The final survey footprint covers 1.4255 square degrees, giving a total survey co-moving volume of $\rm \sim3.7 \times10^6~ Mpc^3$ between z=0.6 and z=1.0.
The VIMOS observations  produced 3029 high signal-to-noise spectra of primary targets in the 6300-8800 $\rm \AA$ 
wavelength range with a resolution of R=2500. Additional targets have been observed at either higher 
redshifts or at fainter magnitudes in the primary z$=$0.6-1 redshift band.
The strength of the Balmer absorption lines and the spectral region around
the calcium break at $\rm4000 ~\AA$ provide the most sensitive diagnostics of the 
stellar population age \citep[e.g.][]{Kauffmann2003b}. The determination of the age distribution of stellar
populations at a redshift of about 7 Gyr is one of the main products of the survey. 

\subsection{The COSMOS Chandra Legacy survey} \cite{Civano2016} present  a 4.6 Ms Chandra survey that covers 
2.2 $\rm deg^2$ of the COSMOS field. The central area has been observed with an exposure time of $\approx$ 160 ksec while the 
remaining area has exposure time of 
$\approx$ 80 ksec. The  limiting depths are 
$\rm 2.2 \times 10^{-16}$, $1.5 \times 10^{-15}$  
$\rm erg~cm^{-2}~s^{-1}$ in the soft (0.5-2 keV), hard (2- 10 keV) bands, respectively. 
The catalogue contains 4016 sources. 
\cite{Lanzuisi2017} and \cite{Marchesi2016b} provide X-ray spectral fits 
for all  sources with over 30 counts. 
For the remaining sources hardness ratios are given defined as
$HR=(H-S)/(H+S)$ where H and S are the net (background subtracted) count rates  
in the 2-7 keV and 0.5-2 keV respectively. These provide a good proxy 
of the intrinsic column density of the source. 
 As most sources have a low number of counts, \cite{Civano2016} used 
 the Bayesian Estimation 
 HR code  \citep{Park2006} which is particularly effective 
 in the low-count regime.  

\cite{Marchesi2016a} matched
the X-ray sources with optical/infrared counterparts, using the
likelihood ratio technique \cite{Sutherland_and_Saunders1992}. 97\%
of the sources have an optical/IR counterpart.
  Finally, a cross-match with the COSMOS photometric catalogue produced by the HELP collaboration \cite{Shirley2019} has been performed. 
HELP includes data from 23 of the premier extragalactic survey fields, imaged by the {\it Herschel} Space Observatory which form 
the Herschel Extragalactic Legacy Project (HELP). The catalogue provides homogeneous and calibrated multi-wavelength data. 
The cross-match with the HELP catalogue is done using 1 arcsec radius and the optical coordinates of the counterpart of each X-ray source.

  \section{Sample Selection}\label{thesample}
  
  \subsection{Optical Spectra}
  We cross-correlate the LEGA-C DR-3 spectroscopic catalogue \citep{VanderWel2021} with the coordinates of the optical
  counterparts of the COSMOS {\it Chandra} Legacy sample.
   We find 173 common sources of which 163 have optical spectra with good signal-to-noise ratio
  $>$3 per pixel (0.6$\rm \AA$). Out of these only 107 have reliable
  measurements in both the $\rm H_\delta$ and the calcium break $\rm D_n$ spectral 
  region. 
   \cite{Kauffmann2003b} have studied the $\rm H_\delta- D_n$ diagram for SDSS 
   optically selected AGN. They find that the continuously star-forming galaxies define a sequence 
   from old galaxies having large values of $\rm D_n$ and low values of $\rm H_\delta$
   moving progressively to young galaxies which have low values of $\rm D_n$ combined with high values of $\rm H_\delta$. 
   Galaxies which experienced recent star-formation bursts present high $\rm H_\delta$ values above the main sequence. 
   However, eight of our sources present low values of $\rm D_n$, placing them in the young galaxy regime, combined with 
   low values of $\rm H_\delta$ which are characteristic of old galaxies. 
   The location of these sources on the $\rm H_\delta-D_n$ diagram cannot be easily interpreted  using the star-formation history models
   in \cite{Kauffmann2003b}. Inspection of their optical spectra suggests that the Lick indices may be contaminated by broad emission features. 
   The details of these galaxies are given in table \ref{excluded}. 
   These eight sources lie outside the 99\% contours of the LEGA-C galaxy population (see the top left panel in Fig. \ref{HdD}). 
    A further eight sources have been discarded because of  poor quality spectral energy distribution fits 
    having $\rm \chi^2>5$, see section 4.1 for details. 
    
    \subsection{Photometry}
     In our analysis, we need reliable estimates of the galaxy properties via SED fitting. The vast majority  of our X-ray AGN have 
been detected in the following photometric bands u, g, r, i, z, J, H, Ks, IRAC1, IRAC2 and MIPS/24. IRAC1, IRAC2 and MIPS/24 are 
the [3.6]$\rm \mu m$, [4.5]$\rm \mu m$ and 24 $\rm \mu m$, photometric bands of {\it Spitzer}.
 However, there are seven sources which have photometric information available only in the optical band and obviously these are excluded from further analysis. Finally, we select only the sources in the redshift range z=0.6-1 and X-ray luminosity range $\rm \log L_X(erg~s^{-1})=42.6-44$. The luminosity cut 
   is dictated by the need to create obscured and unobscured AGN samples in the same luminosity range. 
   Our final sample is comprised of 73 sources. All have detections in the {\it Herschel} PACS 100 and 160$\rm \mu m$ bands while
  57 have detections in the {\it Herschel} SPIRE bands. 
    We plot the redshift and luminosity 
  distribution of our final sample in Fig. \ref{Lzdistr} while 
  in Fig. \ref{weights} we show the distribution on the $\rm L_x -z $ plane. 
  
  \subsection{X-ray absorption}
  Out of our 73 sources, 41 have reasonable quality X-ray spectra with over 30 counts while the remaining sourfsee aslces have only hardness ratios available.
  We first used the column densities given in \citet{Lanzuisi2017, Lanzuisi2018} while for the remaining sources we used the estimates 
  in \cite{Marchesi2016b}. 
  In some cases the AGN X-ray spectra may be contaminated by a non-negligible
  SFR component.
 This component originates from gas heated by supernova remnants to temperatures 
 of $\sim0.8$keV \citep[e.g.][]{Mineo2012}. The luminosity of this component could reach a few times $10^{41}$ $\rm erg~s^{-1}$ in the $\rm 0.5-2$ keV band \citep[e.g.][]{Ranalli2003}. Obviously in the case of low-luminosity AGN 
 this component may contribute a significant part of the soft X-ray emission
 and thus it could affect  the column density  estimates. 
 We have performed detailed simulations to check the effect of the SFR component 
 using {\sc XSPEC} \citep{Arnaud1996}. We model the SFR component using the {\sc APEC} model. We find that for a column density of $\rm 10^{23}$ $\rm cm^{-2}$,
 an AGN X-ray luminosity of $\rm L_{2-7keV}\approx 10^{43}$ $\rm erg~s^{-1}$
 and a SFR luminosity of $\rm L_{0.5-2keV}\sim 3\times10^{41}$ $\rm erg~s^{-1}$ the column density  estimate is not affected. This holds well even in the case of  lower column densities of $\rm N_H=10^{22} cm^{-2}$. However, in the case of even  lower column densities $\rm N_H<10^{21.5}~cm^{-2}$ in combination 
  with low X-ray AGN luminosities $\rm L_{2-7keV}< 10^{43}$ $\rm erg~s^{-1}$ and/or high  SFR component luminosities in excess of  $\rm L_{0.5-2 keV} \sim 3\times10^{41}$ $\rm erg~s^{-1}$ the column density estimates could be underestimated.
  
  We divide our sample according to their column density to unobscured or mildly obscured and heavily obscured sources with $\rm \log N_H(cm^{-2})>23$. 
  The choice of this high threshold column density is dictated by the need 
  to select only bona-fide obscured AGN where the absorption originates in the torus. 
  It has been demonstrated that lower column densities could often  be associated with large scale absorption within 
  the galaxy \citep[e.g.][]{Maiolino1995, Malkan1998, Buchner2017, Circosta2019, Malizia2020, DAmato2020, Gilli2022}. The obscuration associated with interstellar medium of the host galaxy may be more pronounced at higher redshifts \citep{Scoville2017, Tacconi2018}.
   Interestingly, \cite{Gilli2022} predicts that even at a redshift of z$\sim$1 the Galactic column density could be as high as
   $\rm \log N_H(cm^{-2})=22$. 
  There are 23 heavily obscured  and 50 unobscured or mildly obscured sources. 
Hereafter, we refer to these  as the obscured and unobscured AGN subsamples.
 Note that in the remaining of the paper when the classification is 
 based on X-ray spectroscopy or hardness ratio we refer to the objects as obscured and unobscured,
 whereas if the classification is based on optical spectroscopy we refer to the 
 objects as type-2 and type-1 respectively. 
 
 \begin{table*}
\centering
\caption{Excluded sources lying well below the main $\rm H_\delta- D_n$ sequence}
\begin{tabular}{rrrr}
\hline
 LEGA-C ID &  redshift   & $\rm D_n$ & $H_\delta$  \\
            \hline& \\[-1.5ex]
 3795     &   1.005 &  1.26 & -1.30  \\ [0.1cm]
 3589     &    0.6921     & 1.23  & -3.29   \\
 3247     &     0.9585    & 1.14  & -0.66 \\
 2236     &     0.7076    & 1.05  &  -4.23 \\
 2015     &  0.7025 & 1.20 &  -4.24   \\
 1718     &  0.7283 &  1.13 &  -2.93 \\
 939      &  0.6643 &  1.14 &  -3.48 \\
 37       &  0.9044 &  1.07  & 0.58 \\
 \hline
\end{tabular}
    \label{excluded}
\end{table*}

 \begin{figure*}
   \begin{tabular}{c c}
       \includegraphics[width=0.38\textwidth]{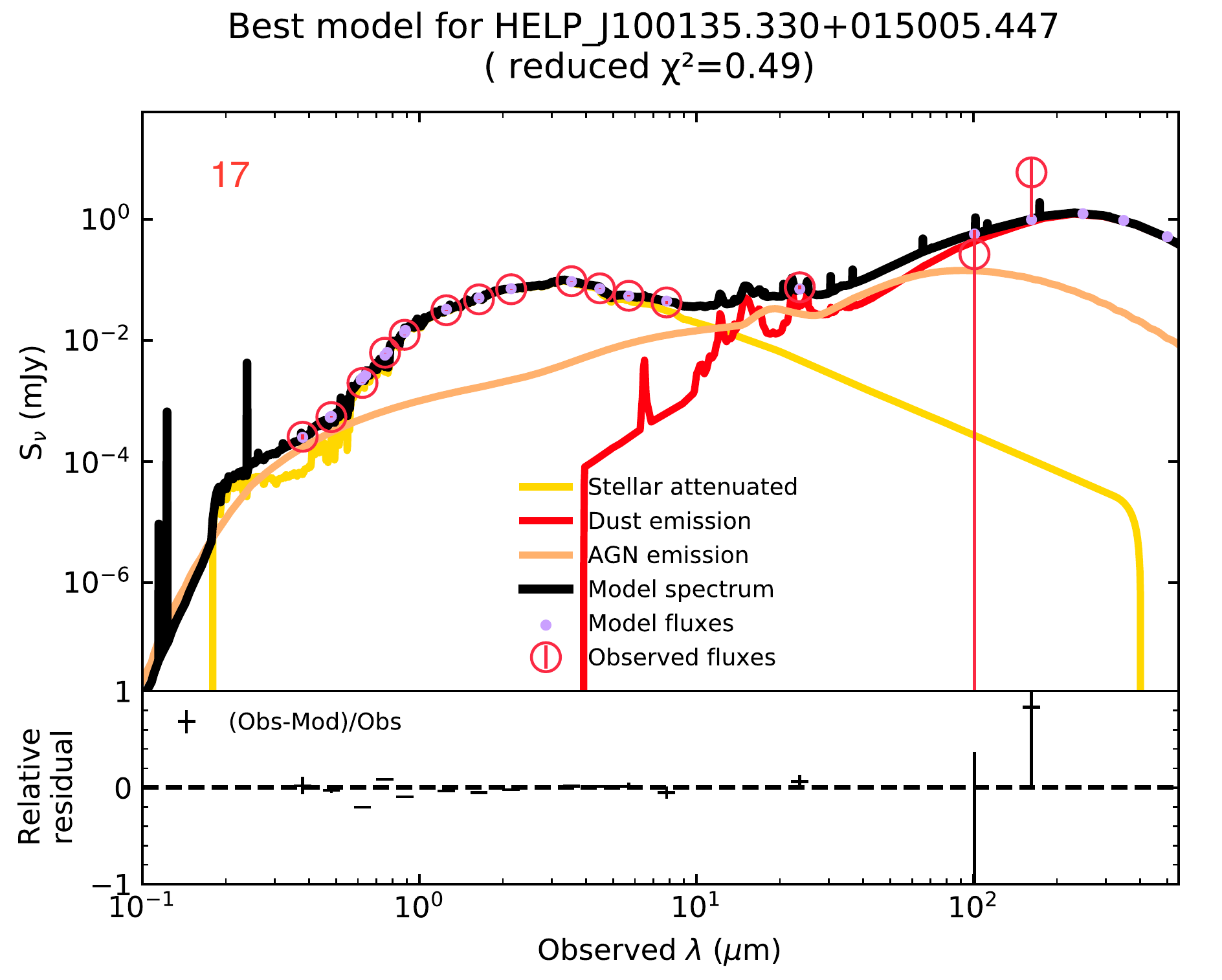} &
      \includegraphics[width=0.38\textwidth]{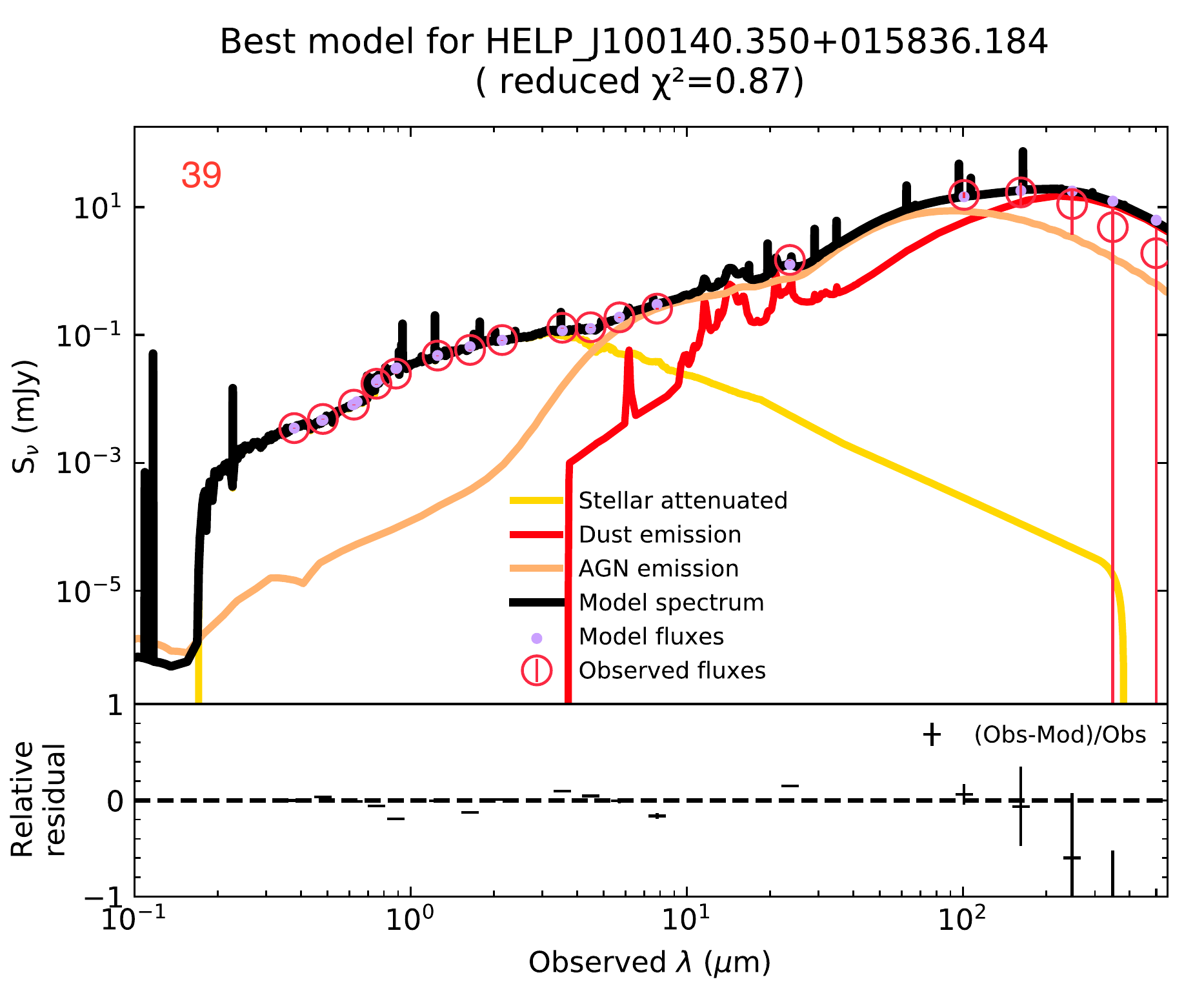} \\
        \includegraphics[width=0.38\textwidth]{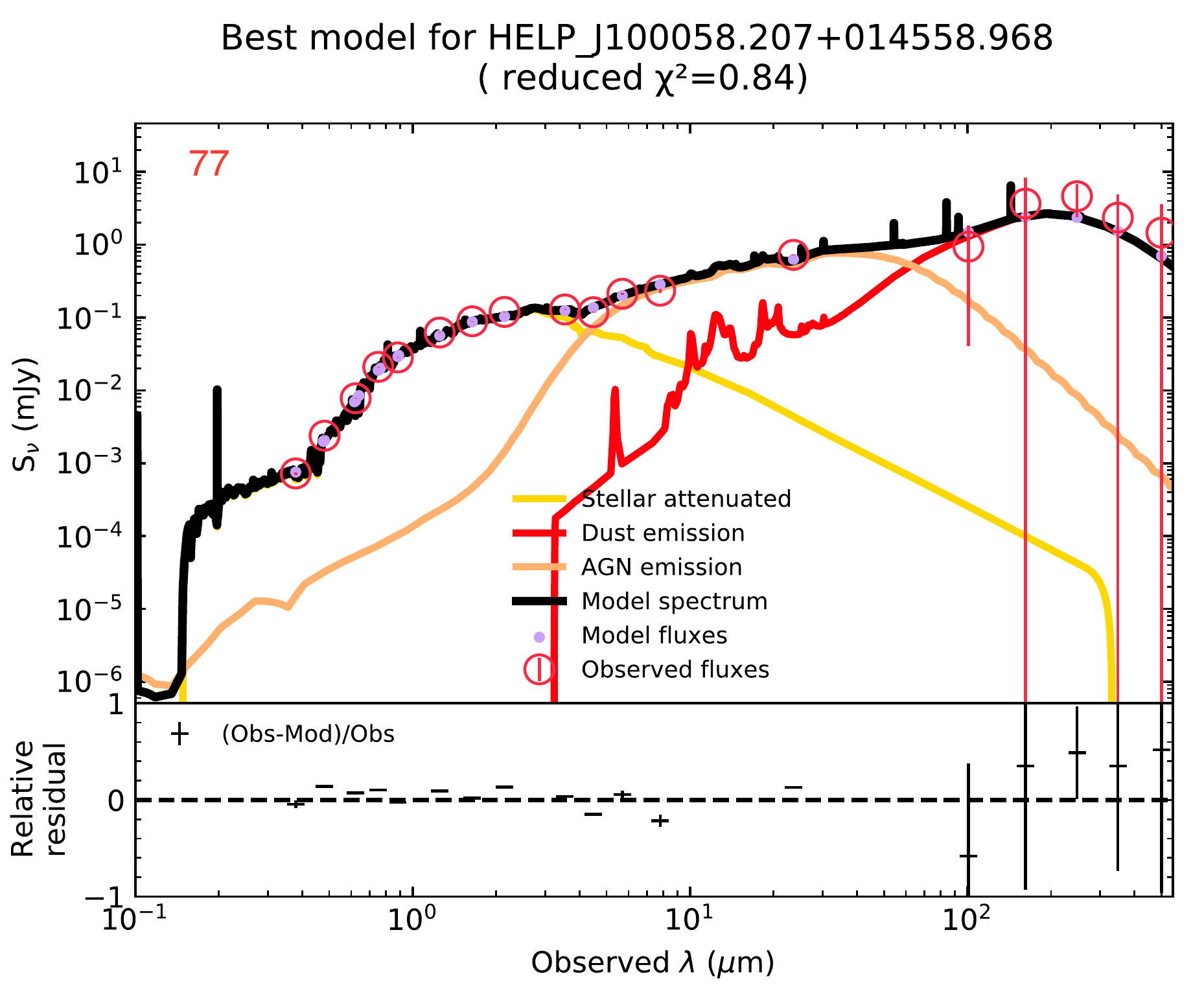} &
       \includegraphics[width=0.38\textwidth]{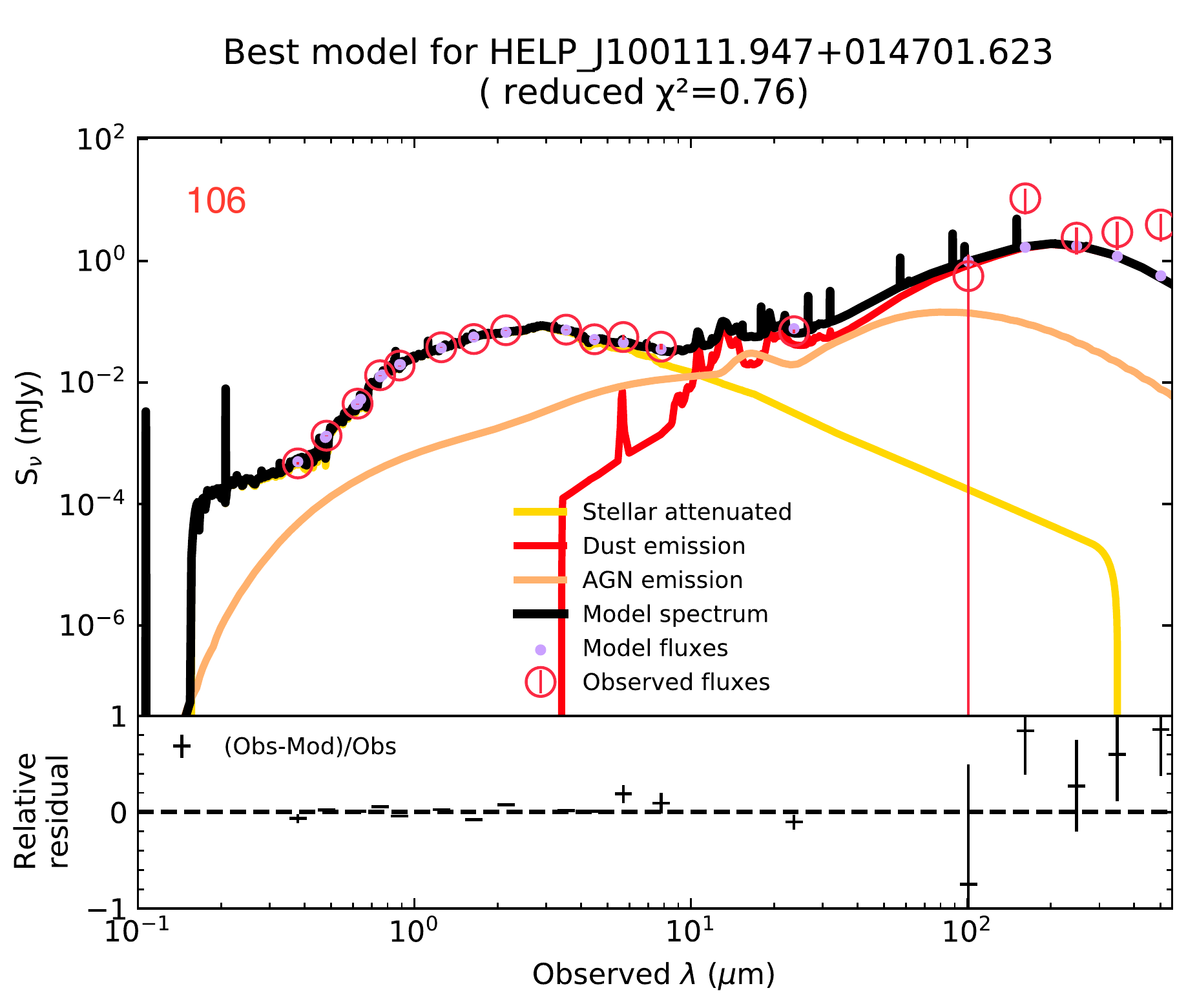} \\
       \includegraphics[width=0.38\textwidth]{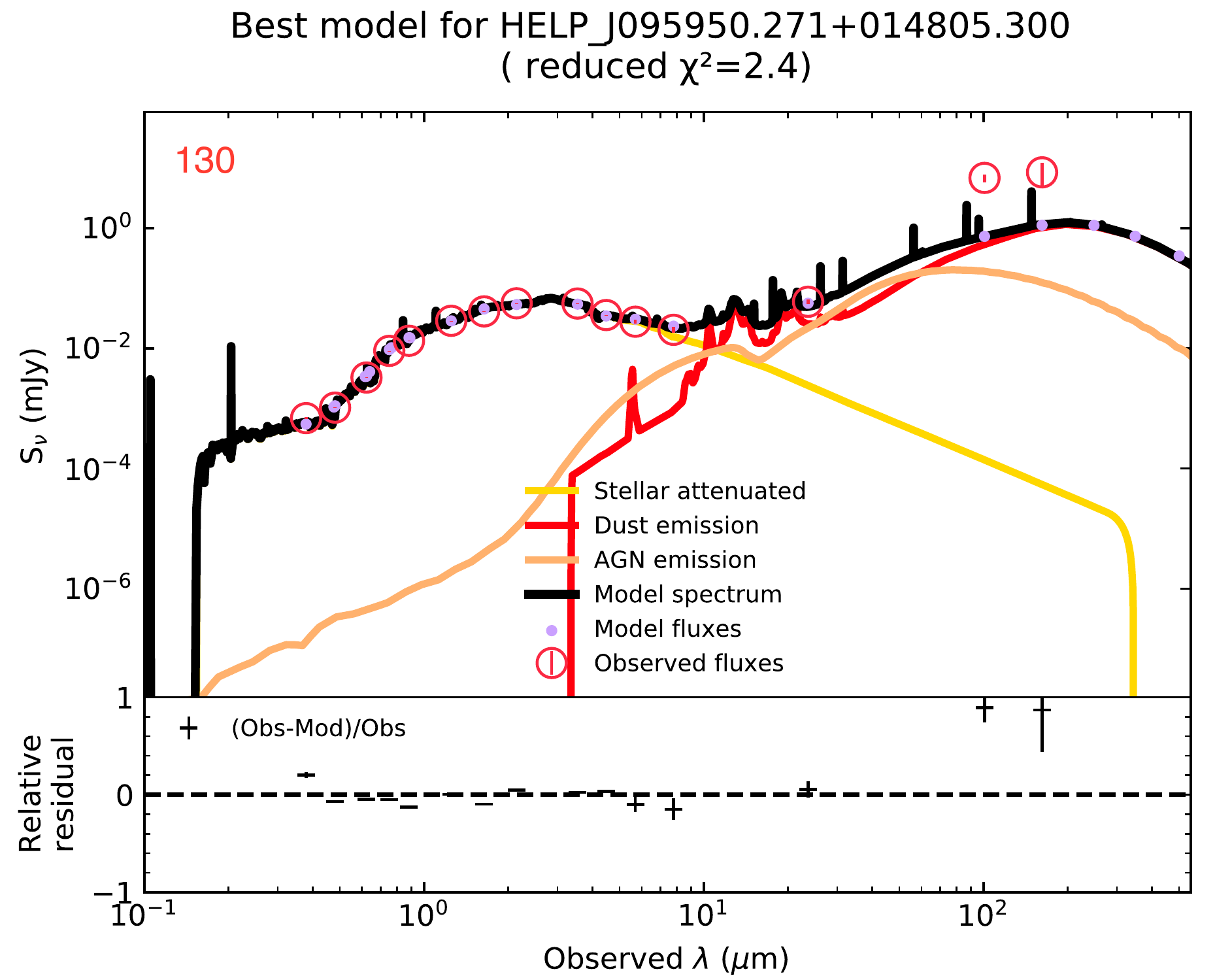} &
       \includegraphics[width=0.38\textwidth]{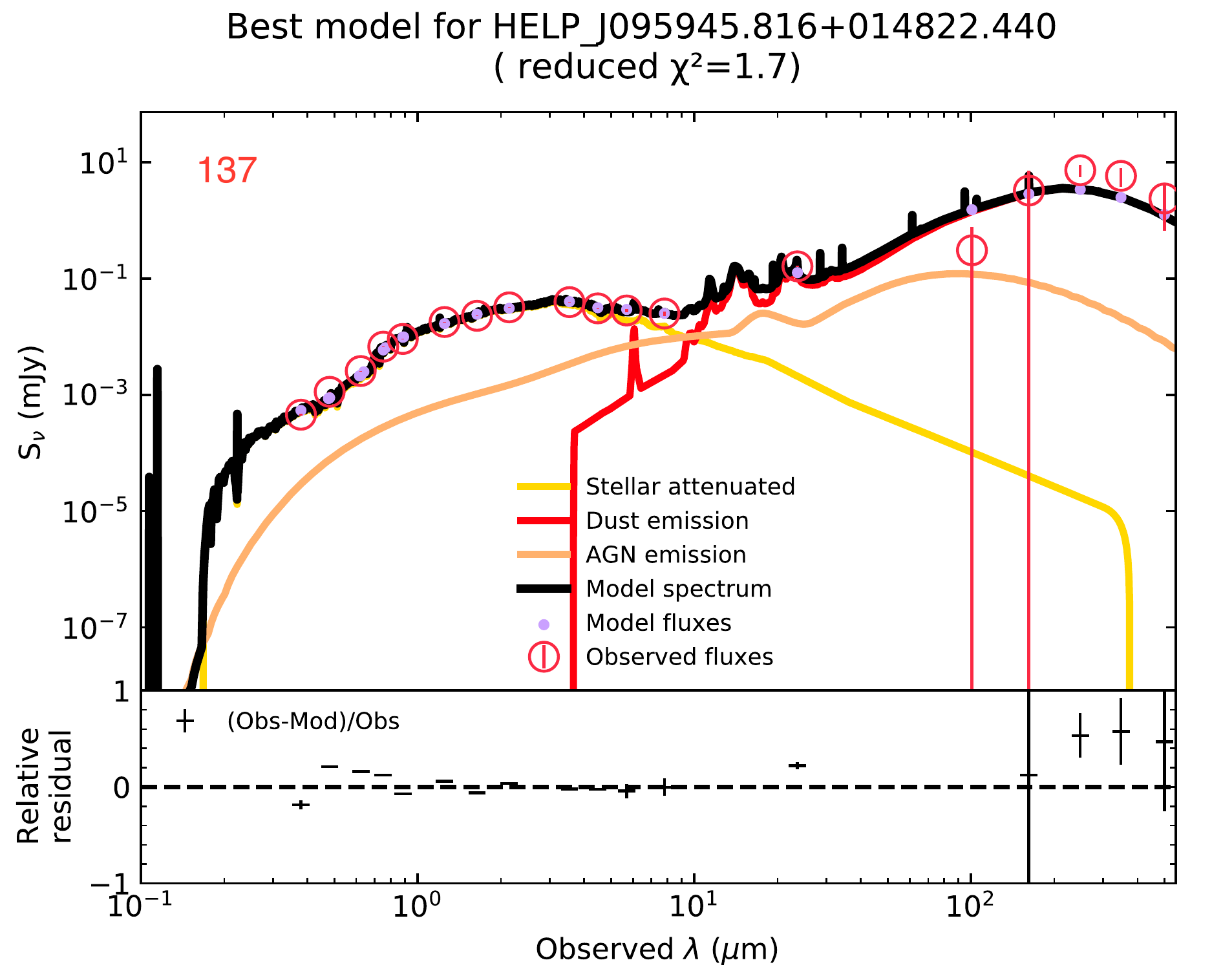} \\
       \includegraphics[width=0.38\textwidth]{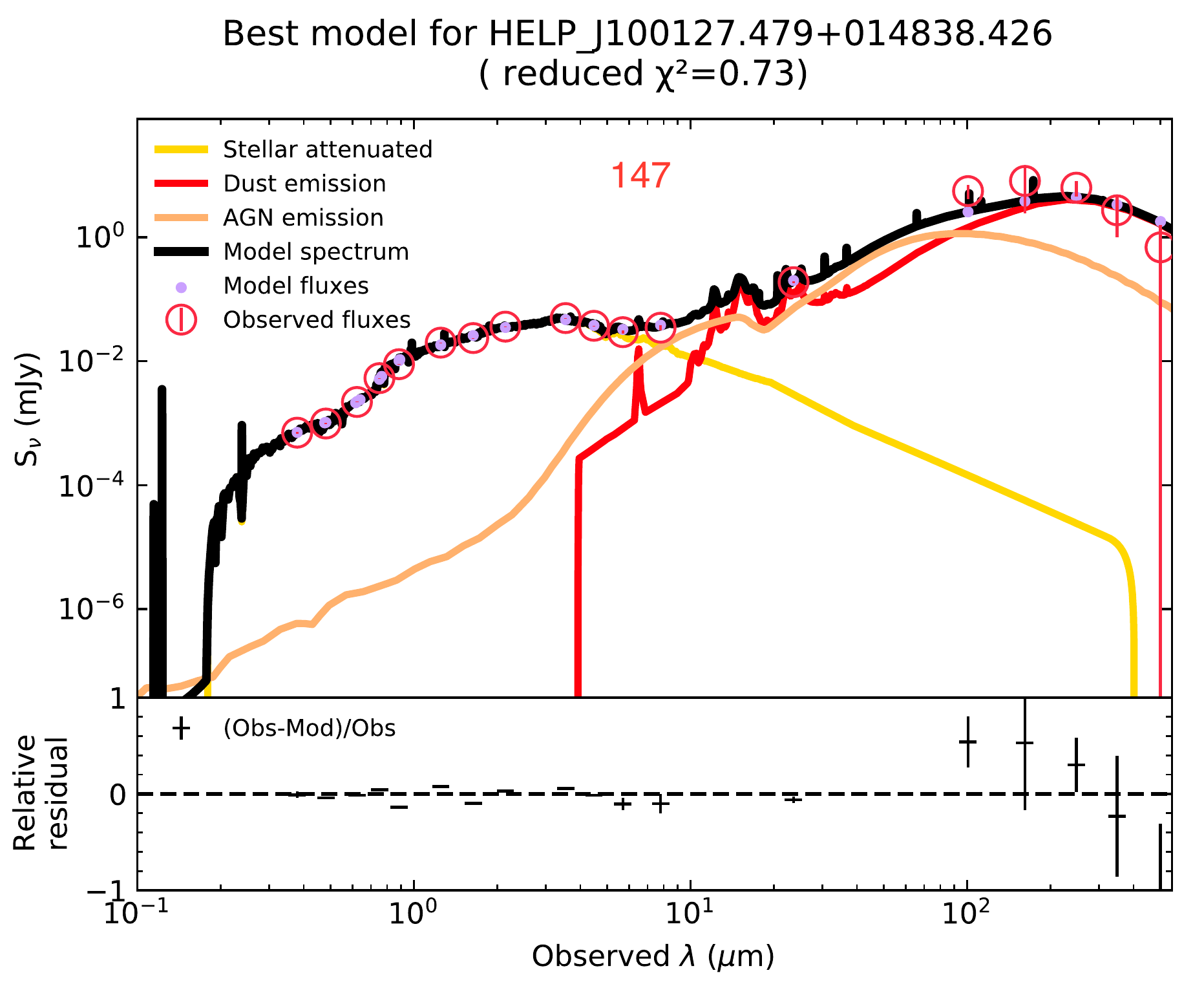} &
       \includegraphics[width=0.38\textwidth]{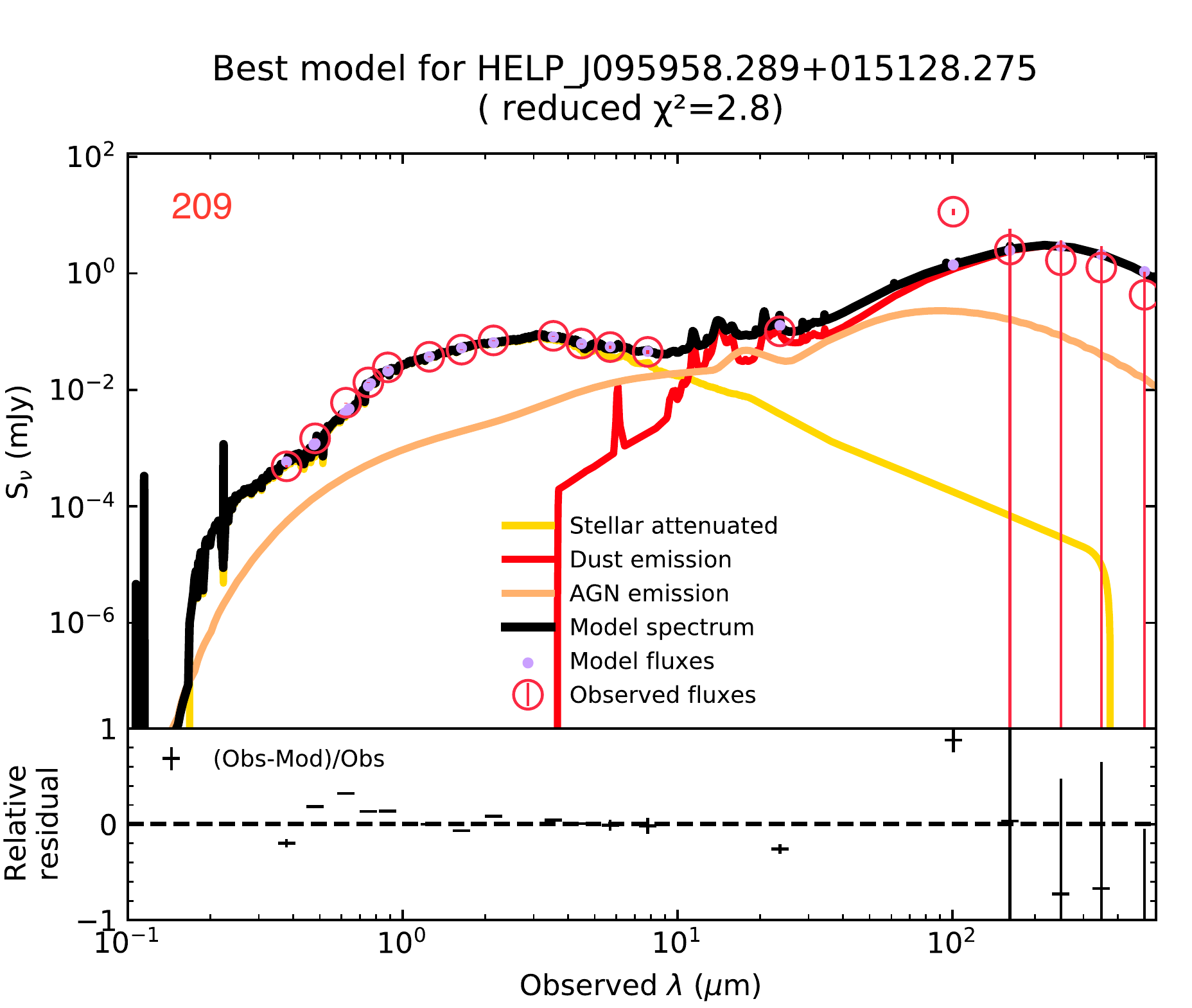}
    \end{tabular}
\caption{Example Spectral Energy Distribution fits of both obscured (17, 39, 130, 147) and unobscured (77, 106, 137, 209) AGN.}\label{exampleSED}. 
\end{figure*}
 
 \begin{figure*}
   \begin{tabular}{c c}
       \includegraphics[width=0.40\textwidth]{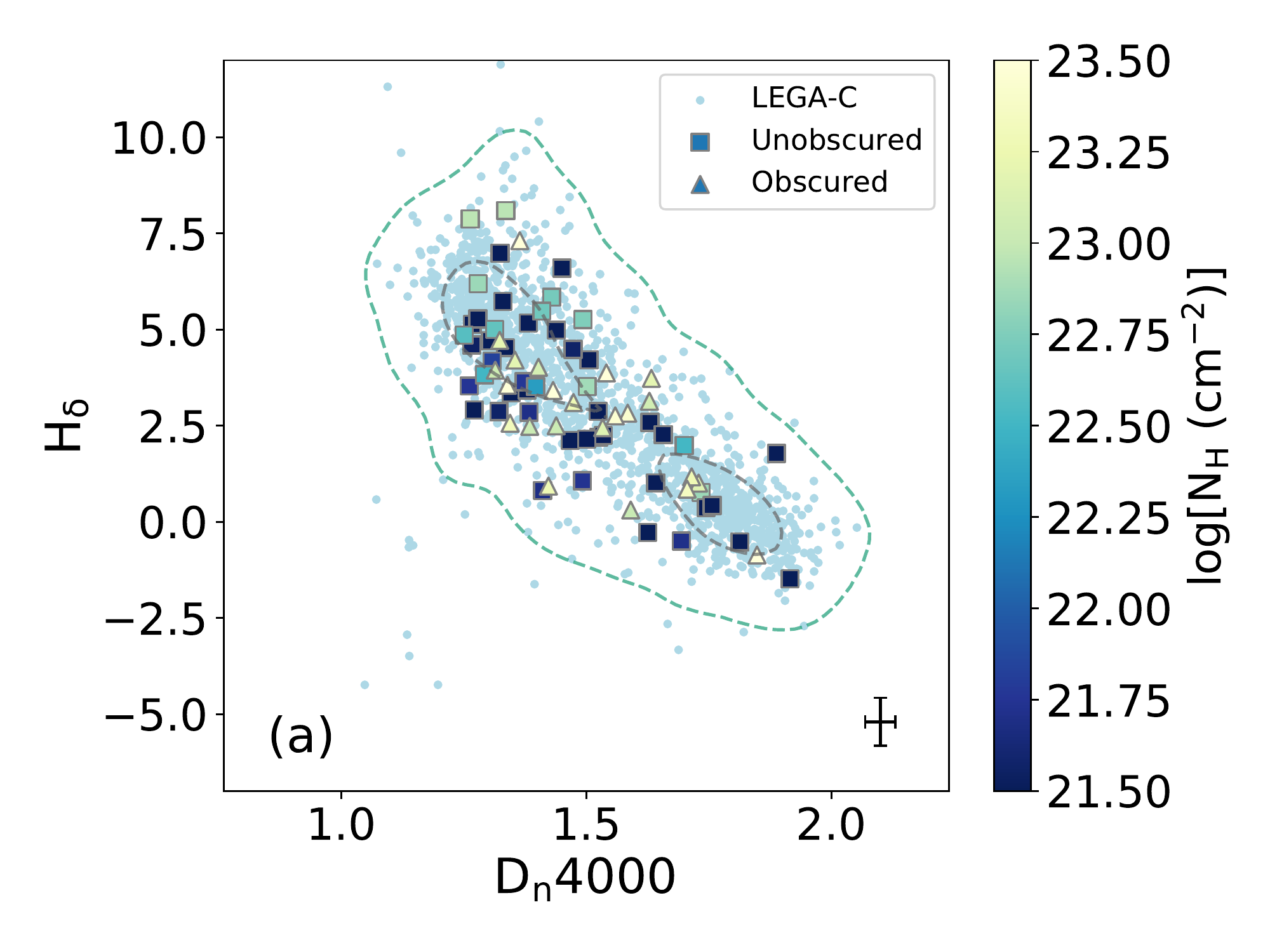} &
              \includegraphics[width=0.40\textwidth]{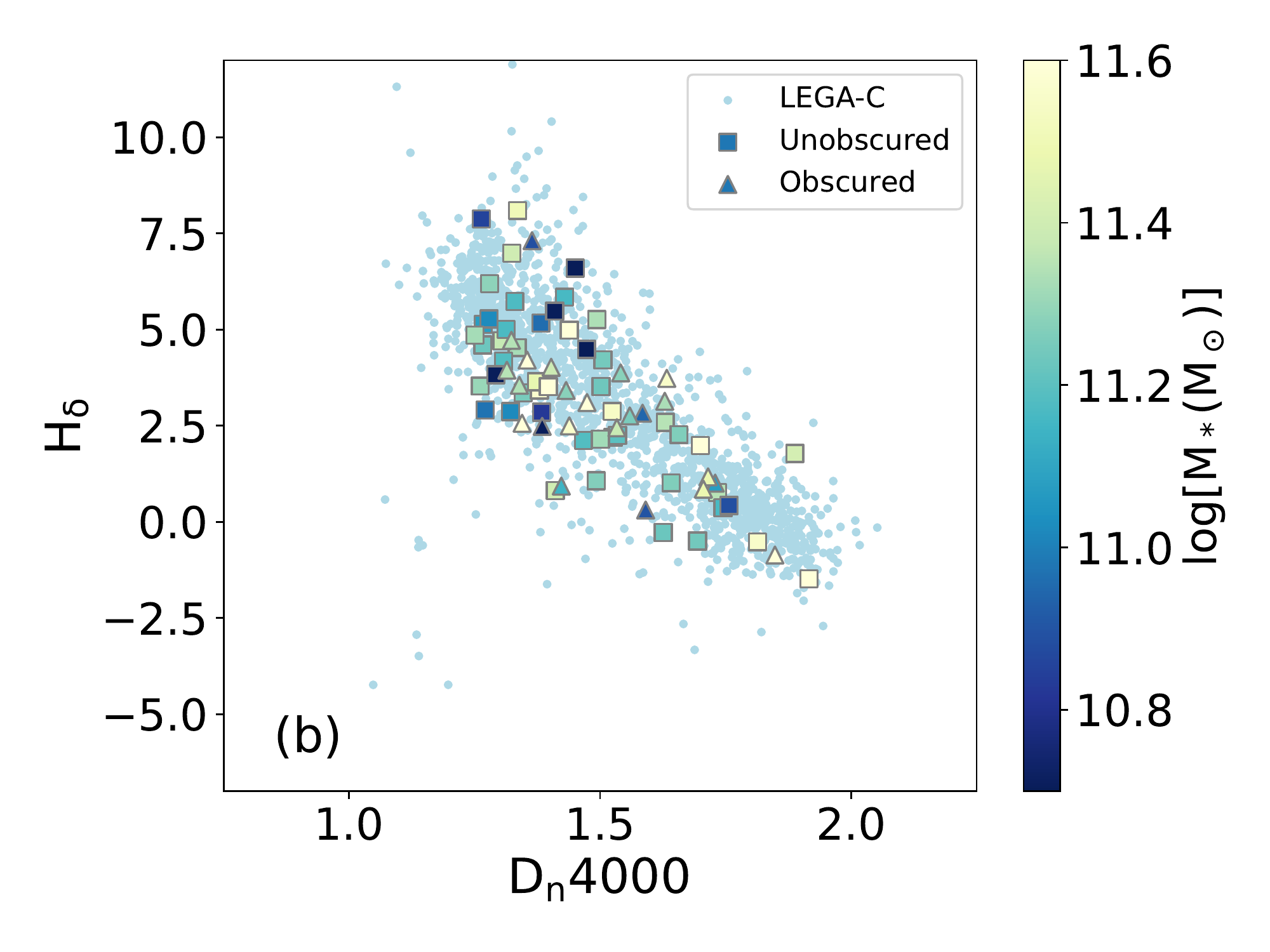} \\

       \includegraphics[width=0.40\textwidth]{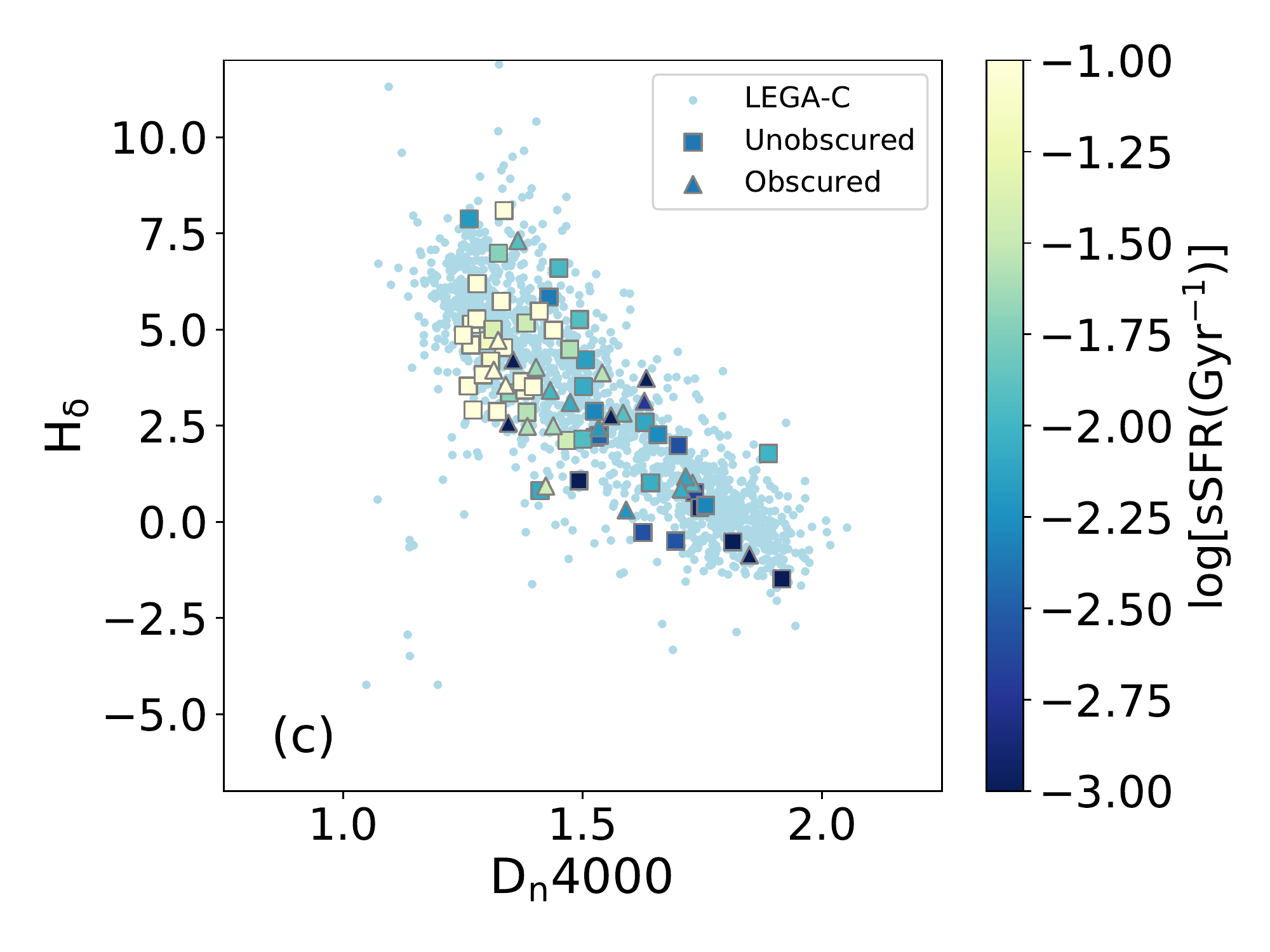} &
              \includegraphics[width=0.40\textwidth]{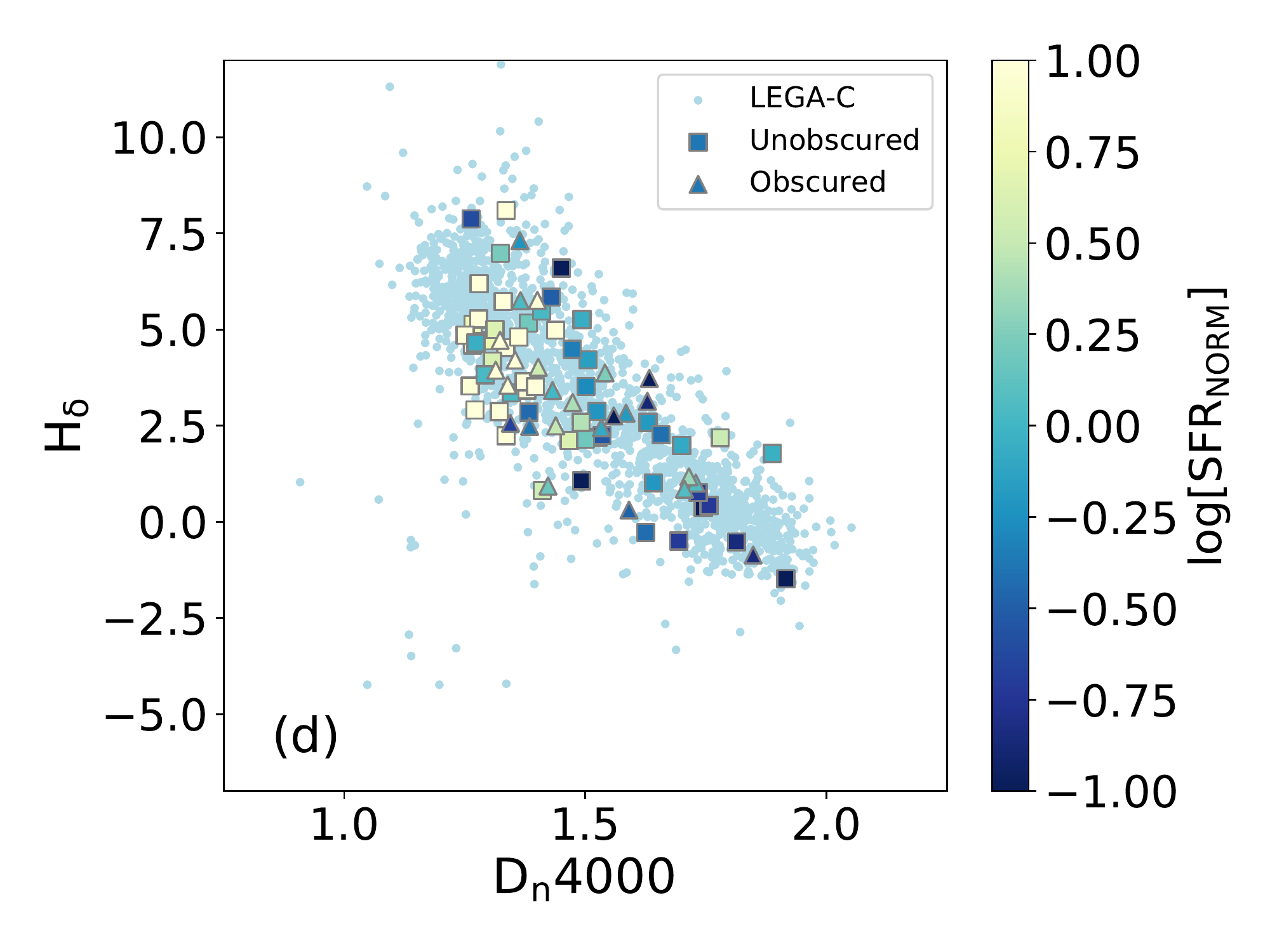} \\

       \includegraphics[width=0.40\textwidth]{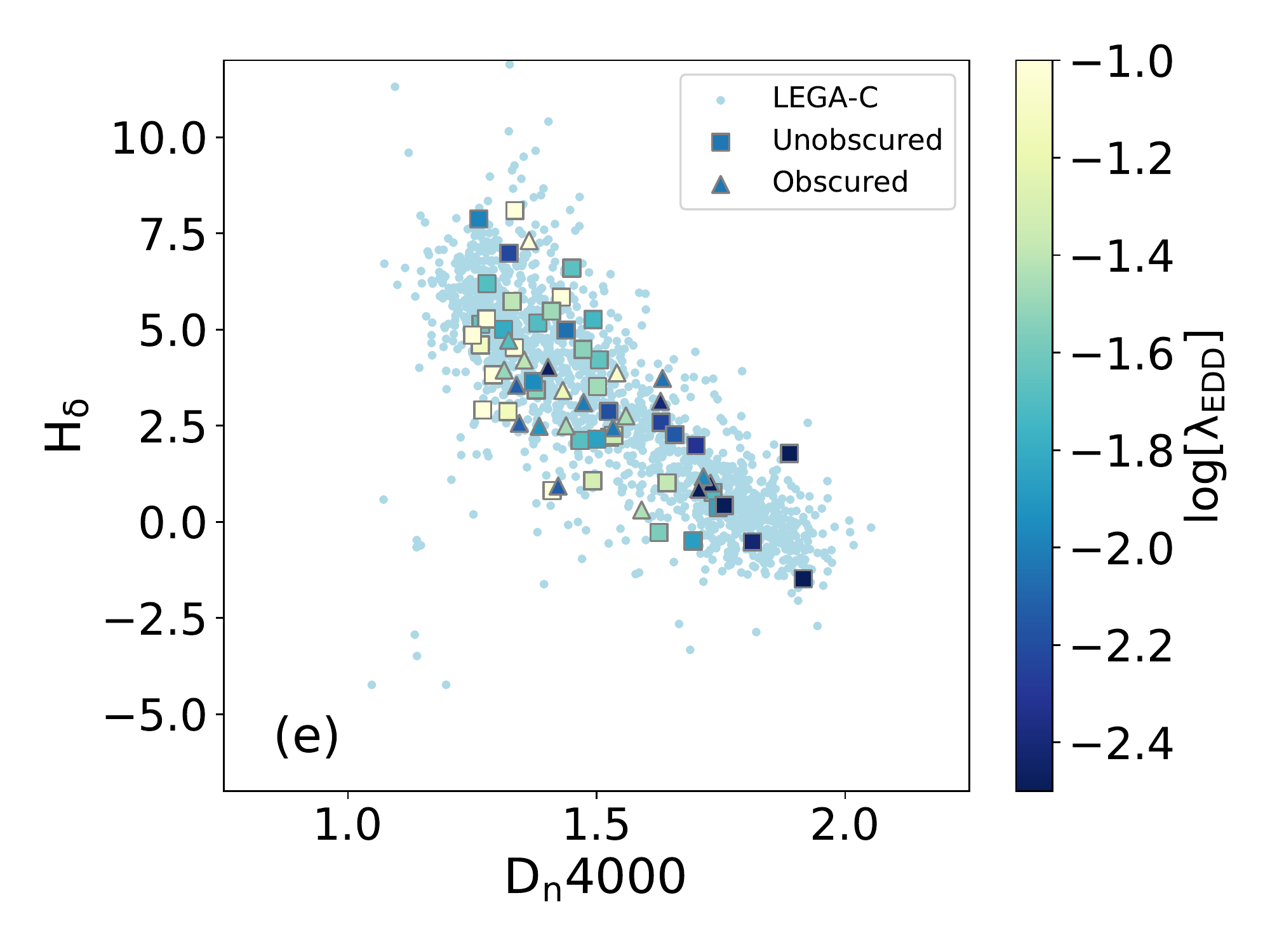} &
              \includegraphics[width=0.40\textwidth]{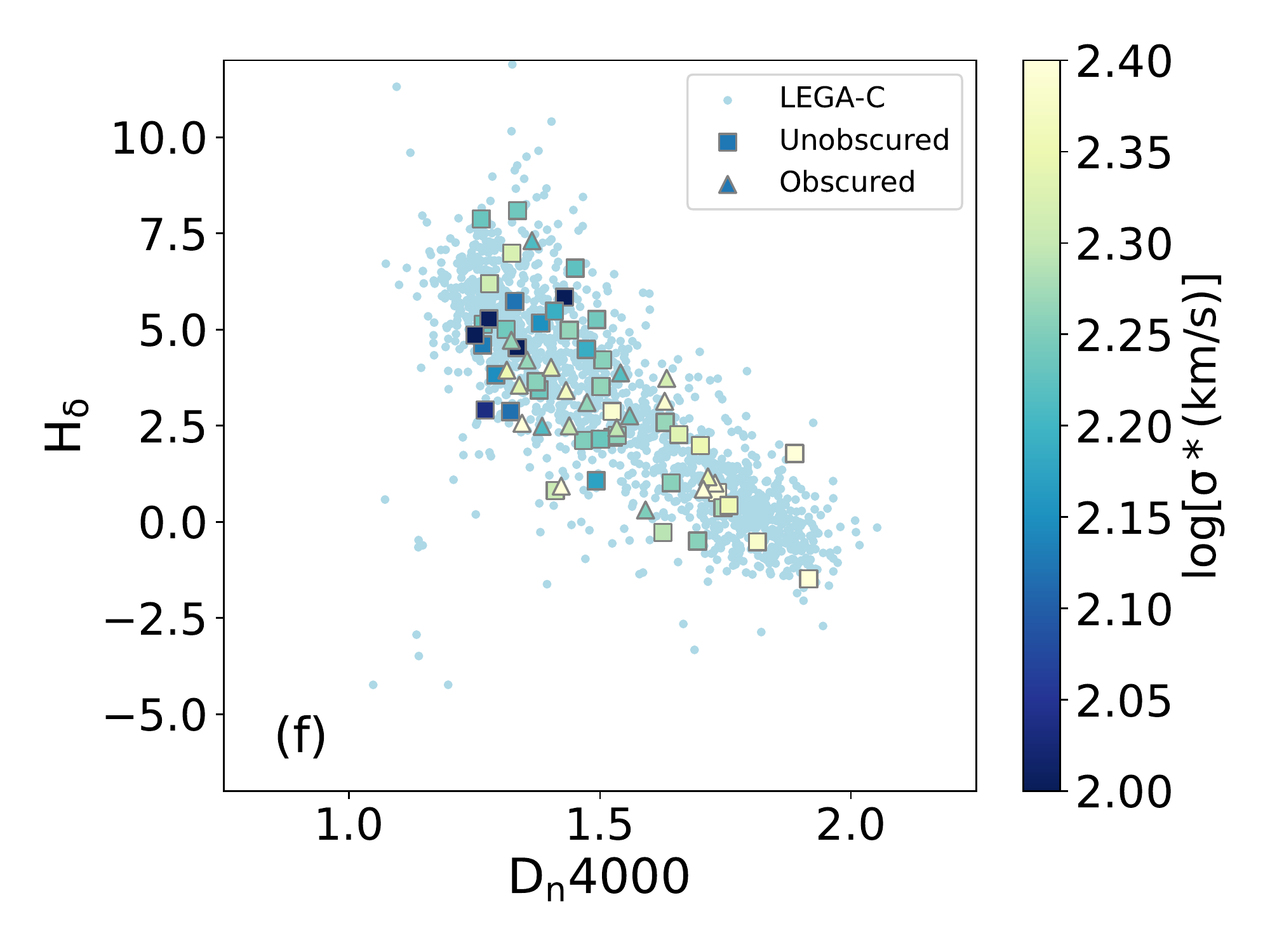} \\
    \end{tabular}
\caption{The age indicators $\rm H_\delta - D_n(4000)$ plots for our sample.
The triangles and the squares denote the obscured and the unobscured AGN respectively.
The colour-coded key gives the value of the  a) intrinsic AGN column density. 
The dotted lines gives the contours that include 68\% and 99\% of the LEGA-C galaxy population b)
stellar mass, c) specific SFR, d) normalised SFR, e) Eddington ratio, and f) stellar velocity dispersion. The obscured and unobscured AGN samples have not been corrected for differences in their luminosity and redshift distribution. }\label{HdD}. 
\end{figure*}

	\section{Analysis}
	\subsection{Spectral Energy distribution fitting using X-CIGALE}\label{sec_cigale}
	In this section we estimate the host galaxy properties, the stellar mass and the star-formation rate. 
	We measure the host galaxy properties of X-ray AGN in our sample, by applying SED fitting using the
	{\sc X-CIGALE} code \citep{Boquien2019}. In its latest version \citep{Yang2020, Yang2022}, CIGALE accounts for extinction of the ultraviolet (UV) and optical emission in the poles of the AGN. At the same time
	it takes into account the X-ray emission of the AGN in order to better constrain the torus emission. This is performed by means 
	of the $\alpha_{ox}$ index which is the spectral slope that connects the 2 keV and the 2500$\AA$ monochromatic emission. 
	For the estimation of the above index, the code requires the intrinsic X-ray fluxes, i.e., X-ray fluxes corrected for X-ray absorption.
	The improvements that these new features add in the fitting process are described in detail in \cite{Yang2020}, \cite{Mountrichas2021a}
	and \cite{Buat2021}.

For the SED fitting process, we use the intrinsic X-ray fluxes estimated in \cite{Marchesi2016b} . 
{\sc X-CIGALE} uses the $\rm \alpha _{ox}-L_{2500\AA}$ relation of \cite{Just2007} to connect the X-ray flux with the AGN emission at 2500\AA. 
We adopt a maximal value of $|\Delta \alpha _{ox}|_{max}=0.2$ that accounts for a $\approx 2\,\sigma$ scatter in the above relation. 
%Photon index, $\Gamma$, i.e., the slope of the X-ray spectrum, is set to 1.4 i.e. the value assumed in the \cite{Marchesi2016b} 
%catalogue for their luminosity estimations.

In the SED fitting analysis, we use the same grid used in \cite{Mountrichas2022a} in the COSMOS field. 
This allows us to provide  a better comparison with the results presented in the above study.
Here, we only summarise the modules included in that work.
A delayed star formation history (SFH) model with a function form  is used to fit the galaxy component. The model includes a star formation burst in the form of an ongoing star formation no longer than 50 Myr \citep{Buat2019}. The 
\cite{Bruzual_Charlot2003} single stellar population template is used to model the stellar emission. Stellar emission 
is attenuated following \cite{Charlot_Fall_2000}. The dust heated by stars is modelled following \cite{Dale2014}. 
The {\sc SKIRTOR}  template \cite{Stalevski2012}, \cite{Stalevski2016} is used for the AGN emission. 
{\sc SKIRTOR} assumes a clumpy two-phase torus model, based on 3D radiation-transfer. 
All free parameters used in the SED fitting process and their input values, are presented in Table 
\ref{table_cigale}. In Fig. \ref{exampleSED}, we present example SED fits for four obscured and four unobscured AGN.

\subsection{Quality examination}
\label{quality}
\subsubsection{Poor SED fits}
We exclude  SEDs with poor fits, in order to constrain our analysis only to the sources with reliable host galaxy measurements. For that purpose, we consider only sources for which the reduced $\rm \chi ^2$, $\rm \chi ^2_r<5$. This value has been used in previous studies \citep[e.g.][]{Masoura2018, Buat2021} and is based on visual inspection of the SEDs. Following this criterion, we  have excluded seven sources from further analysis.

\subsubsection{SFR}
As \cite{Zou2019} pointed out, 
for sources undetected by {\it Herschel}, SFRs derived from SED fitting may be affected by contamination from AGN emission at UV to optical wavelengths. This contamination could be more significant in the case of type-1 AGN. In the \cite{Zou2019} sample, only 32\% of the sources had detection in at least one {\it Herschel} band. They derived SFR estimates by using single band {\it Herschel} photometry and they compared with the SED SFR. They found that although the single band SFRs are overestimated by a factor of a few compared to the SED fits, there is no systematic difference between type-1 and type-2 AGN.
 
\cite{Mountrichas2022a}, used data from the COSMOS  fields and showed that the lack of far-IR {\it Herschel} photometry (both PACS and SPIRE) does not affect the SFR calculations of {\sc X-CIGALE}.  They used 742 AGN (60\% of their total X-ray sample) that have been detected by {\it{Herschel}}. For these sources, they performed SED fitting with and without {\it{Herschel}} bands, using the same parametric space. The results are shown in their Fig. 4. The mean difference of the log(SFR) measurements is 0.01 and the dispersion is $\sigma=0.25$. 

In our case, the far-IR photometry available is way more solid. 
As described in section \ref{thesample},  all of our sources have PACS photometry available, while there are only 23 sources which lack {\it Herschel} SPIRE photometry. As
an additional check we examine the CIGALE SFR Bayesian errors. 
 We found that the median relative SFR errors (error/measurement) are 0.53 and 0.78 for the unobscured and obscured AGN, respectively.  
This further supports the fact that our SFR measurements 
are not contaminated by UV and optical light in the case of unobscured AGN.
 We note that the median logSFR is 0.41 and 0.18 for the unobscured and unobscured AGN. The corresponding median SFR errors are 0.36 and 0.21 respectively.

\begin{table*}
\caption{The models and the values for their free parameters used by {\sc X-CIGALE} for the SED fitting.} 
\centering
\setlength{\tabcolsep}{1.mm}
\begin{tabular}{cc}
       \hline
Parameter &  Model/values \\
	\hline
\multicolumn{2}{c}{Star formation history: delayed model and recent burst} \\
Age of the main population & 1500, 2000, 3000, 4000, 5000, 7000, 10000, 12000 Myr \\
e-folding time & 200, 500, 700, 1000, 2000, 3000, 4000, 5000 Myr \\ 
Age of the burst & 50 Myr \\
Burst stellar mass fraction & 0.0, 0.005, 0.01, 0.015, 0.02, 0.05, 0.10, 0.15, 0.18, 0.20 \\
\hline
\multicolumn{2}{c}{Simple Stellar population: Bruzual \& Charlot (2003)} \\
Initial Mass Function & Chabrier (2003)\\
Metallicity & 0.02 (Solar) \\
\hline
\multicolumn{2}{c}{Galactic dust extinction} \\
Dust attenuation law & Charlot \& Fall (2000) law   \\
V-band attenuation $A_V$ & 0.2, 0.3, 0.4, 0.5, 0.6, 0.7, 0.8, 0.9, 1, 1.5, 2, 2.5, 3, 3.5, 4 \\ 
\hline
\multicolumn{2}{c}{Galactic dust emission: Dale et al. (2014)} \\
$\alpha$ slope in $dM_{dust}\propto U^{-\alpha}dU$ & 2.0 \\
\hline
\multicolumn{2}{c}{AGN module: SKIRTOR)} \\
Torus optical depth at 9.7 microns $\tau _{9.7}$ & 3.0, 7.0 \\
Torus density radial parameter p ($\rho \propto r^{-p}e^{-q|cos(\theta)|}$) & 1.0 \\
Torus density angular parameter q ($\rho \propto r^{-p}e^{-q|cos(\theta)|}$) & 1.0 \\
Angle between the equatorial plan and edge of the torus & $40^{\circ}$ \\
Ratio of the maximum to minimum radii of the torus & 20 \\
Viewing angle  & $30^{\circ}\,\,\rm{(type\,\,1)},70^{\circ}\,\,\rm{(type\,\,2)}$ \\
AGN fraction & 0.0, 0.1, 0.2, 0.3, 0.4, 0.5, 0.6, 0.7, 0.8, 0.9, 0.99 \\
Extinction law of polar dust & SMC \\
$E(B-V)$ of polar dust & 0.0, 0.2, 0.4 \\
Temperature of polar dust (K) & 100 \\
Emissivity of polar dust & 1.6 \\
\hline
\multicolumn{2}{c}{X-ray module} \\
AGN photon index $\Gamma$ & 1.4 \\
Maximum deviation from the $\alpha _{ox}-L_{2500 \AA}$ relation & 0.2 \\
LMXB photon index & 1.56 \\
HMXB photon index & 2.0 \\
\hline
Total number of models (X-ray/reference galaxy catalogue) & 313,632,000/60,984,000 \\
\hline
\label{table_cigale}
\end{tabular}
\tablefoot{For the definition of the various parameter see section \ref{sec_cigale}.}
\end{table*}

%\begin{figure}
%   \begin{tabular}{c}
%       \includegraphics[width=0.47\textwidth]{plot_hist_HdNORM.pdf} 
%    \end{tabular}
%\caption{The distribution of $\rm H_\delta$ age %indicator for the obscured and unobscured AGN %samples as indicated in the legend. The %distributions have been weighted on order to %follow the same X-ray luminosity and redshift %distribution (see section \ref{sec_cigale} for %details). We over-plot the distribution of the %LEGA-C galaxies for reference}\label{Hd}
%\end{figure}

\subsection{Derivation of the normalised star-formation}
 A widely applied method to compare the SFR of AGN with that of galaxies, is to use analytical expressions from the literature. \cite{Schreiber2015} describe the SFR-$\rm M_\star $correlation, known as the main sequence \citep{Elbaz2007, Speagle2014}. 
 The estimated parameter is the $\rm SFR_{NORM}$, defined as the ratio of the SFR of an AGN with $\rm M_\star$
 at redshift z relative to the SFR of normal galaxies at the same stellar mass and redsift. 
 Results from studies that followed this approach \cite[e.g.][]{Mullaney2015, Masoura2018, Bernhard2019, Masoura2021, Torbaniuk2021} 
 may suffer from systematics. These could be 
 caused by the fact that different methods are applied for the estimation of the host galaxy properties (SFR, $\rm M_\star$) of AGN 
 and  galaxies.

 A more advanced approach, is to compare the SFR of AGN with that from a control galaxy, i.e., a non-AGN sample that has been
 selected by applying the same criteria (e.g. photometric coverage) as the AGN sample and for which the galaxy properties 
 have been calculated following the same method (e.g. SED fitting). This method has been applied by \cite{Shimizu2015, Shimizu2017} 
 in hard (14-195 keV) X-ray selected AGN from the Swift Burst Alert Telescope (BAT) and more recently by \cite{Mountrichas2021c, Mountrichas2022a, Mountrichas2022b}.
  The limitation of this method is the available number of reference galaxies at a given 
  $\rm M_\star, z$ bin \citep{Pouliasis2022}.
  Here, we compare the SFR of X-ray AGN with that of normal galaxies. We apply the same SED fitting analysis in both datasets and we require the availability of the same photometric bands. The reference sample  is drawn from the HELP catalogue \citep{Shirley2019}.  
  There are about $\sim 2.5$ million galaxies in the COSMOS field of which $\sim 500,000$ are in the 1.38\,deg$^2$ of UltraVISTA \citep[see also][]{Laigle2016}. There are $\sim 230,000$ galaxies, after we exclude X-ray sources, that meet the photometric requirements we have set on the X-ray sample.

  We compare the SFR of X-ray AGN with that of galaxies, using the SFR$_{norm}$ parameter. We derive the SFR$_{norm}$ following the method described in \cite{Mountrichas2022a}. In more detail, the SFR of each X-ray source is divided by the SFR of galaxies from the reference catalogue. These galaxies are selected to  have stellar mass that differs $\pm 0.1$\,dex from the stellar mass of the AGN and is found within $\pm 0.075 \times (1 + \rm z)$ from the X-ray source. The median value of these ratios is used as the SFR$_{norm}$ of each AGN. In these calculations, each source is weighted based on the uncertainty on the SFR and M$_*$. 
  The most reliable estimates refer to X-ray sources for which the SFR$_{norm}$ has been derived using at least 20 galaxies from the reference catalogue. However, this necessarily lowers to less than ten galaxies when we consider only the most massive systems ($\rm 11.5< log\,[M_*(M_\odot)] < 12.0)$.
 
 \subsection{Black Hole Masses and Eddington Ratios}
 One of the main measurements coming from the high resolution LEGA-C spectra is the velocity dispersion
 \citep{VanderWel2016}. These have an error of only 0.08 dex. Eight  
 sources (6 unobscured and two obscured AGN) do not have 
  a measurement of the velocity dispersion available and hence are excluded from the Eddington ratio analysis.   The velocity dispersions can provide 
  a good proxy of the black hole mass $\rm M_{BH}$. 
   \cite{Ferrarese_Merritt2000, Gebhardt2000} found  a strong correlation between the black hole mass
   and the stellar velocity dispersion. 
  \cite{Merritt_Ferrarese2001} proposed the following relation between the above quantities.
\begin{equation}\label{eq_bhmass}
 \rm M_{BH}= 1.3\times10^{8} ~\sigma_{200}^{4.72} ~M_\odot
\end{equation}
  where $\sigma_{200}$ is the stellar velocity dispersion in units of 200$\rm ~km~s^{-1}$.

  The Eddington luminosity is the maximum luminosity that can be emitted by the AGN. This is defined by the balance between the radiation pressure and the gravitational force exerted  by the black hole. 
\begin{equation}\label{eq_edd}
\rm L_{EDD} \approx 1.3\times 10^{38} ~(M_{BH}/M_\odot)~erg~s^{-1}.
\end{equation}

The Eddington ratio is defined as the ratio 
  of the bolometric luminosity and the Eddington luminosity, $\rm \lambda_{EDD}= L_{BOL}/L_{EDD}$. 
  The bolometric luminosity 
  is usually determined from the X-ray luminosity \citep{Marconi2004, Vasudevan2013, Lusso2012}.
  We use here the most recent relation calibrated in \cite{Duras2020}. These bolometric corrections take into account both the X-ray luminosity and they differentiate between obscured and unobscured AGN.

\begin{figure*}
\center
   \begin{tabular}{c c}
    \includegraphics[width=0.47\textwidth]{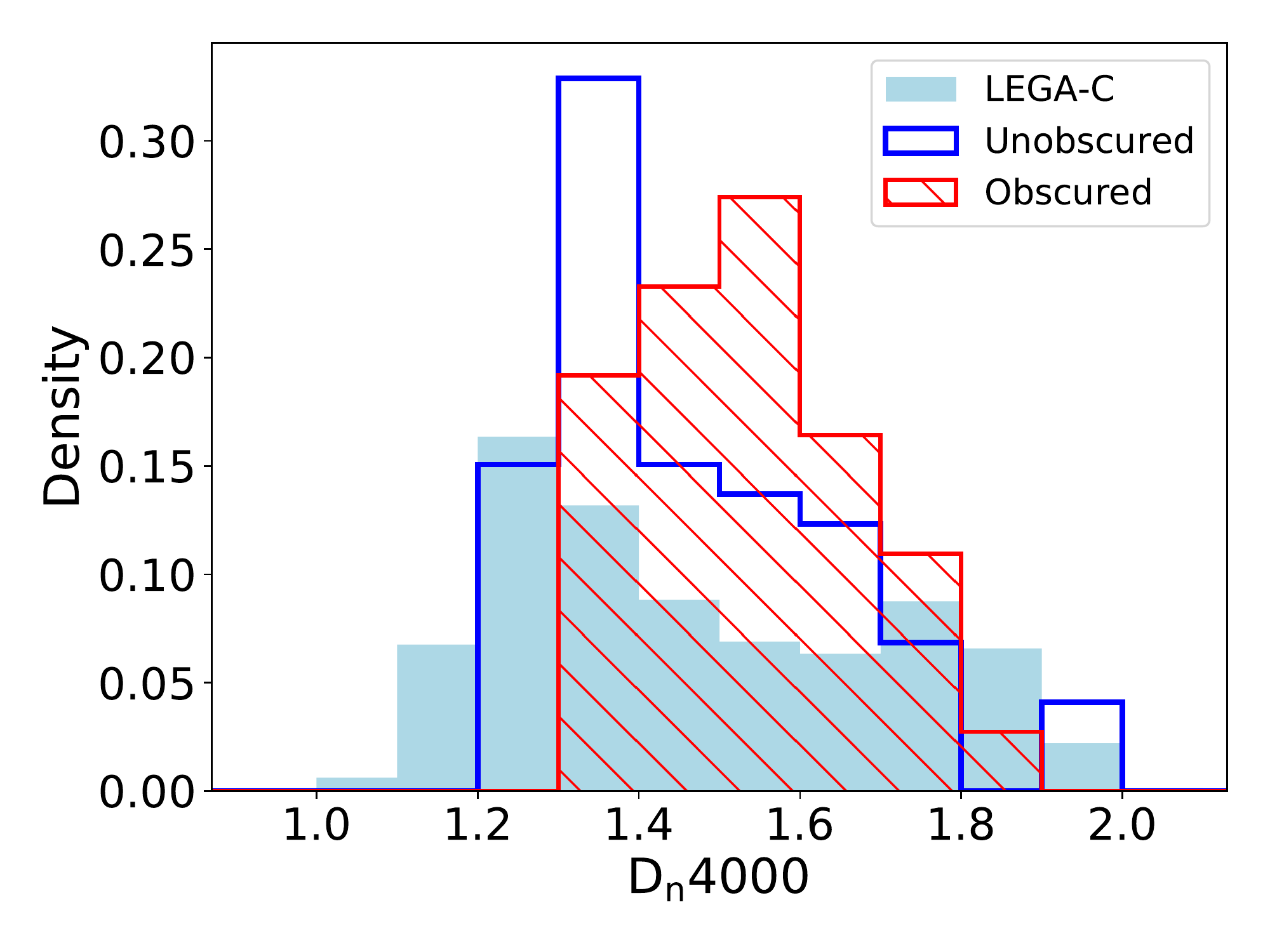} &

    \includegraphics[width=0.47\textwidth]{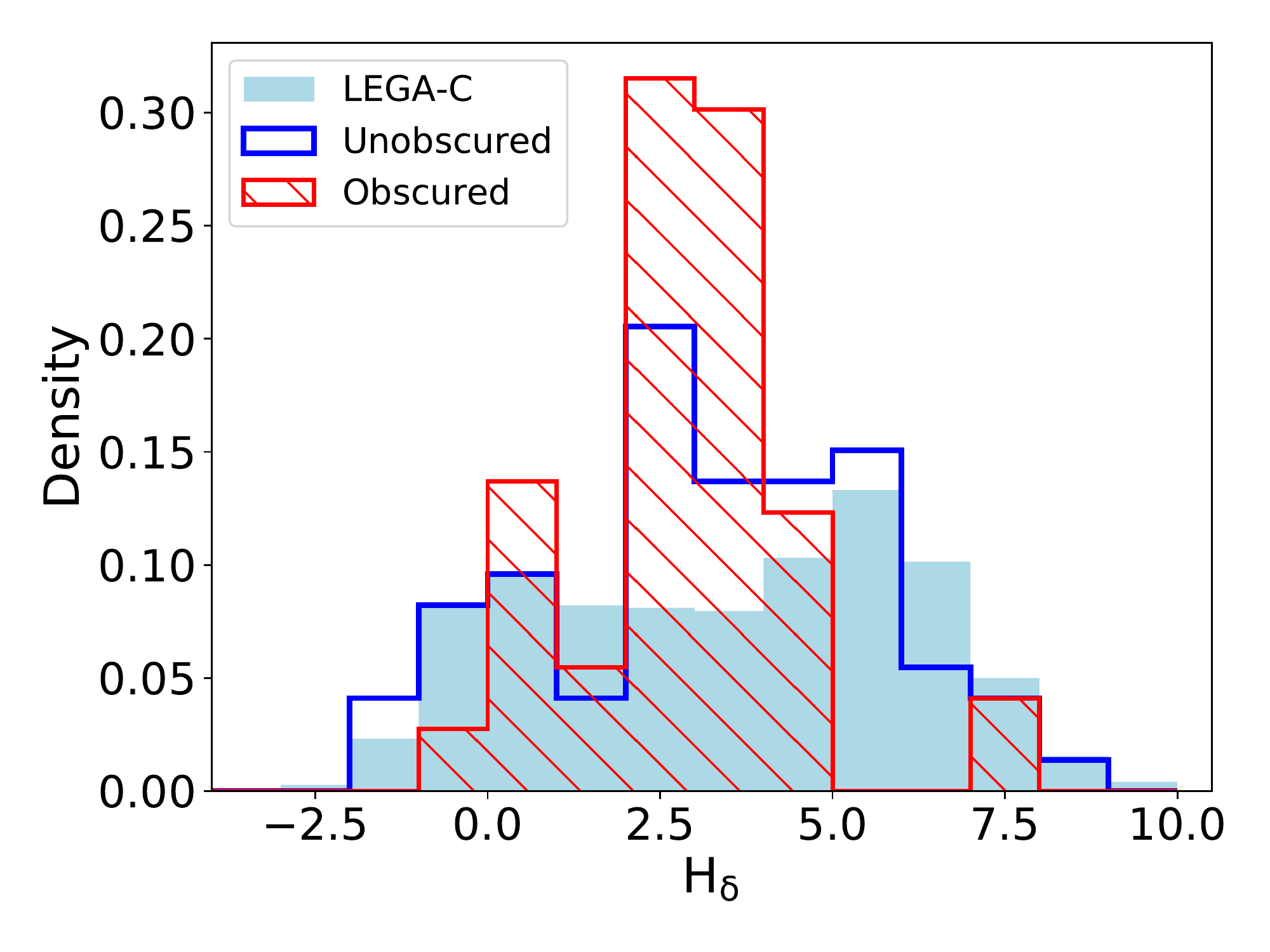}
    \end{tabular}
\caption{The distribution of the $\rm D_n$ (left panel) and $\rm H_\delta$ (right panel) age indicators  for the obscured and unobscured AGN samples. The samples have been normalised in order to follow the same luminosity and redshift distribution (see section \ref{weights} for details). We over-plot the distribution of the LEGA-C galaxies for reference.}
\label{ages}
\end{figure*}

% \begin{figure}
 %  \begin{tabular}{c}
%       \includegraphics[width=0.47\textwidth]{plot_hist_DnNORM.pdf} 
%    \end{tabular}
%\caption{The distribution of $\rm D_n(4000)$ for the obscured and unobscured AGN samples as indicated in the legend. The distributions have been weighted to take into account the redshift and luminosity distribution of the two subsamples (see section \ref{sec_cigale} for details). We over-plot the distribution of the LEGA-C galaxies for reference.}\label{Dn}
%\end{figure}

%\begin{figure}
 %  \begin{tabular}{c}
 %      \includegraphics[width=0.47\textwidth]{plot_hist_HdNORM.pdf} 
 %   \end{tabular}
%\caption{The distribution of $\rm H_\delta$ age indicator for the obscured and unobscured AGN samples as indicated in the legend. The distributions have been weighted on order to follow the same X-ray luminosity and redshift distribution (see section \ref{sec_cigale} for details). We over-plot the distribution of the LEGA-C galaxies for reference.}\label{Hd}
%\end{figure}

\begin{figure}
   \begin{tabular}{c}
       \includegraphics[width=0.47\textwidth]{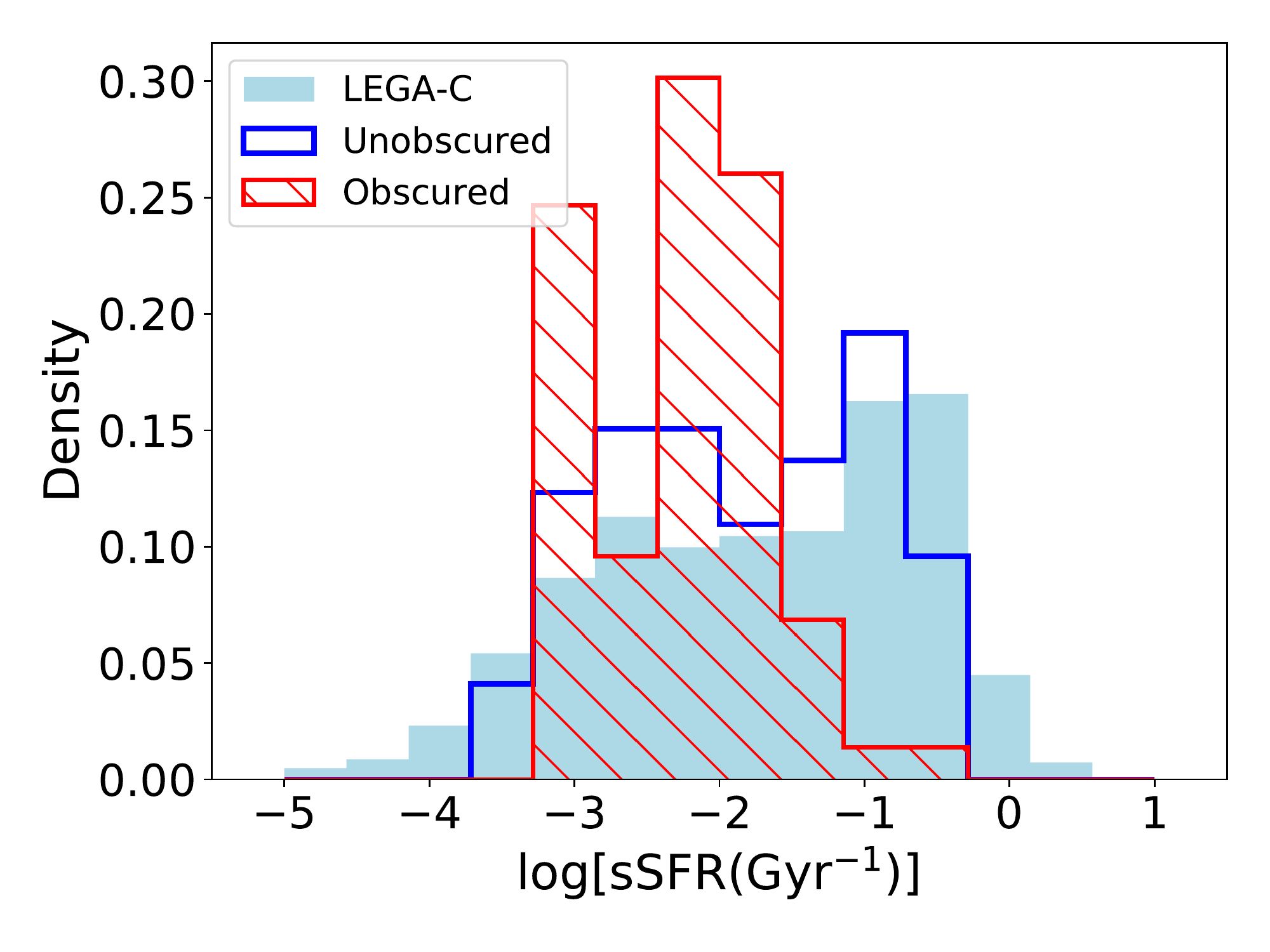} 
    \end{tabular}
\caption{The distribution of the specific SFR, sSFR, for the obscured and unobscured AGN samples. The samples have been weighted in order to follow the same X-ray luminosity and redshift distribution (see section \ref{weights} for details). The sSFR distribution of the LEGA-C galaxies is overplotted for reference.}\label{sSFR}
\end{figure}

\begin{figure}
   \begin{tabular}{c}
       \includegraphics[width=0.47\textwidth]{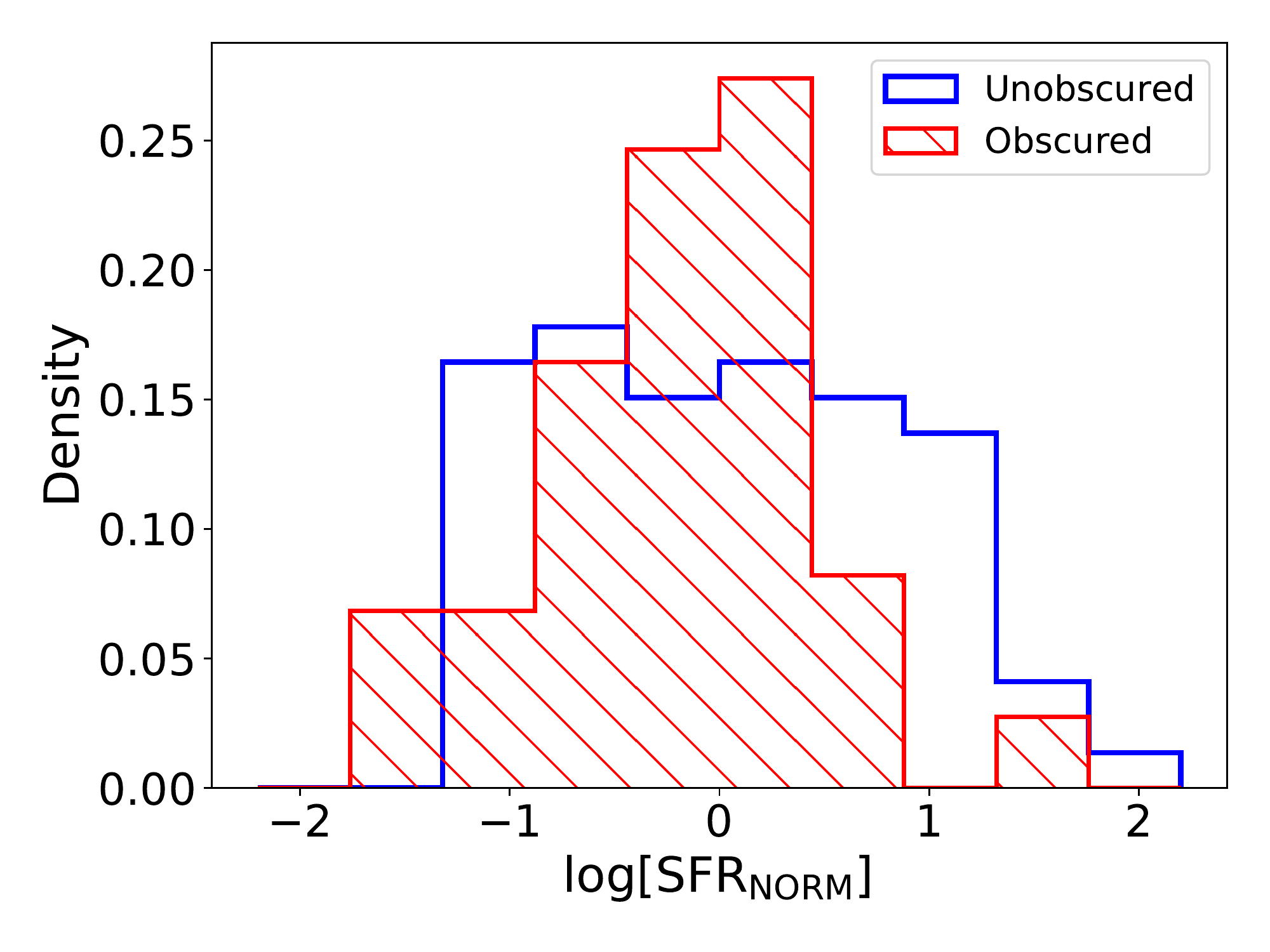} 
    \end{tabular}
\caption{The distribution of $\rm SFR_{NORM}$ for the obscured and unobscured AGN samples.
 The samples have been weighted in order to follow the same X-ray luminosity and redshift distribution (see section \ref{weights} for details)}\label{nSFR}
\end{figure}

\begin{table*}
\centering
\caption{Median values and K-S statistics for the obscured and unobscured sample}
\begin{tabular}{lcccc}
\hline
 Property  &  $\rm \mu$ (Obscured)   & $\rm \mu$ (Unobscured)  & Distance  &  p-value   \\
            \hline& \\[-1.5ex]
 $\rm D_n$     &    $1.53^{+0.025}_{-0.059}$    &  $1.40^{+0.084}_{-0.062}$  &  0.34  &  $3.4\times10^{-4}$  \\ [0.1cm]
 $\rm H\delta$  &   $2.75^{+0.380}_{-0.270}$    & $3.42^{+1.15}_{-0.70}$  &  0.34  &   $3.4\times10^{-4}$   \\ [0.1cm]
 log[$\rm M_\star$]   &   $11.37^{+0.010}_{-0.048}$   &  $11.24^{+0.07}_{-0.045}$  &  0.32  & $1.3\times10^{-3}$            \\ [0.1cm]
 log[sSFR]       &  $-2.19^{+0.124}_{-0.05}$     & $-1.72^{+0.26}_{-0.42}$  &  0.31  &  $1.3\times10^{-4}$    \\ [0.1cm]
 log[$\rm SFR_{NORM}$] &   $-0.09^{+0.10}_{-0.13}$    & $0.025^{+0.16}_{-0.44}$  &  0.28  &   $4.7\times 10^{-3}$    \\ [0.1cm]
 log[$\rm \lambda_{EDD}$] &   $-2.02^{+0.011}_{-0.026}$    & $-1.68^{+0.11}_{-0.13}$  &  0.44  &   $3.36\times 10^{-6}$    \\  [0.1cm]
 \hline
\end{tabular}
    \label{KS}
\end{table*}

\begin{figure*}
\center
   \begin{tabular}{c c}
    \includegraphics[width=0.47\textwidth]{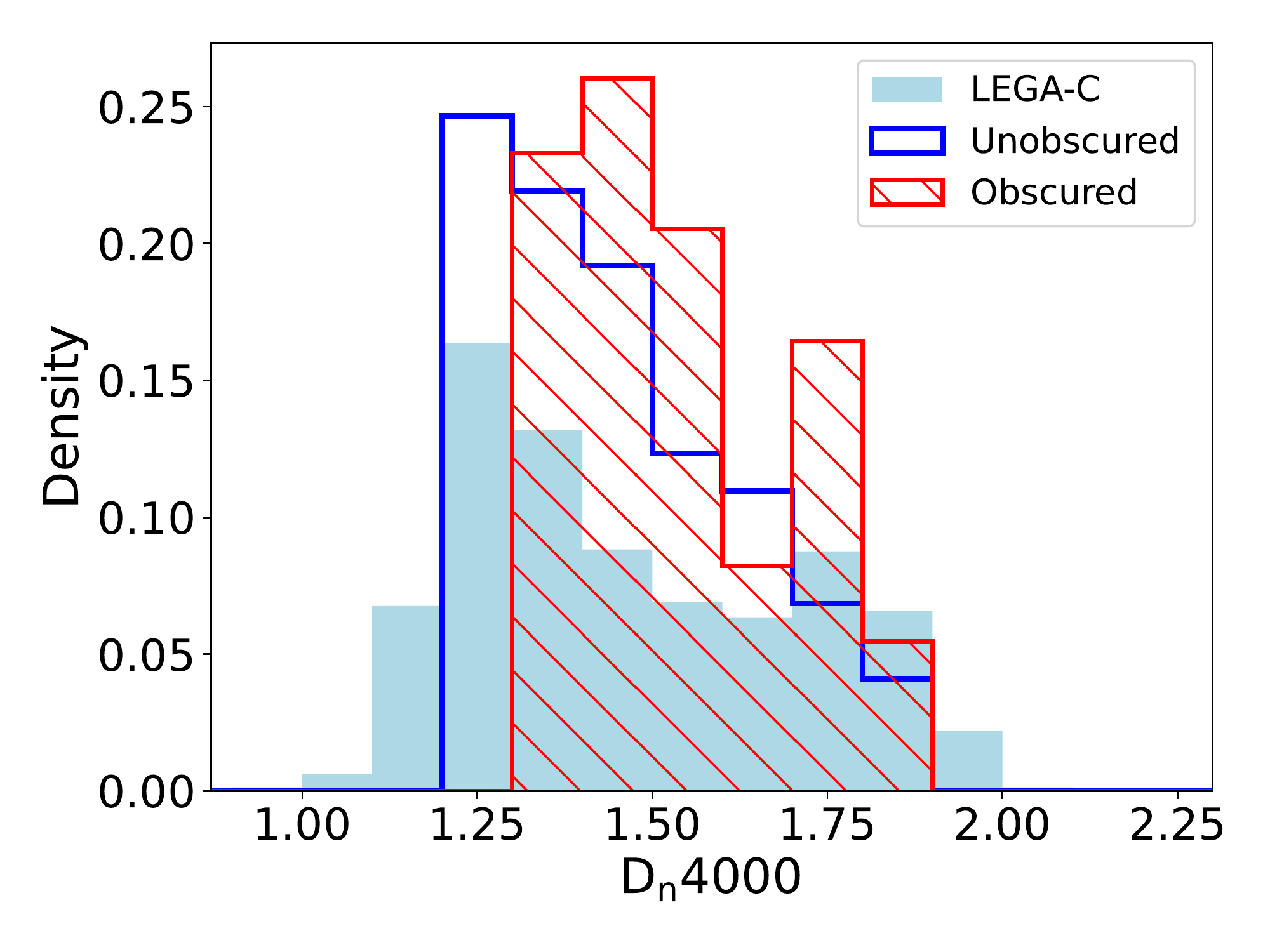} &

    \includegraphics[width=0.47\textwidth]{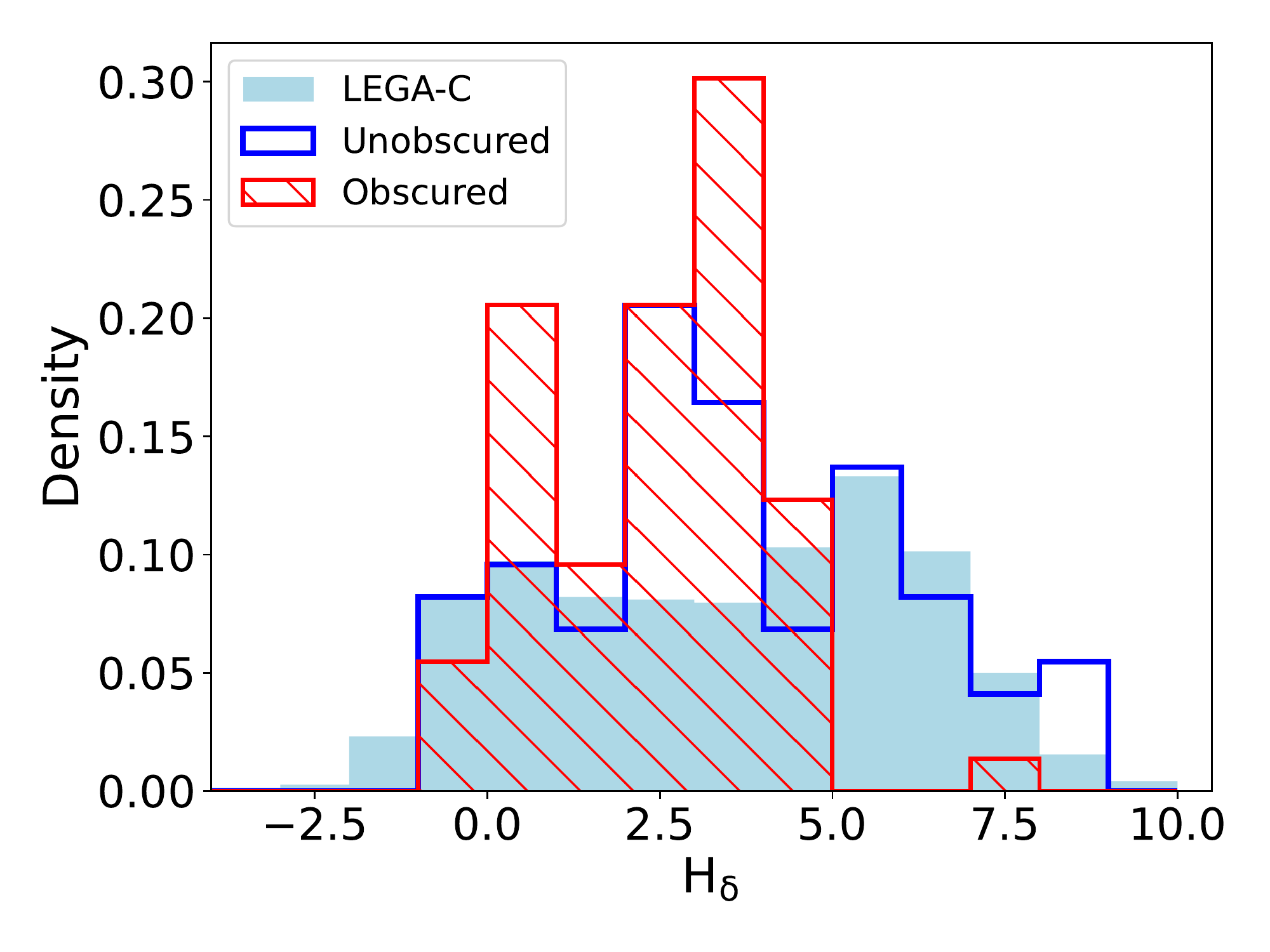}
    \end{tabular}
\caption{The distributions of the $\rm D_n$ (left panel) and $\rm H_\delta$ (right panel) age indicators for the obscured and unobscured AGN samples. The samples have been weighted in order to follow the same redshift and stellar mass  distribution (see sections \ref{weights} and \ref{mstarnorm} for details). We over-plot the distribution of the LEGA-C galaxies for reference.}
\label{Dn_Mweight}
\end{figure*}

\section{Results \& Discussion}
In this section we present the distributions of the age 
of the host galaxy, the SFR, the stellar mass  the Eddington ratio and velocity dispersion 
separately for the  obscured and unobscured AGN populations. 

\subsection{Normalisation of the obscured and unobscured samples} \label{weights}
Since the z and $\rm L_X$ distributions are different for 
obscured and unobscured AGN (Fig. \ref{Lzdistr}), we need to compare 
 the host galaxy stellar mass and age distributions 
 by normalising on luminosity and redshift. This is done  with the use of  control samples.
  We follow the method of \cite{Zou2019} 
for the creation of the control samples, dividing the $\rm \log L_X \--  z$ plane into a grid
with $\rm \Delta z = 0.1$ and 
$\rm \Delta log L_X = 0.5~dex$ (see Fig. \ref{grid}). 
 We randomly select $\rm N_1(z_i,L_j)$ unobscured sources as well as the same number of obscured sources $\rm N_2(z_i,L_j)$ in the considered i,j grid element. After repeating the procedure in each redshift, luminosity bin, we end up with 
 new unobscured and obscured samples with similar distributions of z and $\rm L_X$.  The obscured and unobscured control samples contain 73 sources each i.e. the total number of sources in our sample.
 
 \subsection{Age Indicators}
We have used the amplitude of the 4000-$\rm \AA$ break 
and the $\rm H_\delta$ absorption line \citep{Worthey1997}
as the age indicators of the host galaxies. 
The $\rm D_n(4000)$ index quantifies the strength of the 
calcium break at 4000$\rm \AA$ \citep{Balogh1999}. 
This is defined as $\rm D_n=F_{4050}/F_{3900}$. This break is prominent in older and metal-rich stellar populations. 
It is caused by the {\sc CaII} absorption doublet and by line blanketing of metal lines in stellar atmospheres. Metals in the outer layer of a star’s atmosphere absorb some of the star’s radiation and re-emit it at redder wavelengths. Then galaxies which experienced  
a recent episode of star-formation have a smaller $\rm D_n$ index because of the presence of young stars. 
Another commonly used age indicator is provided by the Balmer lines which are most 
prominent in the youngest  stars. Absorption takes place mostly in 
A stars which have large  amounts of neutral  material available. 
This has been interpreted by \cite{Dressler_Gunn1983}
as an indication that
that the SFR burst ended 0.5-1.5Gyr before the observation.
At the redshifts probed here  higher order Balmer lines such as $\rm H_\delta$, are observed as the redder lines shift out of frame. 
The strength of absorption features is often quantified using the Lick indices.
The Lick indices measure the flux in the absorption line 
comparing it with the flux in a nearby pseudo-continuum. This defines the strength of the line:
$\rm I= (\lambda_2-\lambda_1)(1-F_{L}/F_C$) where where $\rm \lambda_2-\lambda_1$ is the width of the passband  used to measure the index and $\rm F_C$ and $\rm F_L$ are the fluxes of the continuum and the line respectively \cite[for details see][]{Worthey1997}.
The units of $\rm H_{\delta}$ are given in $\rm \AA$. For more details on the extraction of the $\rm D_n$ and $\rm H_{\delta}$
indices are given in \cite{Straatman2018} and \cite{VanderWel2021}. 
We note that the presence of a strong AGN may affect the $\rm D_n$ and $\rm H_{\delta}$ values. For example, regarding the Calcium break, this could happen if the relative contribution of the AGN to the galaxy $\rm R=F^{AGN}/F^{Gal}$ is different in the passbands centred at $\rm 3900\AA$ and $\rm 4050 \AA$. The ratio $\rm R$  above can be determined from our SED fitting.
We found that the ratios are very close to unity and then the resulting corrections are very small. 

In Fig. \ref{HdD} we present the plot with the age indicators of our sample compared to the 
LEGA-C galaxy sample.
We colour-code our sources using 
a) the column density b) the stellar mass c) the specific star-formation rate d) the normalised star formation rate and e) Eddington ratio f) velocity dispersion.
Note that this plot has not being weighted for the different luminosity and redshift distributions 
of the obscured and unobscured objects.

The locus of galaxies is forming a sequence moving from high (low) values of $\rm D_n\sim 2$ 
 ($\rm H_\delta  \sim -3$) to low (high) values of 
$\rm D_n\sim1.2$  ($\rm H_\delta \sim 5$).
Large deviations from this locus  are often used as an indicator of recent bursts of star formation \citep{Kauffmann2003b}. 
 It can be seen that the galaxies are separated in two groups. The old group clusters 
around $\rm (D_n,H_\delta) \approx (1.8,0)$ 
 while the young group populates the area around  $\rm (D_n,H_\delta) \approx (1.3, 5.5)$. The AGN occupy both the old and young cloud but with  some tendency towards the young cloud. 
 In particular, the unobscured AGN  populate more frequently the young cloud compared to the obscured AGN. 
 The obscured AGN appear to have cc intermediate ages populating the region 
 $\rm (D_n,H_\delta) \approx (1.5, 2.5)$. 
 The above differences can be more clearly seen in the histogram of $\rm D_n$ (Fig. \ref{ages}).  We note that in this figure the  samples have been normalised in order to follow the same luminosity and redshift distributions (see section \ref{weights}). 
 It is apparent that the unobscured population is associated mainly with young galaxies. Instead the obscured population occupies an area in the middle 
 of the young and old cloud.
  More specifically, the obscured and unobscured AGN population have a median $\rm D_n$ value  of 1.53 and 1.40 respectively. 
  The K-S test shows that the two populations have different distributions in the $\rm D_n$ at a statistically significant level 
  (see Table \ref{KS}). Following standard practice, we adopt $\rm 2\sigma$ (p-value=0.05) as the threshold for a “statistically significant” difference in host-galaxy properties.
  The distribution of $H_\delta$  (Fig. \ref{ages}) 
  shows that both populations present a peak  around $H_\delta \sim$3 albeit  the host galaxies of the unobscured AGN present also a large 
  tail at younger ages $H_\delta \approx 5$. 

  \cite{Silverman2009} have investigated the ages of the host galaxies of {\it XMM-Newton} selected AGN in the {\it XMM-Newton} COSMOS field using the calcium break
  $\rm D_n(4000)$. They find  an increased AGN fraction among the galaxies with younger populations with $\rm 1<\rm D_n(4000)<1.4$.
  These young galaxies are actively star-forming as indicated by their blue rest frame U-V colour as well as the strength of the 
  [OII] line. 
  Finally, \cite{Hernan2014} analyse the stellar populations in the host galaxies of 53 X-ray selected optically dull active galactic nuclei (AGN) at 
   a redshift range of $\rm 0.34 < z < 1.07$ from the Survey for High-z Absorption Red and Dead Sources (SHARDS). They find a highly significant excess of AGN hosts with $\rm D_n(4000) \approx 1.4$, as well as a deficit of AGN in intrinsically red galaxies. 
  Therefore our results are in good agreement overall with those of \cite{Hernan2014} as far as the total AGN population is concerned.
  An additional important result of our analysis is that we separate for the first time the X-ray selected AGN in obscured and unobscured objects finding a significant difference in their host galaxy ages.

  \subsection{Stellar ages}
  From the $\rm D_n$ and $H_{\delta}$ values derived above it
  becomes evident that the obscured AGN population is associated on average with older host galaxies, compared to the unobscured AGN population.  It  is instructive to derive the {\it approximate} stellar ages of these populations. 
  The {\sc CIGALE} code provides stellar ages estimates. However, \cite{Mountrichas2022c} caution that the {\sc CIGALE}  estimates have limited accuracy if they are not combined with $\rm D_n$ and $H_\delta$ measurements. Hence, they developed a variant of the CIGALE code which uses both these indices to estimate the mass weighted galaxy ages.
   As this code is not publicly available, we can rely on their galaxy age methodology. This is possible because
  the LEGA-C COSMOS sample used by \cite{Mountrichas2022c} is practically identical to our sample and 
  moreover they use the same parameter grid for their SED fitting.

  \cite{Mountrichas2022c} derive the spectral age index using 
  both the $\rm D_n$ and $EW(H_\delta)$ measurements. This index combines 
  $\rm D_n$  and $\rm H_\delta$ to construct the distribution of galaxies 
  along the diagonal distribution  on the $\rm D_n - H_\delta$ plane 
  \citep[see][]{Wu2018}.
 They derive  a spectral age index  (S.A.I.) defined as 
  $\rm -2.40 \times D_n -EW(H_\delta) + 4.36 $   (priv. communication). 
  They combine the S.A.I.  with the {\sc CIGALE}  mass weighted age estimates
  to create the S.A.I - age plot. Using this plot, the combination of EW($H_\delta)$ and 
  $\rm D_n$ translates to a mass-weighted galaxy age. 
  We use the median $\rm H_\delta$ and $\rm EW(H_\delta)$ estimates to 
  obtain the exact S.A.I. for our obscured and unobscured samples. 
  We obtain values of 3280 and 2740
  Myr for the mass-weighted ages of the host galaxies of the obscured and unobscured populations respectively.

\subsection{Distribution of stellar mass}
Here, we examine whether there is any difference in  the stellar mass distributions
 of the obscured and the unobscured AGN population.
 In Fig. \ref{HdD} (top right panel) we show the distribution of the stellar mass as a function of the two age indicators $D_n$ and $H_\delta$. There is a preponderance of AGN associated with low mass galaxies with $\sim \log M_\star (M_\odot)
 <11.2$ in the young cloud 
 ($\rm D_n<1.5$). Most of these low mass galaxies are associated with unobscured AGN. 
  The difference between the galaxy masses of the two populations can be better quantified 
 in the histograms of the stellar mass shown Fig. \ref{Mstar}. 
  In this figure and subsequent histogram figures the y-axis (density) refers to the frequency histogram so that the sum of all bins is equal to one.
 The two distributions  are different with the obscured AGN population being skewed towards higher  stellar mass ($\rm 
p-value=1\times10^{-3})$.   \cite{Zou2019} first noticed that the stellar masses of type-2 AGN are 
 on average higher that that of type-1 AGN \citep[but see][]{Suh2019}. \cite{Mountrichas2021b} confirmed this result in the XMM-XXL field. However, the division between type-1 and type-2 in both \cite{Zou2019} and \cite{Mountrichas2021b}
 is based on optical spectroscopy. Here, we find probably for the first time a difference in stellar mass between obscured and unobscured AGN samples based on X-ray spectroscopy and hardness ratios. 
 Previous works failed to identify a difference in stellar mass between the two populations \citep{Merloni2014, Masoura2021, Mountrichas2021c}. 
 However, in the above works  the column density cut-off between obscured and unobscured AGN  was set at much lower column densities $\rm log N_H (cm^{-2})= 22$ 
 or $\rm log N_H (cm^{-2})= 21.5$. We re-examine this apparent controversy in section \ref{differentNH}.
 The difference in stellar mass 
 could possibly signify a difference in the age of the two populations. This is in the sense that older galaxies had more time to 
 increase their mass through galaxy mergers.

\subsection{Star-formation rate}

 The distribution  of the specific SFR, sSFR, defined as the SFR per stellar mass as a function
  of the age indicators is 
 presented in Fig. \ref{HdD} (middle right panel). 
  We notice a  trend where the 
 majority of the AGN with high sSFR  are hosted by galaxies with the youngest ages. Moreover,   these appear to be primarily associated with  unobscured AGN.  This can be more clearly seen in Fig. \ref{sSFR} we plot the histogram  of the sSFR for the obscured vs. the unobscured AGN. 
 The two distributions are 
 different at a statistically significant level yielding $\rm p-value\approx 1\times10^{-4}$. 
 
 The $\rm SFR_{NORM}$ results can provide a more accurate picture 
 of the AGN SFR 
 relative to galaxies of the same stellar mass and redshift.
 Our $\rm SFR_{NORM}$ results are presented in Fig. \ref{HdD}  (middle right panel) as a function of the age of the 
host galaxy. There is a tendency for the highly star-forming systems to be associated
with young galaxies 
regardless of the AGN type. Most AGN host galaxies with $ \rm \log SFR_{NORM}>0.5$ 
have a calcium break value of $\rm D_n<1.5$.
In Fig. \ref{nSFR} we compare the 
$\rm SFR_{NORM}$ histograms for obscured and unobscured AGN. 
Both the obscured and unobscured AGN populations cluster around the main sequence of 
star-formation ($\rm \log SFR =0$). The median $\rm \log SFR_{NORM}$ equals -0.09 and 0.03 for the obscured and unobscured AGN respectively.
The obscured AGN have an $\rm SFR_{NORM}$ skewed towards lower values compared to the unobscured AGN.

There has been no concrete evidence 
in the literature that the SFR properties of X-ray selected obscured and unobscured AGN are different regardless of whether the separation between the two populations is performed using optical spectroscopy \citep{Zou2019, Mountrichas2021b} or X-ray spectroscopy \citep{Masoura2021}. 
\cite{Mountrichas2021b} studied 
 the X-ray selected AGN in the XMM-XXL field, in the redshift range $z=0-1$, classifying them as type-1 or type-2  on the basis of optical spectroscopy. They estimate  $\rm \log SFR_{NORM} \sim 0$ for 
both types of AGN. \cite{Masoura2021} studied the X-ray selected AGN in the same field performing the classification between unobscured  and obscured AGN based on Bayesian X-ray hardness ratios and applying a column density dividing threshold 
of $\rm \log N_H (cm^{-2})= 21.5$. Finally, \cite{Mountrichas2021c} studied the $\rm SFR_{NORM}$ for obscured and unobscured AGN in the XBOOTES field applying a hardness ratio classification using column density dividing threshold 
of $\rm \log N_H (cm^{-2})= 22$.
 Both works find that the $\rm SFR_{NORM}$ distributions are similar for the two populations. 
 
 \begin{figure}
   \begin{tabular}{c}
       \includegraphics[width=0.47\textwidth]{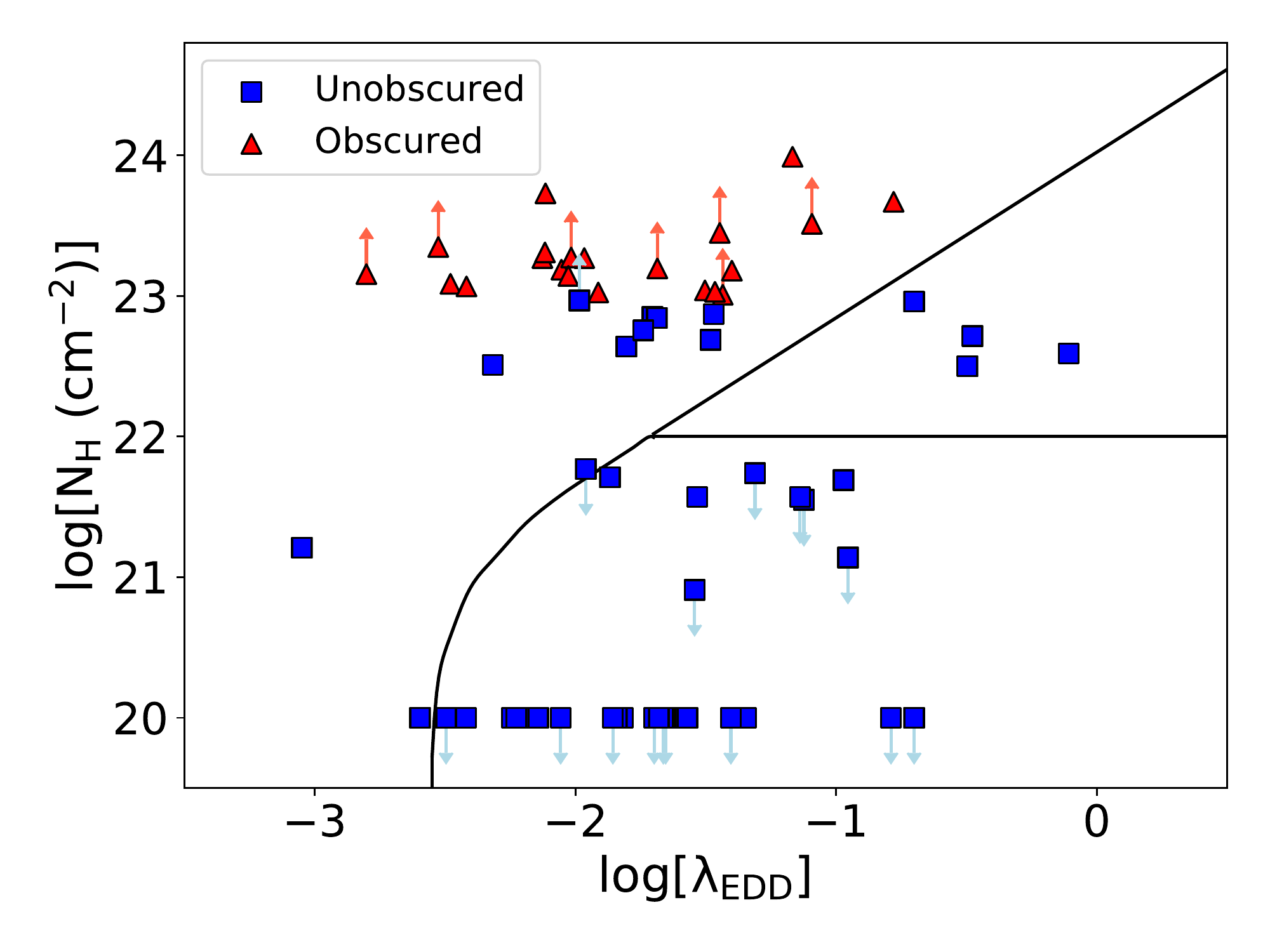} 
    \end{tabular}
\caption{The column density versus the Eddington ratio. The solid curve represents the effective Eddington ratio where the
outward radiation pressure on gas is equal to the inward gravitational
pull. The horizontal line denotes the column density value where the obscuration could come from large scale clouds in the galaxy}\label{ricci}
\end{figure}
 
\begin{figure}
   \begin{tabular}{c}
       \includegraphics[width=0.47\textwidth]{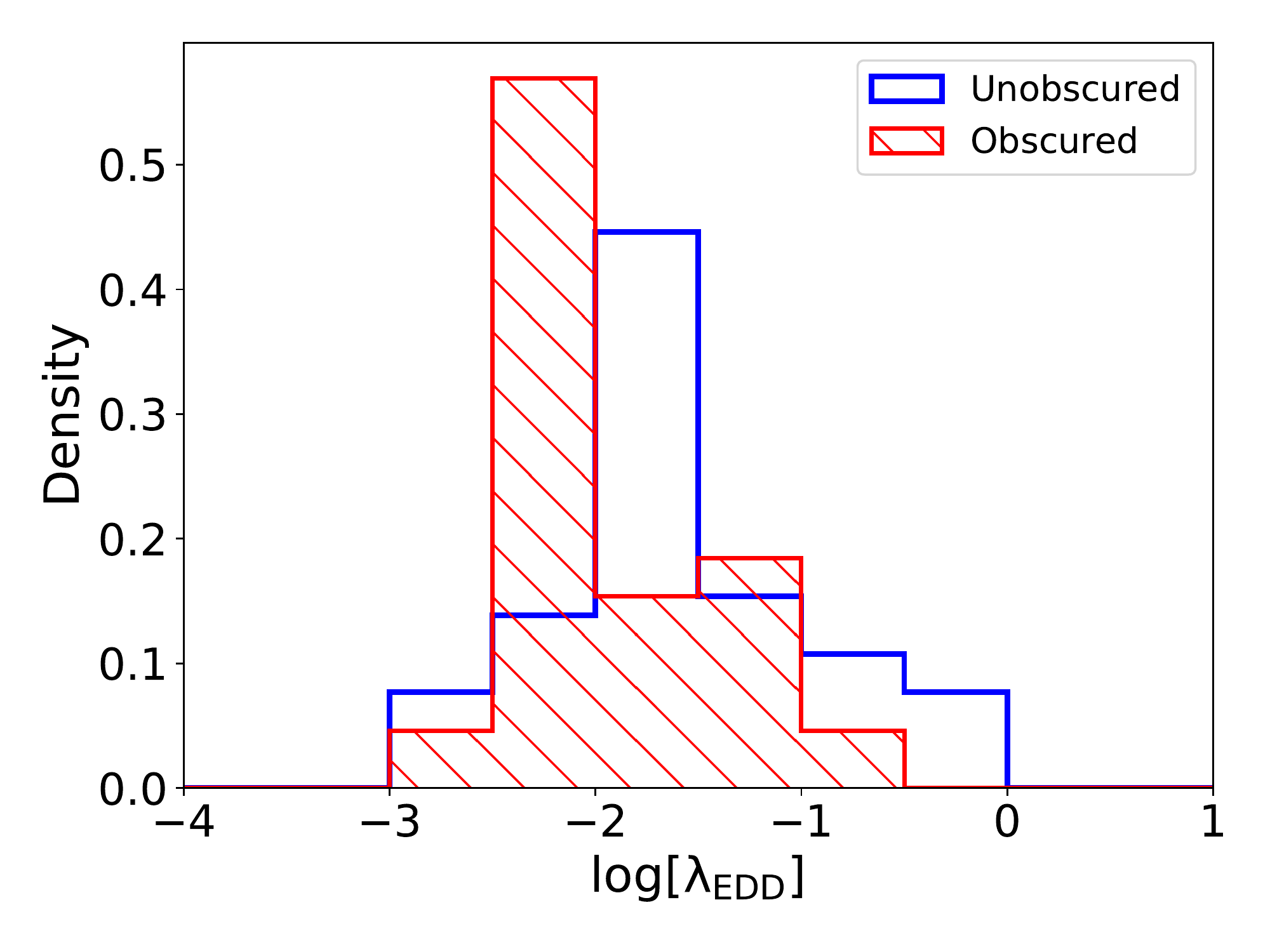} 
    \end{tabular}
\caption{The  distribution of $\rm log[\lambda_{EDD}]$ for the obscured and unobscured AGN samples }\label{lambda}
\end{figure}

\begin{figure*}
   \begin{tabular}{c c}
       \includegraphics[width=0.47\textwidth]{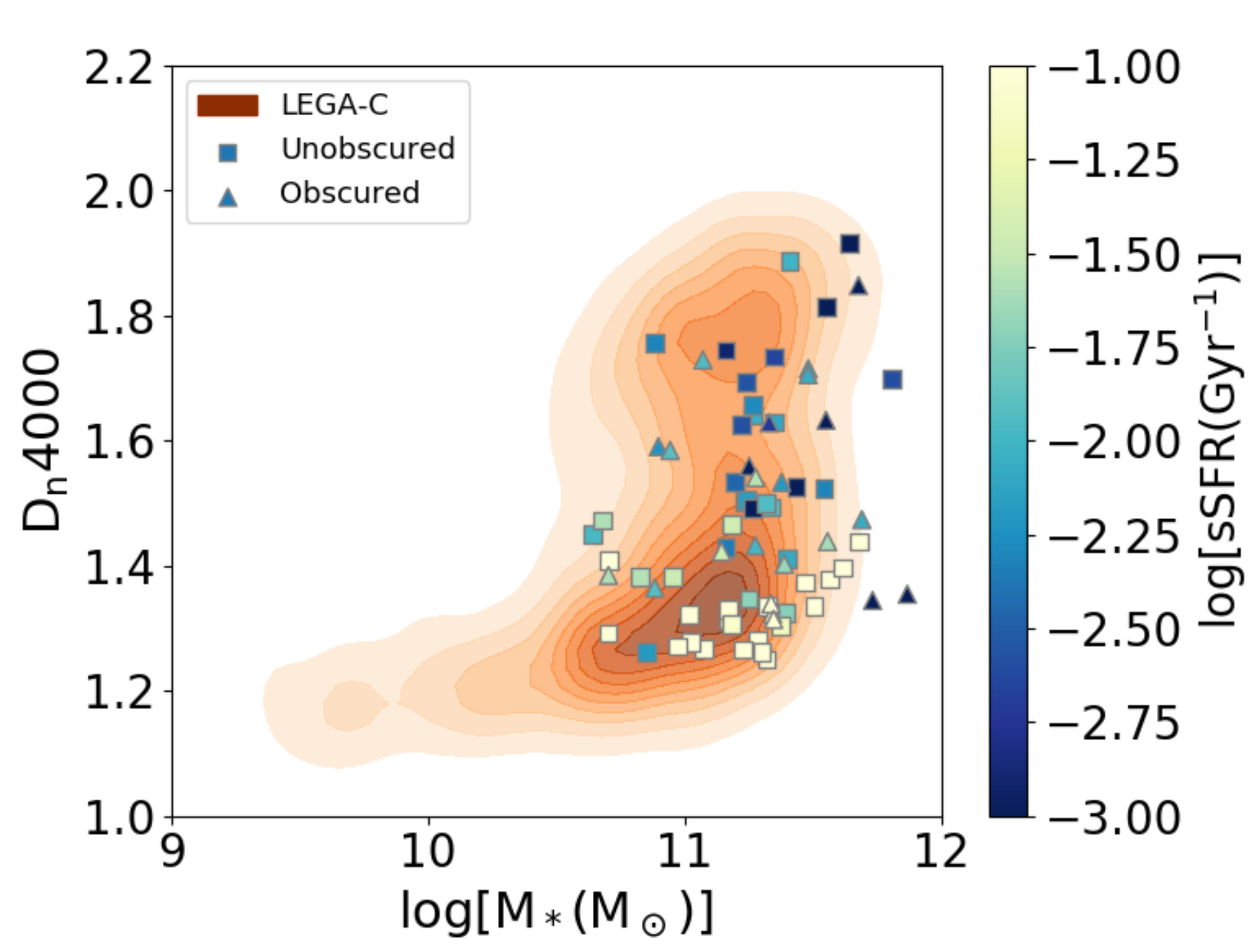} &
       \includegraphics[width=0.47\textwidth]{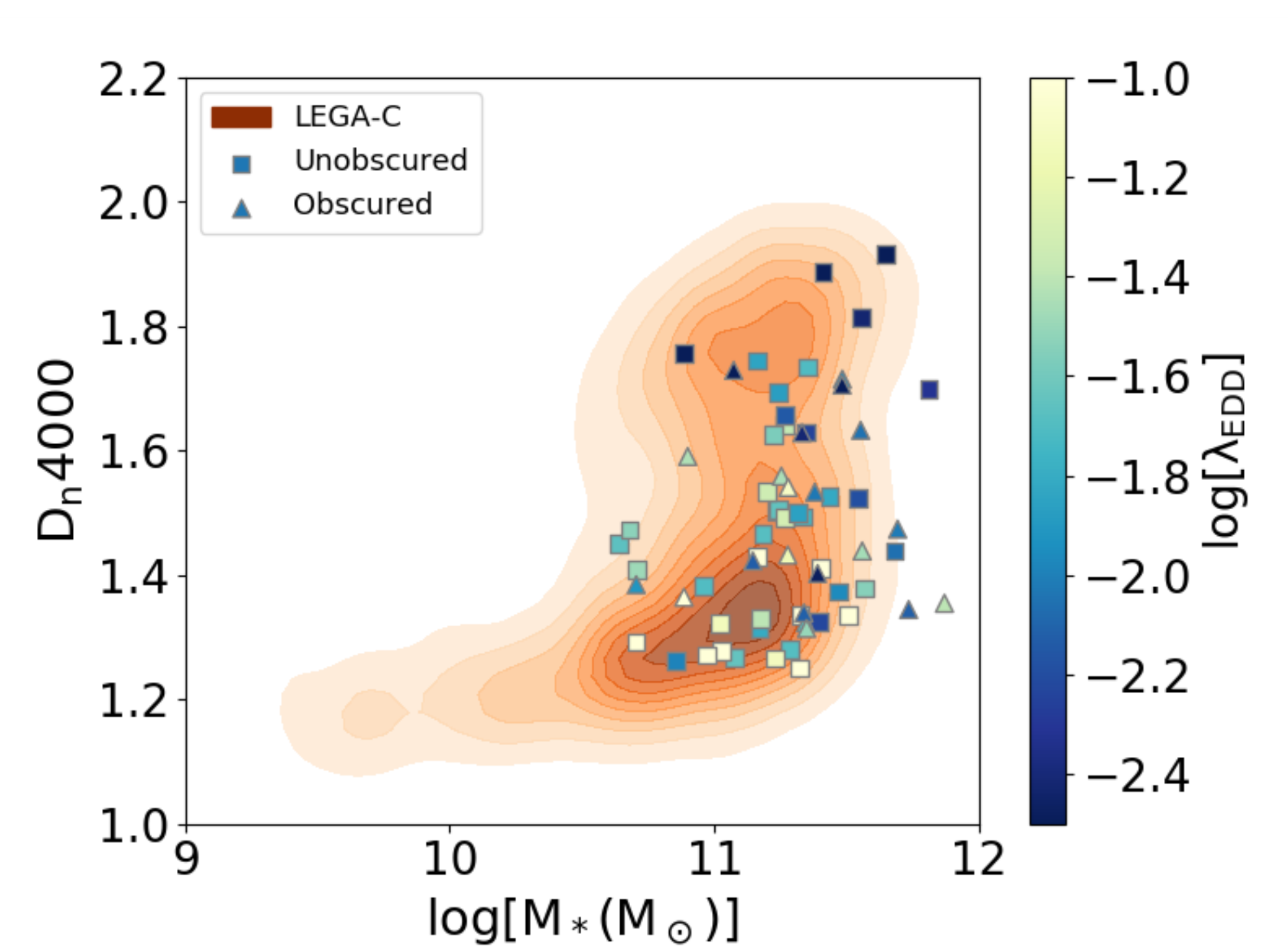} \\
      \includegraphics[width=0.47\textwidth]{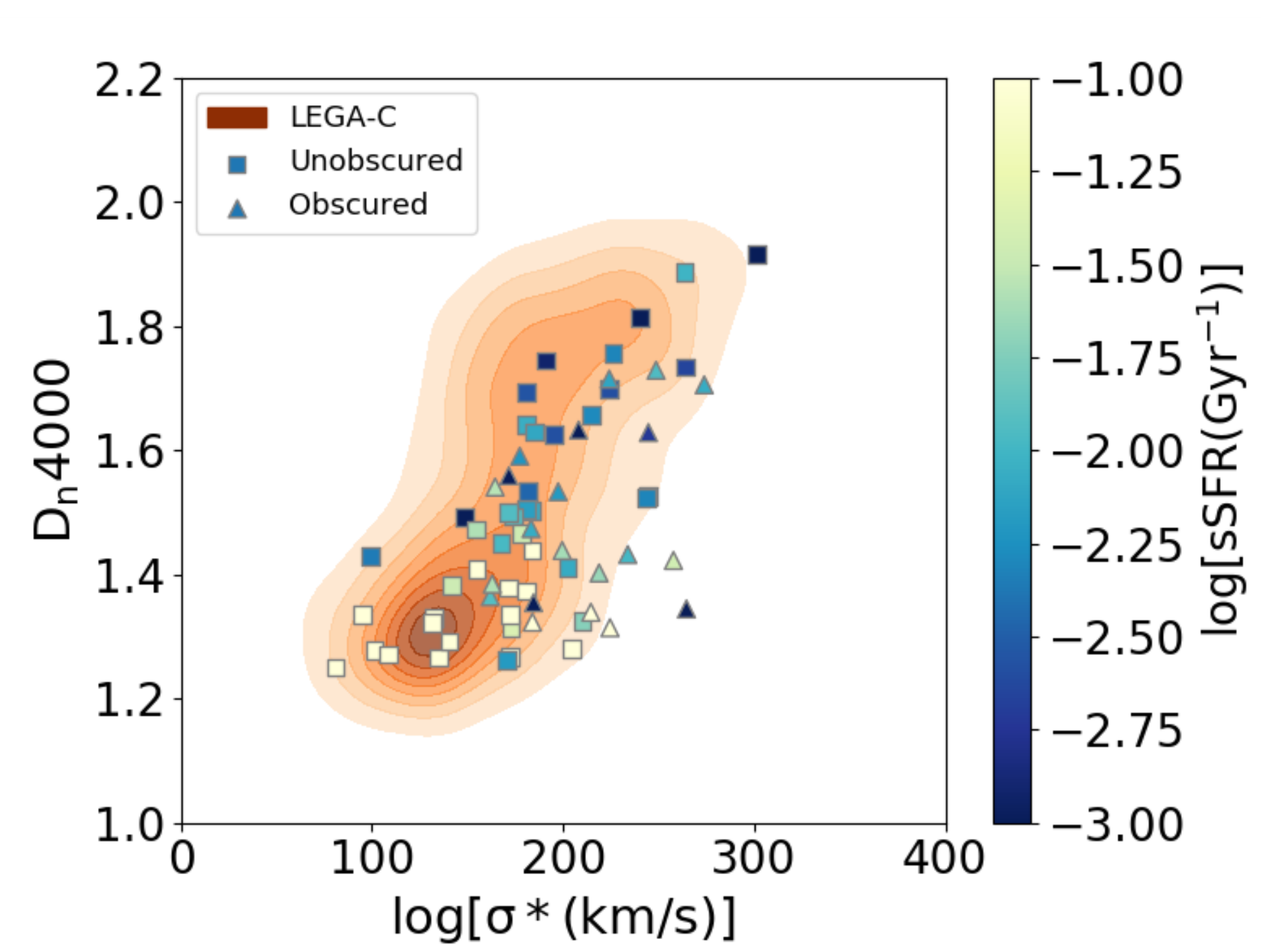} &
       \includegraphics[width=0.47\textwidth]{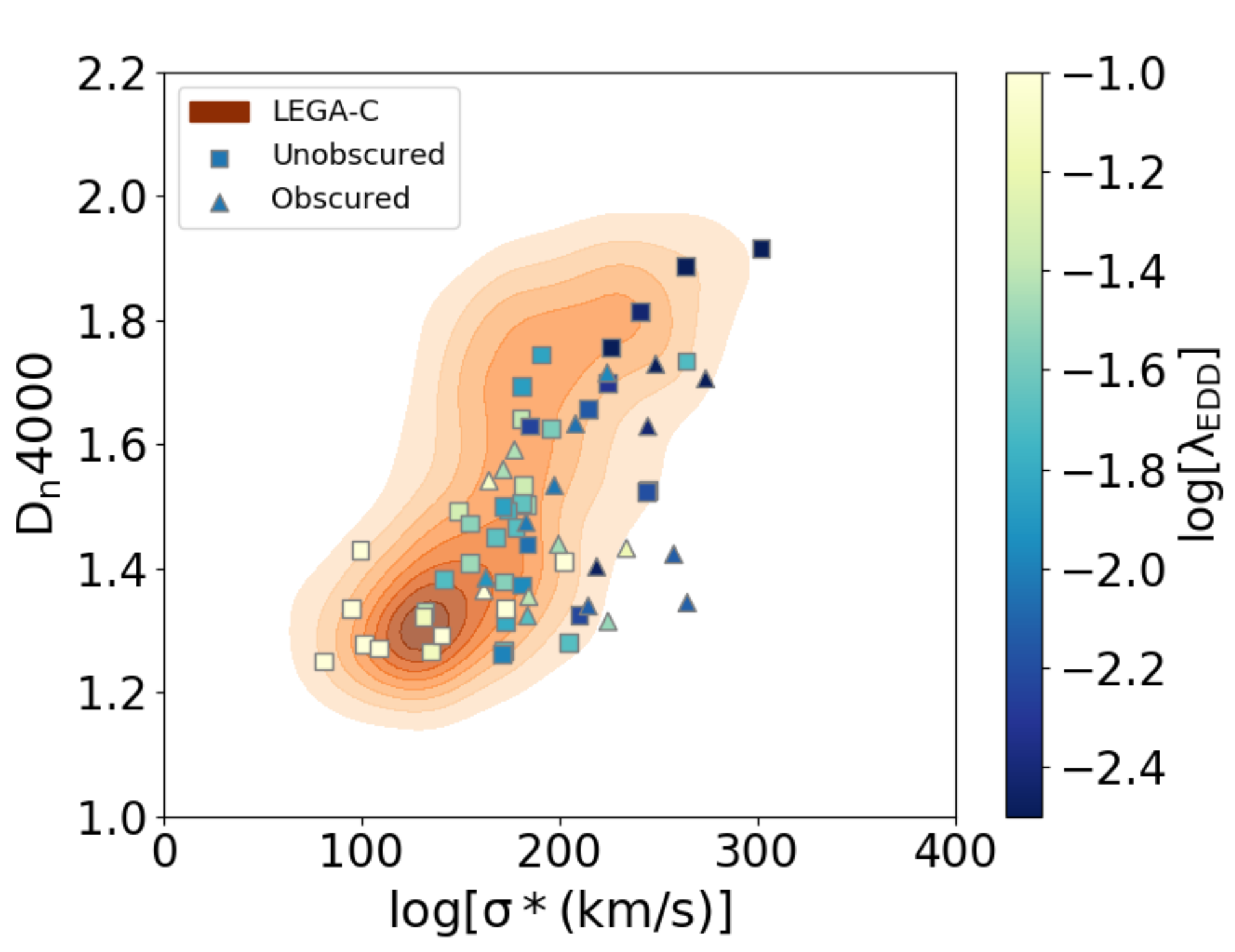} 
    \end{tabular}
\caption{Top panel: The calcium break $\rm D_n$ as  a function of the stellar mass. Lower panel: The calcium break as function of the velocity dispersion $\rm \sigma^{\star}$.
In the right and left hand plots the colour key 
represents the sSFR and the Eddington ratio, respectively.
The contours represent the LEGA-C galaxy population.
 }\label{DnMsigma}
\end{figure*}

\begin{figure*}
   \begin{tabular}{c c c}
       \includegraphics[width=0.30\textwidth]{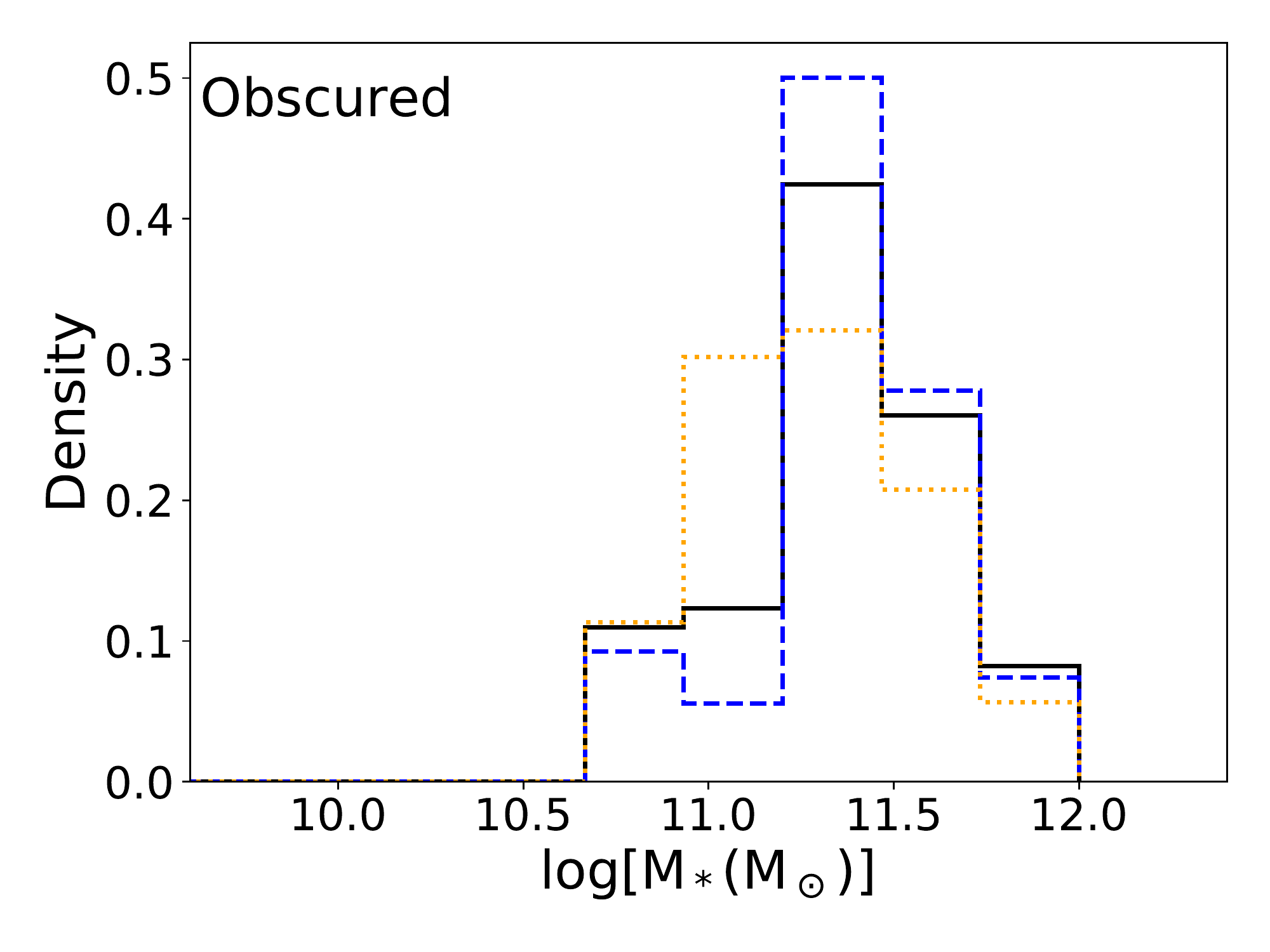} &
       
              \includegraphics[width=0.30\textwidth]{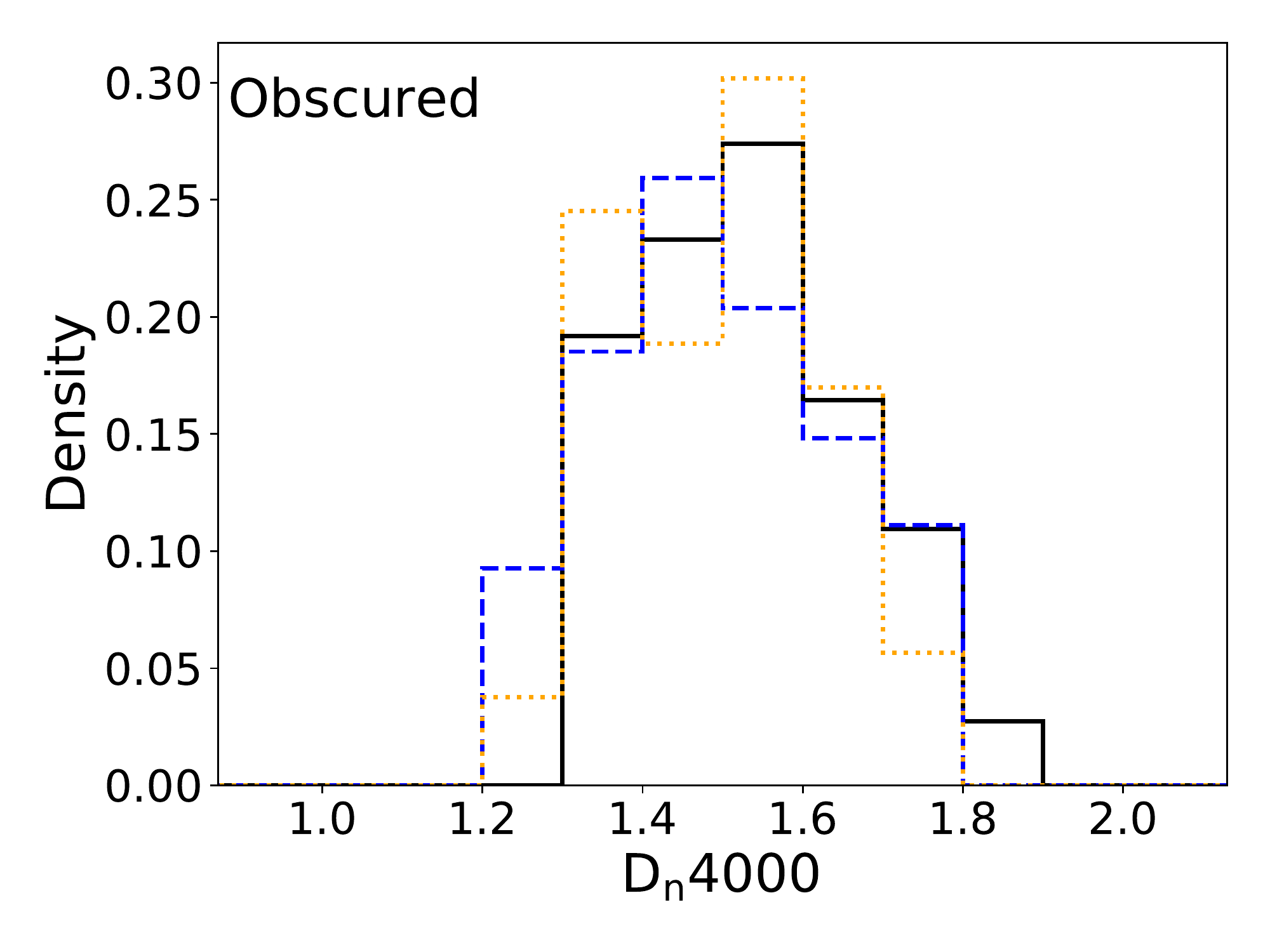} &

              \includegraphics[width=0.30\textwidth]{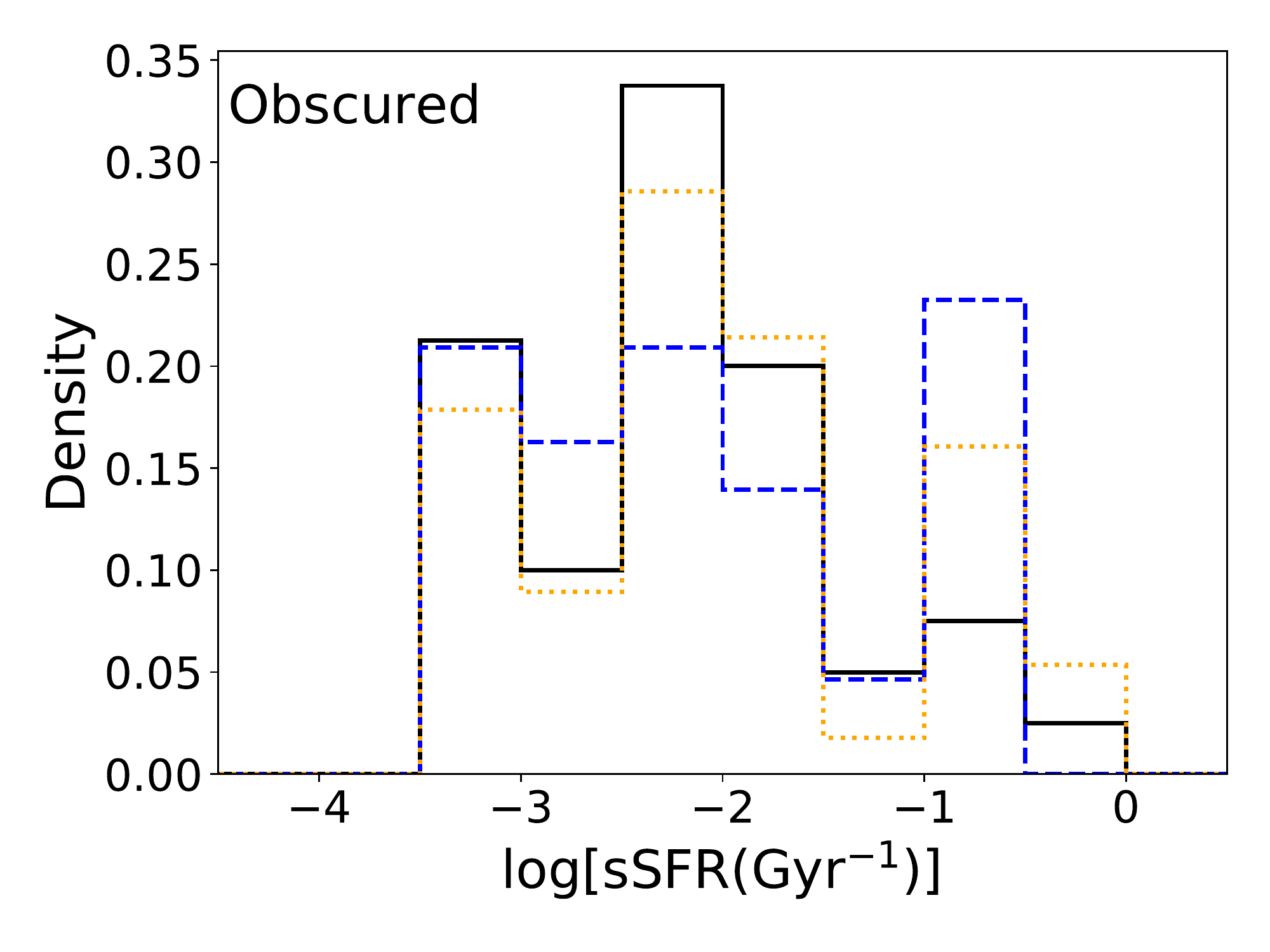} \\

              \includegraphics[width=0.30\textwidth]{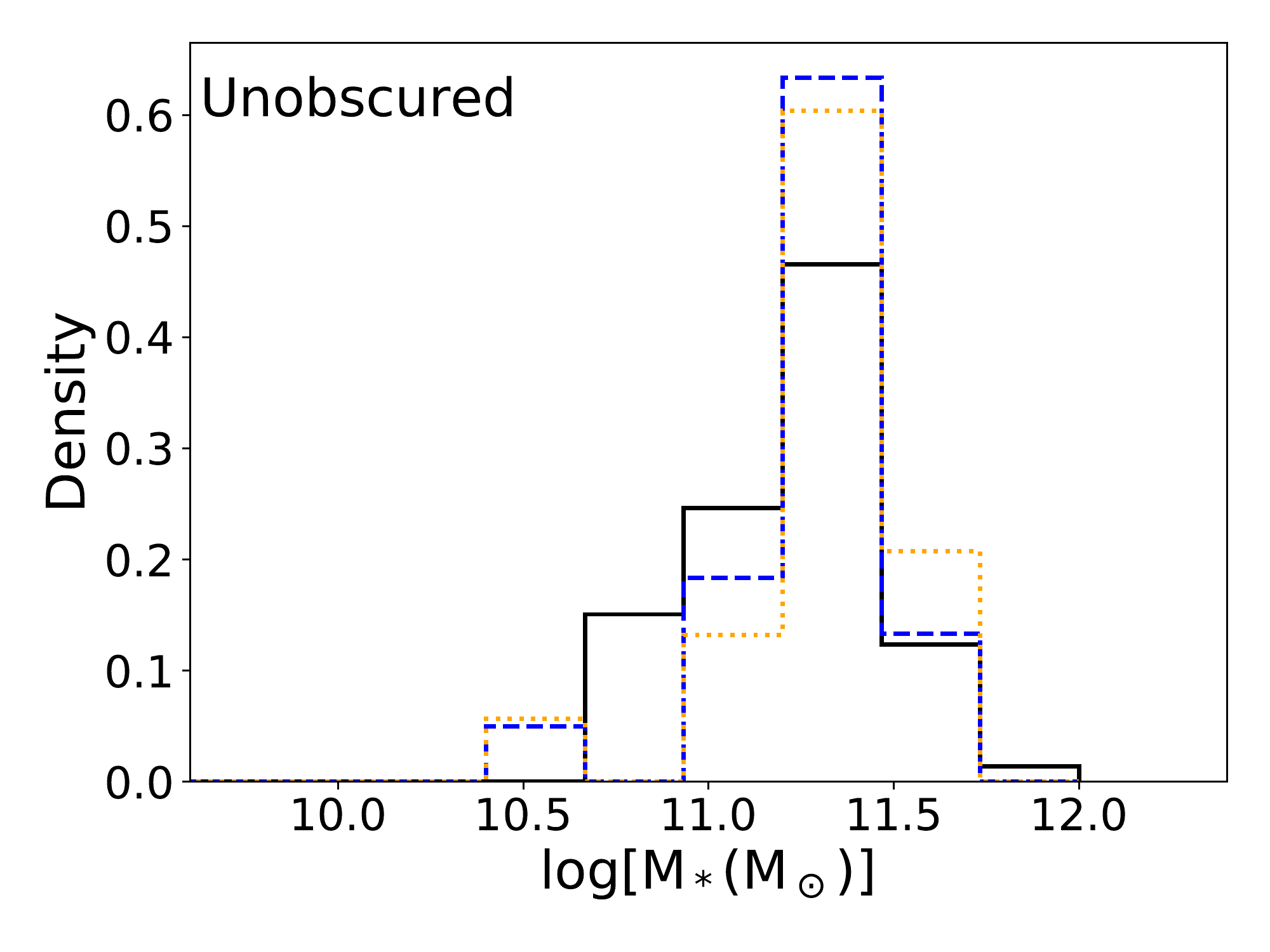} &

              \includegraphics[width=0.30\textwidth]{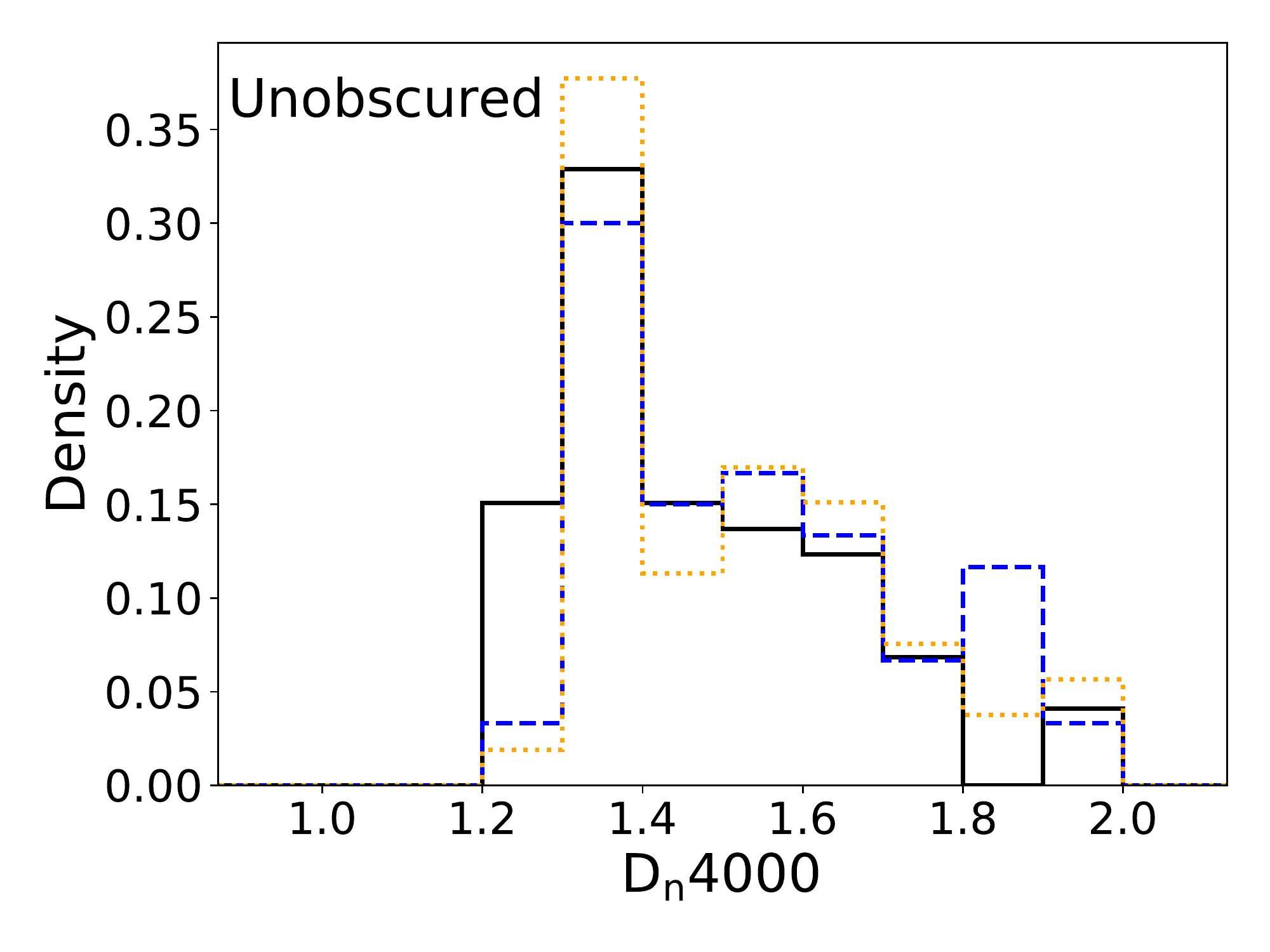} &

              \includegraphics[width=0.30\textwidth]{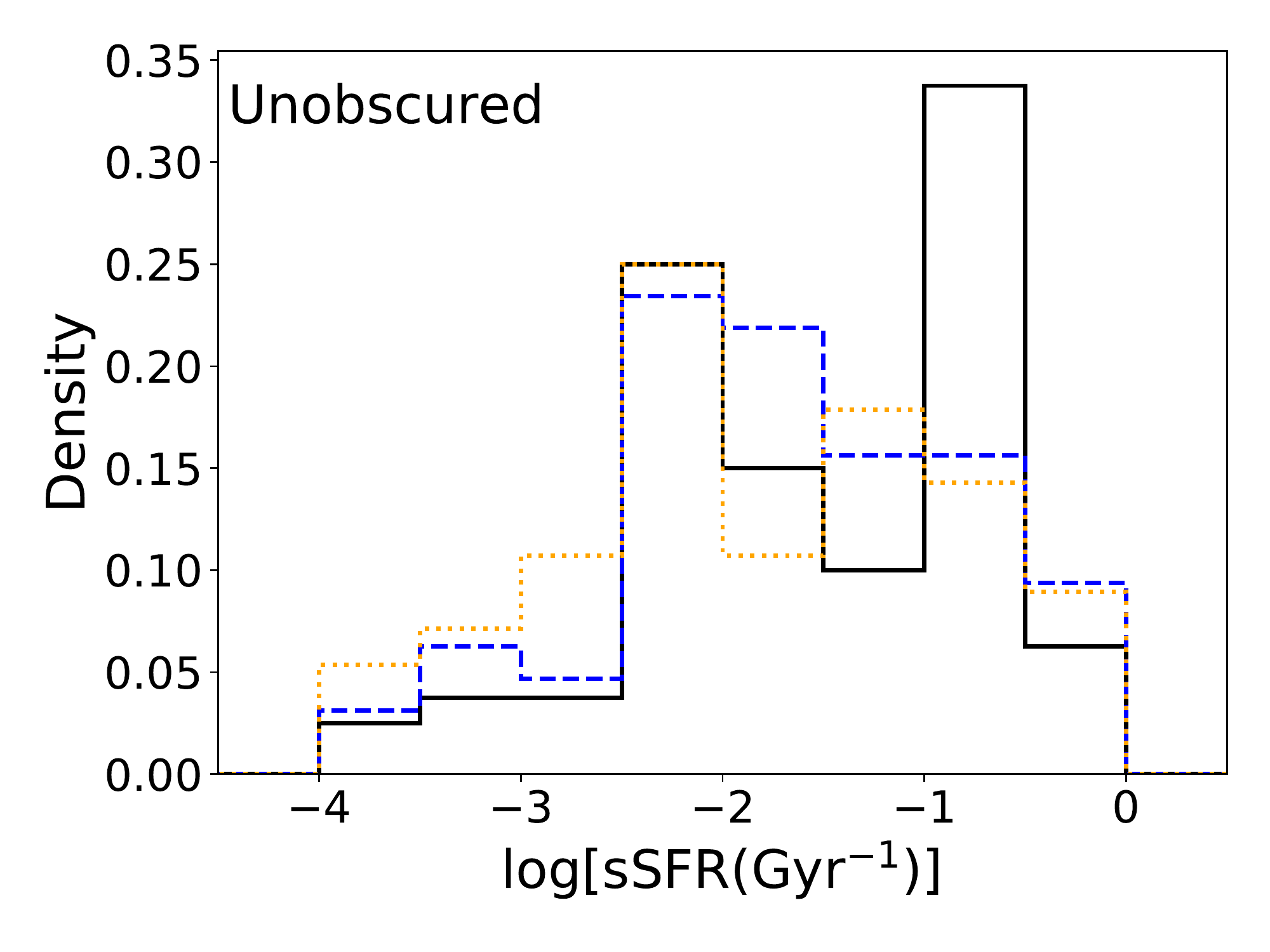} \\
              
           \includegraphics[width=0.30\textwidth]{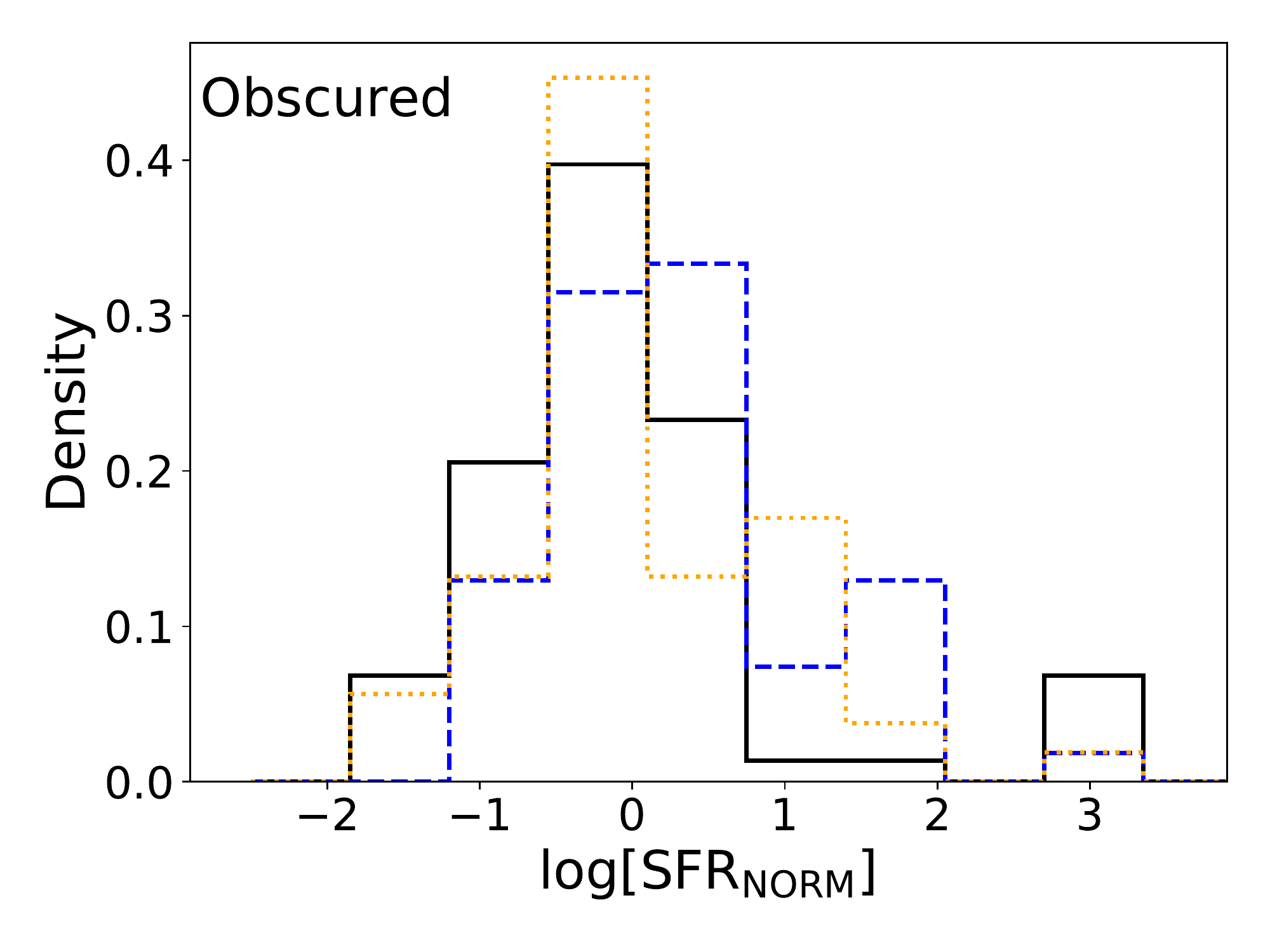} &
                      \includegraphics[width=0.30\textwidth]{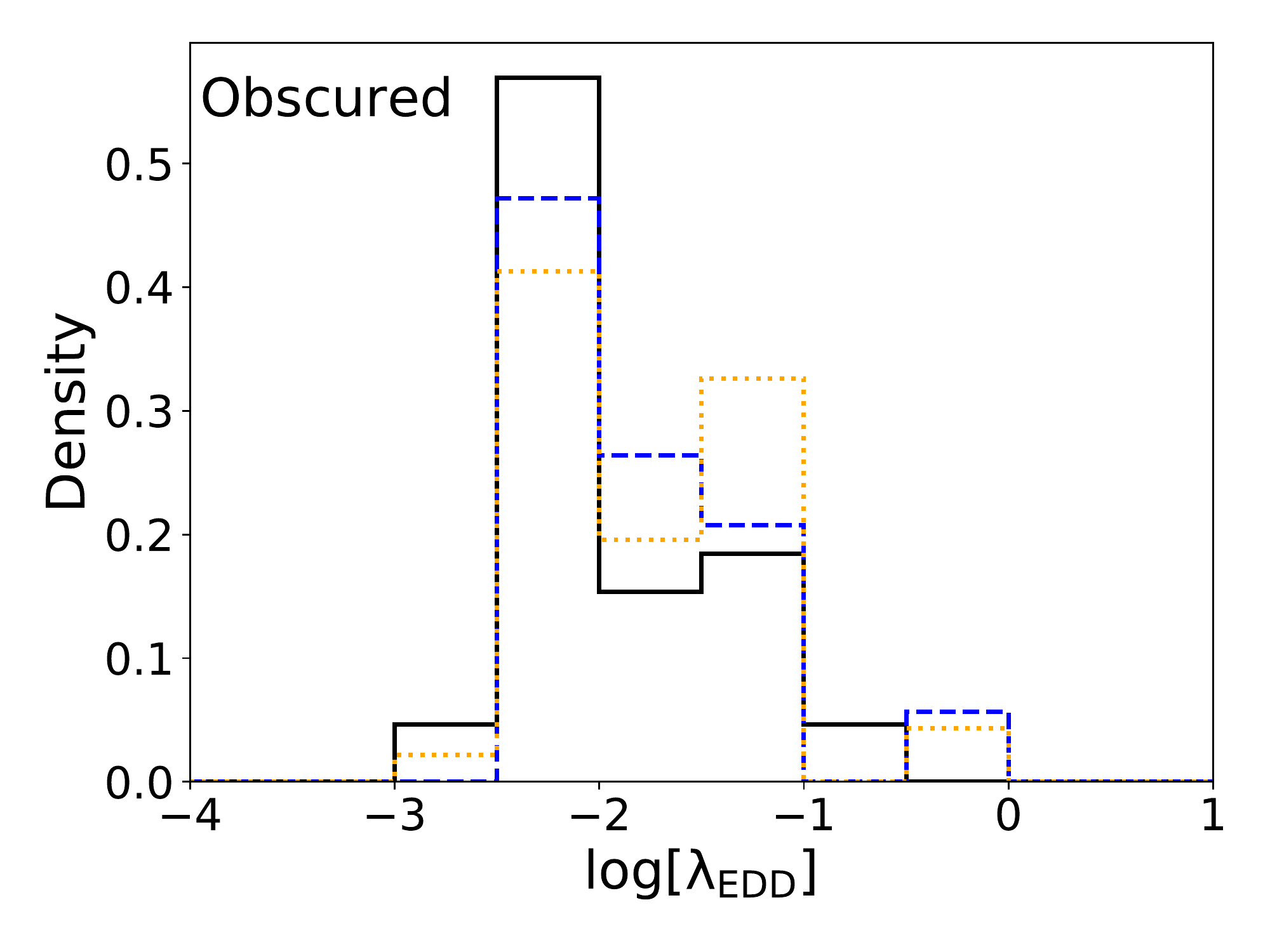} &
                      
                \includegraphics[width=0.30\textwidth]{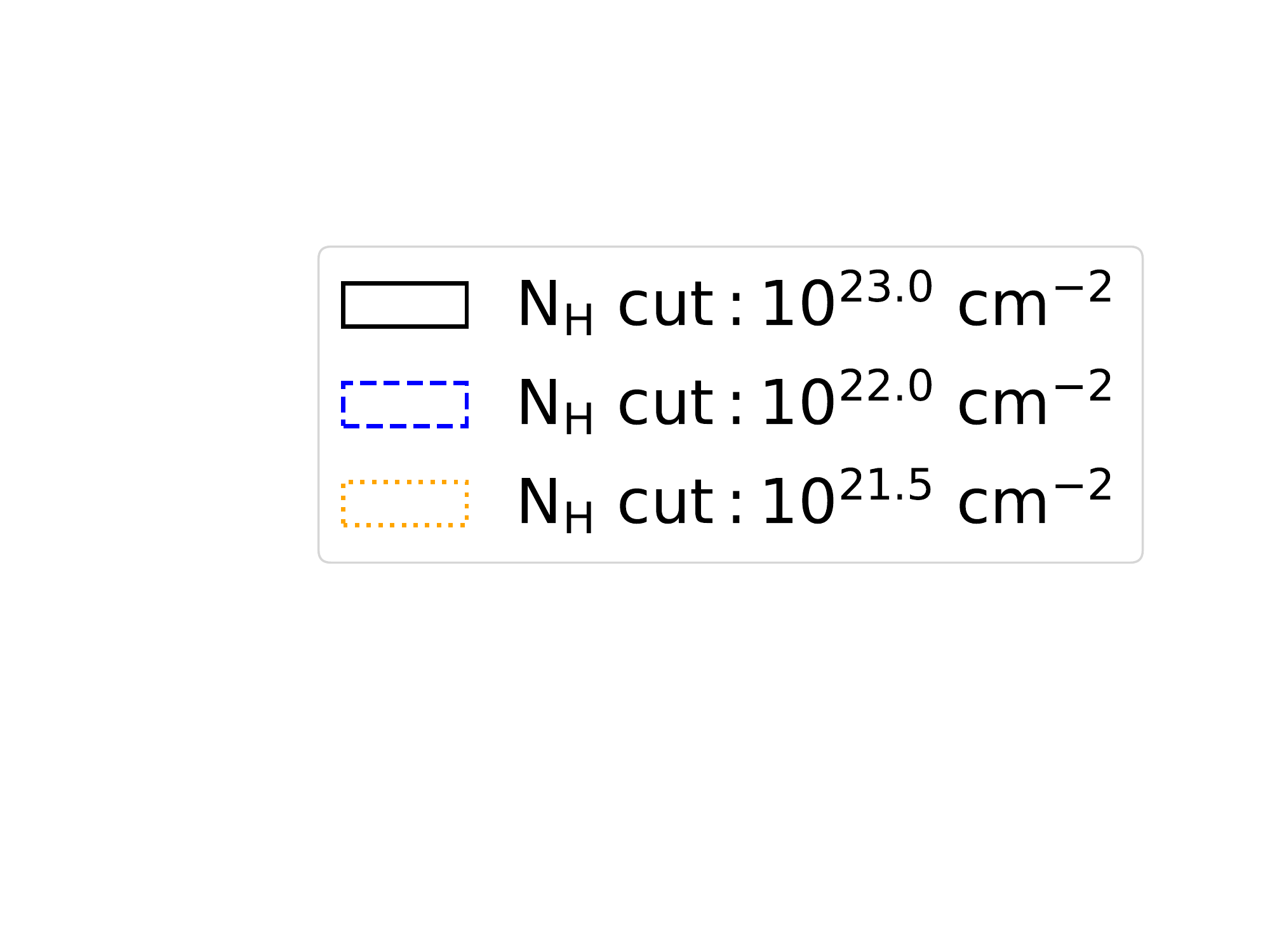}

           \\
              \includegraphics[width=0.30\textwidth]{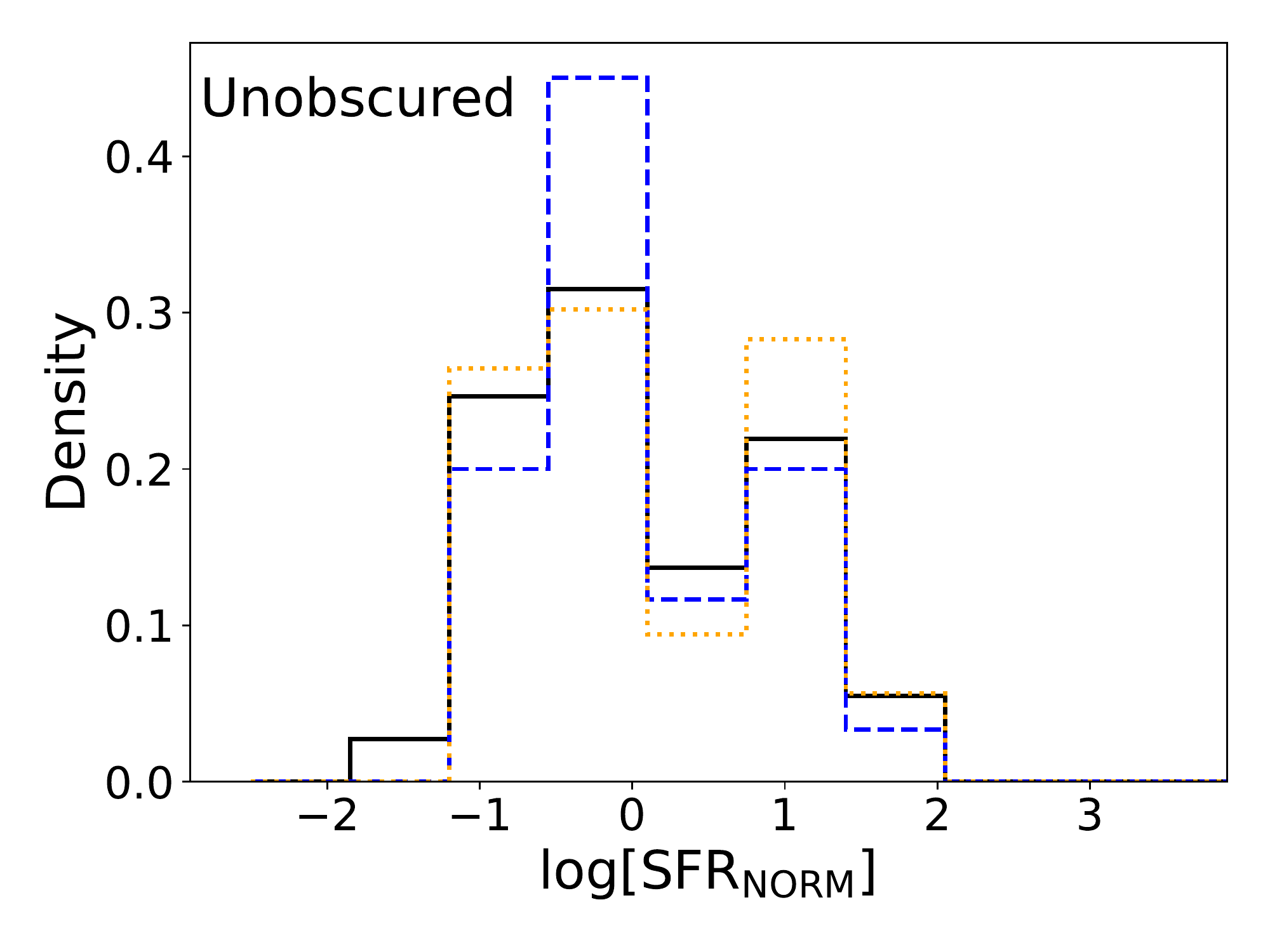} &         
              \includegraphics[width=0.30\textwidth]{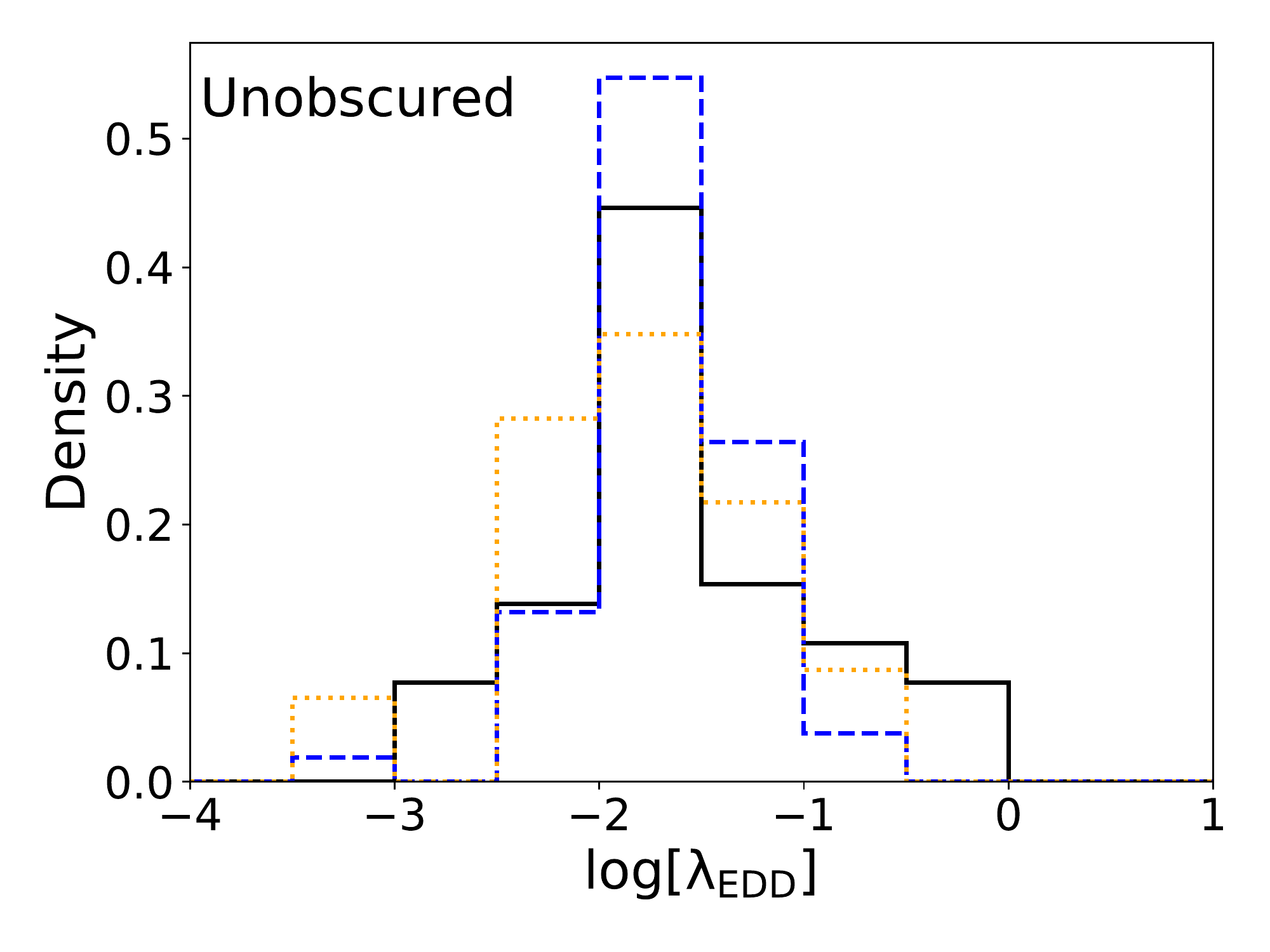} \\          
    
    \end{tabular}
\caption{The distributions of : a) Stellar mass b) calcium break $D_n(4000)$ c) sSFR
d) $\rm SFR_{NORM}$ e) Eddington ratio for the obscured (upper panel) and unobscured AGN (lower panel). 
The distributions are shown for 
three column density $\rm N_H$ cuts: $\rm \log N_H (cm^{-2})= 23$ (sold black line), $\rm \log N_H (cm^{-2})= 22$ (blue dash line) and $\rm \log N_H (cm^{-2})= 21.5$ (yellow dot line). 
}\label{KS-NH}
\end{figure*}
	
\begin{figure}
   \begin{tabular}{c}
       \includegraphics[width=0.47\textwidth]{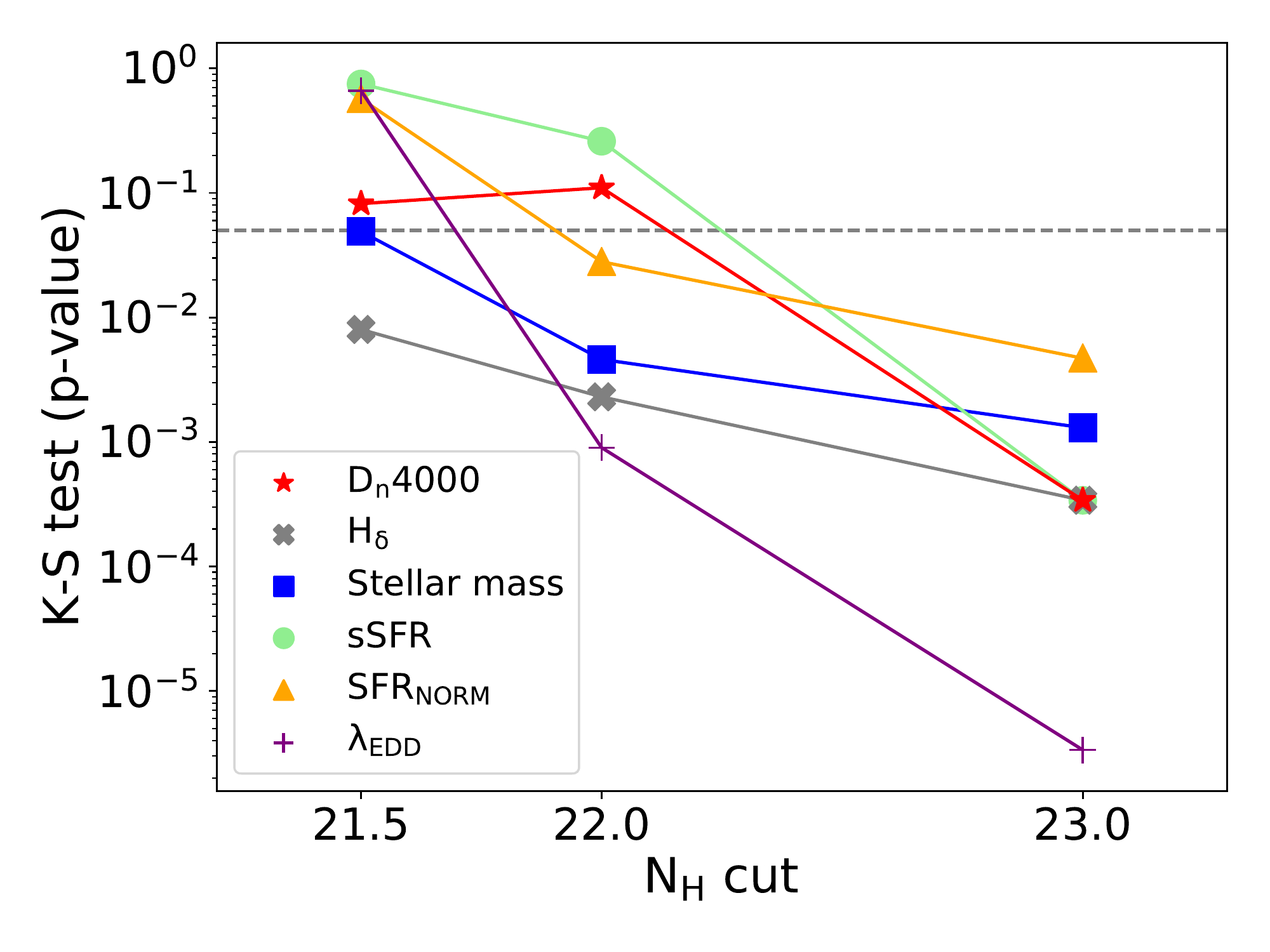} 
    \end{tabular}
\caption{The Kolmogorov-Smirnov p-values for the comparison of the obscured and unobscured population parameters as a function of 
the column density cut: a) stellar mass (blue, square box) b) $\rm D_n(4000)$ (red, star) c) sSFR (green, circle) d) $\rm SFR_{NORM}$ (orange, triangle) e) Eddington ratio (purple, cross). The horizontal line denotes the 95\% confidence level. 
  }\label{NHcut}
\end{figure}

\begin{table*}
\centering
\caption{Median values and K-S statistics for the obscured and unobscured sample in the case of 
normalising the samples with $M_\star$}
\begin{tabular}{lcccc}
\hline
 Property  &  $\rm \mu$ (Obscured)   & $\rm \mu$ (Unobscured)  & Distance  &  p-value   \\
            \hline& \\[-1.5ex]
 $\rm D_n$                & 1.53   $^{+0.05}_{-0.10}$    & 1.41 $^{+0.08}_{-0.028}$  &  0.273  &  $8\times10^{-3}$  \\ [0.1cm]
 $\rm H\delta$            & 2.82  $^{+0.27}_{-0.34}$     & 3.51$^{+0.17}_{-0.64}$   &  0.315  &   $1.3\times10^{-3}$   \\ [0.1cm]
% log[$\rm M_\star$]   &   $11.37^{+0.010}_{-0.048}$   &  $11.24^{+0.07}_{-0.045}$  &  0.32  & $1.3\times10^{-3}$            \\ [0.1cm]
 log[sSFR]                & -1.91 $^{+0.28}_{-0.03}$      & -1.723$^{+0.26}_{-0.35}$   &  0.205  &  0.092    \\ [0.1cm]
 log[$\rm SFR_{NORM}$]    &  0.064 $^{+0.178}_{-0.009}$       & -0.043$^{+0.23}_{-0.67}$     &  0.246  &   0.023    \\ [0.1cm]
 log[$\rm \lambda_{EDD}$] &  -1.91 $^{+0.22}_{-0.14}$     & -1.67$^{+0.13}_{-0.05}$  &  0.290  &   $7.5\times 10^{-3}$    \\  [0.1cm]
 \hline
\end{tabular}
    \label{KS_Mstar}
\end{table*}

\subsection{Eddington ratio distribution}
The Eddington ratio $\rm \lambda_E$ distribution provides  another key element in the 
study of the AGN demographics. It has been suggested that the radiation pressure exerted on 
the obscuring screen regulates the AGN obscuration \citep{Fabian2008,Fabian2009, Ricci2017}
 in the sense that the
 high radiation pressure pushes the obscuring torus further away and therefore decreases the obscuration. 
 According to this model high Eddington ratios should correspond primarily to unobscured AGN while the obscured AGN should 
have lower Eddington ratios. \cite{Ricci2017} present the $\rm N_H - \lambda_{EDD}$ plane for hard X-ray selected AGN 
 where they demonstrate that there is a strong separation between the obscured and unobscured AGN with the former occupying
 primarily the low $\rm \lambda$ high $\rm N_H$ region.
  More recently, \cite{Ananna2022} has provided a comprehensive study of the Eddington ratio distribution for obscured 
 and unobscured AGN in the local Universe using the {\it Gehrels-SWIFT}/BAT AGN sample. 
  They find that the Eddington ratio distribution of obscured AGN is significantly skewed towards lower 
  Eddington ratios. 
  %Our sample offers the opportunity to add the dimension of the galaxy age in such studies.
  In Fig. \ref{ricci}, we present the column density versus the Eddington ratio diagram. Our obscured sources have relatively low $\log[\lambda_{EDD}]<-0.8$. 
   All lie above the effective Eddington limit where the
outward radiation pressure on gas is equal to the inward gravitational pull. 
  We present the distribution of the Eddington ratio as a key in the 
  colour-coded diagram of Fig. \ref{HdD}. The histogram of the $\lambda_E$ is shown in Fig. \ref{lambda}.
   There is a tendency for obscured AGN to populate the lower Eddington ratio bins with a peak
    at $\rm \lambda_E \approx -2$. The two distributions are different at a highly statistically significant level according
    to the K-S test ($\rm p-value\sim 3\times10^{-6}$). Therefore our results  are in reasonable agreement with the findings of \cite{Ananna2022} in the local Universe.

\begin{table*}
\centering
\caption{Median values and K-S statistics for lower column density  $\rm N_H$ cuts}
\begin{tabular}{lcccccccc}
\hline
& \multicolumn{4}{c}{ $\rm \log N_H(cm^{-2})=21.5$} &\multicolumn{4}{c}{$\rm \log N_H(cm^{-2})=22$} \\
 Property            &  $\rm \mu_{ob}$  & $\rm \mu_{unob}$ & Dist  &  p-value   &  $\rm \mu_{ob}$  & $\rm \mu_{unob}$ & Dist  & p-value    \\
            \hline& \\[-1.5ex]
            
 $\rm D_n$     &    $1.53$    &  $1.45$  &  0.24  &  0.082   &   1.48  &  1.50  & 0.21 & 0.11\\ [0.1cm]
 $\rm H\delta$  &   $2.86$    & $4.21$  &  0.32  &   0.008 &3.55 & 2.27 & 0.33 & 0.0023  \\ [0.1cm]
 
 log[$\rm M_\star$]   &   $11.26$   &  $11.37$  &  0.026  & $0.049$   & 11.34 & 11.26 & 0.31 & 0.0046 \\ [0.1cm]
 
 log[sSFR]       &  $-1.92$     & $-1.97$  &  0.13  &  0.75 &  -1.99 & -1.97 & 0.18 & 0.26 \\ [0.1cm]
 log[$\rm SFR_{NORM}$] &   $-0.09$    & $-0.15$  &  0.14  &   $0.57$  & 0.24 & -0.04 & 0.260 & 0.028 \\ [0.1cm]
 log[$\rm \lambda_{EDD}$] &   $-1.83$    & $-1.81$  &  0.15  &   0.66 & -1.98 & -1.66 & 0.37 & $9\times10^{-4}$\\ [0.1cm]
 
 \hline
\end{tabular}
    \label{KSdifferentCut}
\end{table*}

\subsection{The relation between the galaxy age and stellar mass}
In the previous sections we argued that the host galaxies 
of the obscured AGN are on average older than those of the unobscured AGN. The mass distribution of obscured AGN appears  to be skewed towards higher masses. 
  At the same time, the host galaxies of obscured AGN appear to have lower star-formation  rates. 
  All  the above could suggest that the mass of the host galaxy is directly related to its age. 
  We plot the galaxy age $\rm D_n$ vs stellar mass in Fig. \ref{DnMsigma}  (upper panels) using as colour keys 
  the {\rm sSFR} and the Eddington ratio. 
  There does not appear to be an obvious relation between the stellar mass and the age in the sense that
  the host galaxies with a typical stellar masses of $M_{\star}=10^{11-11.5} M_\odot$ 
  cover a very wide range of galaxy ages. 
  The relation between the black hole mass and host galaxy age appears to apply as well to the normal (non-AGN) galaxies which have a mass of 
  $M_{\star}>10^{10.5} M_\odot$. In contrast, lower mass galaxies are 
  associated with young galaxies ($\rm D_n \approx 1.2$).

  \subsection{The relation between the galaxy age and black hole mass}  
In Fig. \ref{DnMsigma} (lower panel), there appears to be a trend for a correlation between the host galaxy age and $\rm \sigma^{\star}$ in the sense that the older the age the largest the velocity dispersion $\sigma^\star$ that is the more massive  the black hole. This trend is weakened by couple of 
 obscured AGN which have young ages $\rm D_n<1.4$ 
relative to their velocity dispersion $\rm \sigma_\star >250~km~s^{-1}$.
Note that the relation between galaxy age and stellar velocity dispersion 
is more prominent in  the underlying galaxy population. 
 Regarding the AGN population, the youngest AGN  host galaxies, are those which show  the highest Eddington ratios and highest star-formation rates.

\subsection{Normalising  with stellar mass}\label{mstarnorm}
In the previous sections we have demonstrated that the ages of the host galaxies of obscured AGN are higher compared to the non-obscured ones. 
The normalisation or weighing between the two populations was based on control samples using the X-ray luminosity and redshift \citep{Zou2019, Mountrichas2022b}. 
As the masses of the obscured AGN are higher than those of the unobscured ones, we explore whether the stellar 
mass  could be the underlying reason for the difference in age. We perform the weighting  method using control samples that has been described in section 
\ref{weights}. The results are presented in Table \ref{KS_Mstar}. The differences in the age indicators 
$\rm D_n$ and $H_\delta$ again imply that the age of the host galaxies of the obscured AGN population is older at a statistically significant level. 
The $\rm D_n$ and $\rm H_\delta$ indices distributions are presented in Fig. \ref{Dn_Mweight}.

\subsection{The effect of the absorbing column density cut} \label{differentNH}
In section \ref{thesample} we argued that we need to adopt a high absorbing column density cut to make sure that our the obscuration of our sources originates
 in the torus rather than the host galaxy. Here, we investigate the effect that a lower column density cut would have on our results. 
 We explore the usually adopted value of $\rm \log N_H(cm^{-2}) =22$
 which is most commonly used as the dividing line between obscured and unobscured AGN. In addition, we investigate the effect of an even lower 
 cut-off of $\rm \log N_H(cm^{-2}) =21.5$. This column density dividing line has been  proposed by \cite{Merloni2014} who argue that this brings the 
 samples of type-1 and type-2 objects (selected on the basis of optical spectroscopy) in relatively  good agreement with the unobscured and unobscured AGN samples (selected on the basis of X-ray spectroscopy). 
 We construct anew the obscured and unobscured AGN samples based on the 
 above criteria and examine their host galaxy properties. In Fig.\ref{NHcut}
 we present the distributions of the calcium break, the stellar mass, the sSFR, Eddington ratio for all three cuts. We present the new p-values for the Kolmogorov-Smirnov tests in Table \ref{KSdifferentCut}.  From the above plot and the table it becomes evident that the differences between the obscured and the unobscured AGN are less pronounced. 
  In particular, it appears that there is no difference in  $\rm D_n$ for 
  obscured and unobscured AGN. The stellar mass appears to be different at
  $\rm \log N_H(cm^{-2}) =22$ while at the dividing threshold of $\rm \log N_H(cm^{-2}) =21.5$ the stellar mass distributions are similar. 
  The effect of column density cut can be visualised in Fig.\ref{KS-NH} where 
 we plot the p-values for the different parameters as a function of the column density.

 \subsection{Comparison with theoretical models}
 Many theoretical models assert that AGN can be triggered by major mergers of massive gas rich galaxies \cite[e.g][]{Hopkins2006a, Hopkins2010}. The mergers cause the gas to lose angular momentum and eventually to fall in the centre of the galaxy. These mergers could resemble the IR luminous galaxies in the local Universe \citep{Sanders1988}. The nucleus should remain heavily obscured in this initial stage of the AGN. Observational evidence suggests that a large fraction of heavily obscured AGN are associated with mergers \citep{Koss2010, Koss2018, Kocevski2015, Lanzuisi2018}.
The nuclear gas will either feed the central SMBH or will be consumed in star-formation.
This phase is marked by co-existence  of copious star-formation and AGN  activity. When the AGN increases its Eddington ratio, powerful outflows push away the surrounding material in a short outburst phase \citep[see e.g.][]{Fabian2012, KingPounds2015}. In the Colour Magnitude Diagram, $ u-r ~vs ~ M_r$, the galaxy should move from the red cloud (absorbed phase) to the blue cloud (intense star formation phase)  to the green valley and eventually back to the red cloud when both the star formation and the AGN turn off \citep{Hickox2009}.

There is observational evidence that the major mergers may not be the dominant AGN triggering mechanism \citep[see][]{KormendyHo2013}.  \cite{Schawinski2011} and \cite{Kocevski2012} find that most moderate luminosity AGN are hosted by disk galaxies in the redshift range z=1.5-3.  Also \cite{Georgakakis2009} find that AGN hosted by disk galaxies contribute an appreciable fraction of the AGN luminosity density at z$\approx$0.8. Then it is likely that secular processes drive gas to the centre of the galaxy and trigger the AGN \citep{DraperBallantyne2012}. These secular processes include supernova winds, minor mergers, interactions with  other galaxies or cold accretion flow \citep{Kormendy2004}.

The current work shows that the host galaxies of the obscured AGN are on average older than those of obscured AGN. This is rather at odds with the major merger scenario where the obscured  phase marks the first stage of the AGN. As the host galaxies of the (moderate luminosity) obscured AGN in the COSMOS field appear to be at a later evolution stage compared to the unobscured AGN, this may be more 
suggestive of secular evolution. This obscured phase comes as a second phase following the high SFR and high Eddington ratio phase. In the obscured phase,    both the star-formation and the AGN accretion are abated. The low  Eddington ratios do not appear to be sufficiently high to clear the surrounding gas.  Alternatively when the Eddington ratios decrease the surrounding gas can fall 
back towards the centre increasing the obscuration.

\subsection{Outstanding questions}
The current work provides some interesting insights on the host galaxy age of the obscured and 
unobscured AGN population. It also suggests the presence of  significant differences between the host galaxy masses  as well as their SFR. 
 These differences may be explained  
by the difference  in galaxy age. 
 However, there are remaining questions that need to be addressed
in order to fully understand the differences of galaxy properties in 
the obscured and unobscured AGN populations. These questions are mostly related to the classification between the obscured (type-2) and unobscured (type-1) AGN.
\begin{itemize}
 \item{a) Column density limit. The differences in the host galaxy age appear to be pronounced  only 
for the heavily obscured sources with $\rm \log N_H(cm^{-2})=23$.
There
are no differences between the age of the obscured and unobscured 
populations when 
the column density threshold becomes $\rm \log N_H(cm^{-2})=22$
or $\rm \log N_H(cm^{-2})=21.5$. 
 The difference in stellar mass probably persists even at lower column densities of $\rm \log N_H(cm^{-2})=22$
 This is in agreement 
with previous results  by \cite{Mountrichas2021c} in the XBOOTES field.  Interestingly, \cite{Lanzuisi2017} and \cite{Buchner2017} show that there is a
strong correlation between the absorbing column density and the stellar mass
in X-ray selected AGN suggesting a difference in stellar mass 
 between the obscured and unobscured AGN.}
\item{b) Optical versus X-ray classification. Similar controversies occur when the classification between obscured and unobscured AGN is based on X-ray spectroscopy (or hardness ratio) and optical 
spectroscopy (i.e. redenning). 
Although we found a significant difference between the ages of obscured and unobscured AGN 
at a redshift of $\rm z\sim0.7$, the work on optically selected SDSS AGN by \cite{Kauffmann2003b} 
finds comparable ages for type-1 and type-2 AGN  at lower redshifts $\rm z<0.3$.}
% The differences in stellar mass found between type-1 and type-2  
% AGN in the COSMOS field \citep{Zou2019} were reproduced by our work
%  in the same field using an X-ray criterion for the obscuration.
%   This result is  confirmed in the XMM-XXL field when the 
%   two populations are identified using optical spectroscopy %\cite{Mountrichas2021b}.
    \item{c) Luminosity. There is a possibility that the differences with previous results 
    may not be attributed to the 
    classification method but rather on the luminosity. Owing to the deeper flux limit, the AGN 
    in the COSMOS field are less luminous compared to those in XBOOTES and XMM-XXL which have been used to 
    compare the host galaxy properties of obscured and unobscured sources \citep[e.g.][]{Masoura2021, Mountrichas2021c}. On the other hand, studies that have classified their AGN based on optical criteria 
    have either used the lower $\rm L_X$ sources of the COSMOS field \citep[e.g.][]{Zou2019} or have restricted their analysis to low 
    luminosity systems at low redshift \citep{Mountrichas2021b}. 
    It is then likely that the differences found in the 
    host galaxy properties of type-1 and type-2 AGN are because of their 
    lower luminosities of the AGN utilised rather
    than the different classification method (X-ray vs optical).}
     \end{itemize}

\section{Summary}
We examine the ages of the host galaxies of {\it Chandra} X-ray selected AGN 
in the COSMOS field in combination with other properties such as the stellar mass, the 
SFR and finally the SMBH Eddington ratio. The ages are explored using the calcium break and the Balmer
$H_{\delta}$ absorption line obtained from the LEGA-C VLT/VIMOS survey. 
  Accurate stellar masses and SFR  are derived using the spectral energy distribution code CIGALE.
  We compare for the first time the ages of the host galaxies of obscured and unobscured X-ray selected AGN. Our sample 
  consists of 50 unobscured or mildly obscured ($\rm \log N_H(cm^{-2} ) < 23$) and 23 
  heavily obscured AGN ($\rm \log N_H(cm^{-2} ) > 23$) in the redshift range $\rm 0.6<z<1$.
  One of our goals  is to test the predictions of evolutionary unification models which assert that
  heavily obscured sources mark the birth phase of the AGN. 
  Our main results can be summarised as follows.
  
  The AGN occupy the full range of ages from old systems to young galaxies. 
   They preferentially lie around a value of the calcium break of $\rm D_n \approx 1.4-1.5 $.
   This implies that AGN are associated with galaxies that have intermediate ages. 
   Our findings corroborate previous results
  by \cite{Hernan2014} but they are rather in tension with other results in the COSMOS field
  which assert that the X-ray selected AGN are preferentially associated with young 
  star-forming systems \citep{Silverman2009}. 
  
  One of the  main results of this work is  that obscured AGN are associated with older galaxies 
  having a median value for the calcium break of $\rm D_n \approx 1.53$ compared to 
  unobscured AGN which have $\rm D_n \approx 1.40$. A Kolmogorov-Smirnov test finds that the $D_n$ distributions of obscured and unobscured 
  AGN are different at a highly significant level (p-value $\rm 3\times10^{-4}$).
  
  There is some evidence that the stellar masses of obscured AGN are skewed towards 
  %(median $\rm \log M_{\star} (M_\odot) = 11.24$) 
  higher values compared to those of unobscured AGN 
  %(median $\rm \log M_{\star} (M_\odot) = 11.37 $), 
  $\rm p-value=1.3\times10^{-3}$. 
   This result  is in line with earlier findings based on the classification between type-1 and type-2 
  AGN following optical spectroscopy \citep{Zou2019, Mountrichas2021b, Koutoulidis2022}. 
  However, it is the first time that this is reported based on classification
  resulting  from X-ray spectroscopy. The difference in stellar mass is consistent with the older ages of 
  obscured AGN found above under the assumption that older galaxies are assocciated with 
  more massive galaxies \citep{Kauffmann2003b,Wu2018,Wilman2008}.
  Our results on the different ages of the host galaxies of the obscured and unobscured 
  AGN populations still apply when we normalise our samples according to stellar mass.
  
  The SFR of obscured AGN appears to be lower than that of unobscured AGN as demonstrated both by the comparison of the sSFR and the more sensitive $\rm SFR_{NORM}$ distributions
  for the obscured and the unobscured AGN. 
   This combined with the finding above that the Eddington ratios  in unobscured AGN are significantly higher than
   those of obscured AGN  implies that 
  high levels of SFR go hand in hand with high accretion rates possibly fed by the same gas depot.
  
  We have demonstrated that the above results regarding the host galaxy ages probably
  do not persevere for  lower column densities  ($\rm \log N_H (cm^{-2}) = 21 .5$ or 22). 
  
  Finally, the Eddington ratio distribution of the obscured AGN 
  are skewed towards lower values compared to the 
  unobscured AGN. This supports a scenario where the shape torus is regulated by the AGN radiation pressure as suggested by \cite{Ananna2022}.  
  
  All the above have  implications for the evolutionary unification models
   which are based on massive galaxy mergers.
  These models postulate that an AGN begins its life in a highly obscured phase. 
  When the Eddington rate increases the obscured screen is pushed away 
  by the radiation pressure and the AGN passes in an unobscured phase which is 
  accompanied by intense star-formation. The results presented in this paper 
  support that the obscured AGN on average are not hosted by the youngest galaxies in contrast to the above scenarios. 
  As the Eddington ratios of obscured AGN are lower than those of unobscured 
  AGN, this supports a scenario where the obscured AGN did not reach 
  sufficiently high levels of radiation pressure to blow away the obscuring screen.
  Instead, the lower SFR in obscured AGN in combination 
  with their lower Eddington ratios
  may be related to limited availability of fuel in the vicinity of the black hole. 
   As there are rather diverging results on the gas content  of AGN \citep{Maiolino1997, Perna2018},
   this hypothesis needs to be systematically explored using future ALMA observations of
   AGN. 
  
\begin{acknowledgements}
The authors are grateful to the anonymous referee for his careful reading 
of the manuscript and  numerous constructive comments. IG acknowledges financial support by the European Union’s Horizon 2020 programme “XMM2ATHENA” under grant agreement No 101004168. The research leading to these results has received funding (EP and IG) from the European Union’s Horizon 2020 Programme under the AHEAD2020 project (grant agreement n. 871158). GM acknowledges support by the Agencia Estatal de Investigación, Unidad de Excelencia María de Maeztu, ref. MDM-2017-0765. We acknowledge the use of LEGA-C data  which have been obtained 
 with the ESO Telescopes at the La Silla Paranal Observatory.
\end{acknowledgements}
 
\bibliography{ref}{}

\begin{thebibliography}{126}
\expandafter\ifx\csname natexlab\endcsname\relax\def\natexlab#1{#1}\fi

\bibitem[{Aird {et~al.}(2019)Aird, Coil, \& Georgakakis}]{Aird2019}
Aird, J., Coil, A.~L., \& Georgakakis, A. 2019, \mnras, 484, 4360

\bibitem[{Aird {et~al.}(2012)Aird, Coil, Moustakas, Blanton, Burles, Cool,
  Eisenstein, Smith, Wong, \& Zhu}]{Aird2012}
Aird, J., Coil, A.~L., Moustakas, J., {et~al.} 2012, \apj, 746, 90

\bibitem[{{Ananna} {et~al.}(2022){Ananna}, {Weigel}, {Trakhtenbrot}, {Koss},
  {Urry}, {Ricci}, {Hickox}, {Treister}, {Bauer}, {Ueda}, {Mushotzky}, {Ricci},
  {Oh}, {Mej{\'\i}a-Restrepo}, {Brok}, {Stern}, {Powell}, {Caglar}, {Ichikawa},
  {Wong}, {Harrison}, \& {Schawinski}}]{Ananna2022}
{Ananna}, T.~T., {Weigel}, A.~K., {Trakhtenbrot}, B., {et~al.} 2022, \apjs,
  261, 9

\bibitem[{Antonucci(1993)}]{Antonucci1993}
Antonucci, R. 1993, \araa, 31, 473

\bibitem[{{Arnaud}(1996)}]{Arnaud1996}
{Arnaud}, K.~A. 1996, in Astronomical Society of the Pacific Conference Series,
  Vol. 101, Astronomical Data Analysis Software and Systems V, ed. G.~H.
  {Jacoby} \& J.~{Barnes}, 17

\bibitem[{{Balogh} {et~al.}(1999){Balogh}, {Morris}, {Yee}, {Carlberg}, \&
  {Ellingson}}]{Balogh1999}
{Balogh}, M.~L., {Morris}, S.~L., {Yee}, H.~K.~C., {Carlberg}, R.~G., \&
  {Ellingson}, E. 1999, \apj, 527, 54

\bibitem[{{Banerji} {et~al.}(2021){Banerji}, {Jones}, {Carniani}, {DeGraf}, \&
  {Wagg}}]{Banerji2021}
{Banerji}, M., {Jones}, G.~C., {Carniani}, S., {DeGraf}, C., \& {Wagg}, J.
  2021, \mnras, 503, 5583

\bibitem[{Bernhard {et~al.}(2019)Bernhard, Grimmett, Mullaney, Daddi,
  Tadhunter, \& Jin}]{Bernhard2019}
Bernhard, E., Grimmett, L.~P., Mullaney, J.~R., {et~al.} 2019, \mnras, 483, L52

\bibitem[{{Blecha} {et~al.}(2018){Blecha}, {Snyder}, {Satyapal}, \&
  {Ellison}}]{Blecha2018}
{Blecha}, L., {Snyder}, G.~F., {Satyapal}, S., \& {Ellison}, S.~L. 2018,
  \mnras, 478, 3056

\bibitem[{Boquien {et~al.}(2019)Boquien, Burgarella, Roehlly, Buat, Ciesla,
  Corre, Inoue, \& Salas}]{Boquien2019}
Boquien, M., Burgarella, D., Roehlly, Y., {et~al.} 2019, \aap, 622, A103

\bibitem[{Brown {et~al.}(2019)Brown, Nayyeri, Cooray, Ma, Hickox, \&
  Azadi}]{Brown2019}
Brown, A., Nayyeri, H., Cooray, A., {et~al.} 2019, \apj, 871, 87

\bibitem[{Bruzual \& Charlot(2003)}]{Bruzual_Charlot2003}
Bruzual, G. \& Charlot, S. 2003, MNRAS, 344, 1000

\bibitem[{Buat {et~al.}(2019)Buat, Ciesla, Boquien, Ma{\l}ek, \&
  Burgarella}]{Buat2019}
Buat, V., Ciesla, L., Boquien, M., Ma{\l}ek, K., \& Burgarella, D. 2019, \aap,
  632, A79

\bibitem[{{Buat} {et~al.}(2021){Buat}, {Mountrichas}, {Yang}, {Boquien},
  {Roehlly}, {Burgarella}, {Stalevski}, {Ciesla}, \& {Theul{\'e}}}]{Buat2021}
{Buat}, V., {Mountrichas}, G., {Yang}, G., {et~al.} 2021, \aap, 654, A93

\bibitem[{{Buchner} \& {Bauer}(2017)}]{Buchner2017}
{Buchner}, J. \& {Bauer}, F.~E. 2017, \mnras, 465, 4348

\bibitem[{Charlot \& Fall(2000)}]{Charlot_Fall_2000}
Charlot, S. \& Fall, S.~M. 2000, ApJ, 539, 718

\bibitem[{{Chen} {et~al.}(2013){Chen}, {Hickox}, {Alberts}, {Brodwin}, {Jones},
  {Murray}, {Alexander}, {Assef}, {Brown}, {Dey}, {Forman}, {Gorjian},
  {Goulding}, {Le Floc'h}, {Jannuzi}, {Mullaney}, \& {Pope}}]{Chen2013}
{Chen}, C.-T.~J., {Hickox}, R.~C., {Alberts}, S., {et~al.} 2013, \apj, 773, 3

\bibitem[{Chen {et~al.}(2015)Chen, Hickox, Alberts, Harrison, Alexander, Assef,
  Brodwin, Brown, Moro, Forman, Gorjian, Goulding, Hainline, Jones, Kochanek,
  Murray, Pope, Rovilos, \& Stern}]{Chen2015}
Chen, C.-T.~J., Hickox, R.~C., Alberts, S., {et~al.} 2015, \apj, 802, 50

\bibitem[{Ciotti \& Ostriker(1997)}]{Ciotti1997}
Ciotti, L. \& Ostriker, J.~P. 1997, \apj, 487, L105

\bibitem[{Circosta {et~al.}(2019)Circosta, Vignali, Gilli, Feltre, Vito,
  Calura, Mainieri, Massardi, \& Norman}]{Circosta2019}
Circosta, C., Vignali, C., Gilli, R., {et~al.} 2019, \aap, 623, A172

\bibitem[{{Civano} {et~al.}(2016){Civano}, {Marchesi}, {Comastri}, {Urry},
  {Elvis}, {Cappelluti}, {Puccetti}, {Brusa}, {Zamorani}, {Hasinger},
  {Aldcroft}, {Alexander}, {Allevato}, {Brunner}, {Capak}, {Finoguenov},
  {Fiore}, {Fruscione}, {Gilli}, {Glotfelty}, {Griffiths}, {Hao}, {Harrison},
  {Jahnke}, {Kartaltepe}, {Karim}, {LaMassa}, {Lanzuisi}, {Miyaji}, {Ranalli},
  {Salvato}, {Sargent}, {Scoville}, {Schawinski}, {Schinnerer}, {Silverman},
  {Smolcic}, {Stern}, {Toft}, {Trakhtenbrot}, {Treister}, \&
  {Vignali}}]{Civano2016}
{Civano}, F., {Marchesi}, S., {Comastri}, A., {et~al.} 2016, \apj, 819, 62

\bibitem[{{Dale} {et~al.}(2014){Dale}, {Helou}, {Magdis}, {Armus},
  {D{\'{\i}}az-Santos}, \& {Shi}}]{Dale2014}
{Dale}, D.~A., {Helou}, G., {Magdis}, G.~E., {et~al.} 2014, ApJ, 784, 83

\bibitem[{{D'Amato} {et~al.}(2020){D'Amato}, {Gilli}, {Vignali}, {Massardi},
  {Pozzi}, {Zamorani}, {Circosta}, {Vito}, {Fritz}, {Cresci}, {Casasola},
  {Calura}, {Feltre}, {Manieri}, {Rigopoulou}, {Tozzi}, \&
  {Norman}}]{DAmato2020}
{D'Amato}, Q., {Gilli}, R., {Vignali}, C., {et~al.} 2020, \aap, 636, A37

\bibitem[{{Di Matteo} {et~al.}(2005){Di Matteo}, {Springel}, \&
  {Hernquist}}]{DiMatteo2005}
{Di Matteo}, T., {Springel}, V., \& {Hernquist}, L. 2005, Nature, 433, 604

\bibitem[{{Draper} \& {Ballantyne}(2012)}]{DraperBallantyne2012}
{Draper}, A.~R. \& {Ballantyne}, D.~R. 2012, \apj, 751, 72

\bibitem[{{Dressler} \& {Gunn}(1983)}]{Dressler_Gunn1983}
{Dressler}, A. \& {Gunn}, J.~E. 1983, \apj, 270, 7

\bibitem[{{Duras} {et~al.}(2020){Duras}, {Bongiorno}, {Ricci}, {Piconcelli},
  {Shankar}, {Lusso}, {Bianchi}, {Fiore}, {Maiolino}, {Marconi}, {Onori},
  {Sani}, {Schneider}, {Vignali}, \& {La Franca}}]{Duras2020}
{Duras}, F., {Bongiorno}, A., {Ricci}, F., {et~al.} 2020, \aap, 636, A73

\bibitem[{{Elbaz} {et~al.}(2007){Elbaz}, {Daddi}, {Le Borgne}, {Dickinson},
  {Alexander}, {Chary}, {Starck}, {Brandt}, {Kitzbichler}, {MacDonald},
  {Nonino}, {Popesso}, {Stern}, \& {Vanzella}}]{Elbaz2007}
{Elbaz}, D., {Daddi}, E., {Le Borgne}, D., {et~al.} 2007, \aap, 468, 33

\bibitem[{{Fabian}(2012)}]{Fabian2012}
{Fabian}, A.~C. 2012, \araa, 50, 455

\bibitem[{{Fabian} {et~al.}(2008){Fabian}, {Vasudevan}, \&
  {Gandhi}}]{Fabian2008}
{Fabian}, A.~C., {Vasudevan}, R.~V., \& {Gandhi}, P. 2008, \mnras, 385, L43

\bibitem[{{Fabian} {et~al.}(2009){Fabian}, {Vasudevan}, {Mushotzky}, {Winter},
  \& {Reynolds}}]{Fabian2009}
{Fabian}, A.~C., {Vasudevan}, R.~V., {Mushotzky}, R.~F., {Winter}, L.~M., \&
  {Reynolds}, C.~S. 2009, \mnras, 394, L89

\bibitem[{{Ferrarese} \& {Merritt}(2000)}]{Ferrarese_Merritt2000}
{Ferrarese}, L. \& {Merritt}, D. 2000, \apjl, 539, L9

\bibitem[{Florez {et~al.}(2020)Florez, Jogee, Sherman, Stevans, Finkelstein,
  Papovich, Kawinwanichakij, Ciardullo, Gronwall, Urry, Kirkpatrick, LaMassa,
  Ananna, \& Wold}]{Florez2020}
Florez, J., Jogee, S., Sherman, S., {et~al.} 2020, \mnras, 497, 3273

\bibitem[{{Fornasini} {et~al.}(2020){Fornasini}, {Civano}, \&
  {Suh}}]{Fornasini2020}
{Fornasini}, F.~M., {Civano}, F., \& {Suh}, H. 2020, \mnras, 495, 771

\bibitem[{{Gebhardt} {et~al.}(2000){Gebhardt}, {Kormendy}, {Ho}, {Bender},
  {Bower}, {Dressler}, {Faber}, {Filippenko}, {Green}, {Grillmair}, {Lauer},
  {Magorrian}, {Pinkney}, {Richstone}, \& {Tremaine}}]{Gebhardt2000}
{Gebhardt}, K., {Kormendy}, J., {Ho}, L.~C., {et~al.} 2000, \apjl, 543, L5

\bibitem[{{Georgakakis} {et~al.}(2009){Georgakakis}, {Coil}, {Laird},
  {Griffith}, {Nandra}, {Lotz}, {Pierce}, {Cooper}, {Newman}, \&
  {Koekemoer}}]{Georgakakis2009}
{Georgakakis}, A., {Coil}, A.~L., {Laird}, E.~S., {et~al.} 2009, \mnras, 397,
  623

\bibitem[{{Gilli} {et~al.}(2022){Gilli}, {Norman}, {Calura}, {Vito}, {Decarli},
  {Marchesi}, {Iwasawa}, {Comastri}, {Lanzuisi}, {Pozzi}, {D'Amato}, {Vignali},
  {Brusa}, {Mignoli}, \& {Cox}}]{Gilli2022}
{Gilli}, R., {Norman}, C., {Calura}, F., {et~al.} 2022, arXiv e-prints,
  arXiv:2206.03508

\bibitem[{{Glikman} {et~al.}(2018){Glikman}, {Lacy}, {LaMassa}, {Stern},
  {Djorgovski}, {Graham}, {Urrutia}, {Lovdal}, {Crnogorcevic}, {Daniels-Koch},
  {Hundal}, {Urry}, {Gates}, \& {Murray}}]{Glikman2018}
{Glikman}, E., {Lacy}, M., {LaMassa}, S., {et~al.} 2018, \apj, 861, 37

\bibitem[{{Harrison} {et~al.}(2018){Harrison}, {Costa}, {Tadhunter},
  {Fl{\"u}tsch}, {Kakkad}, {Perna}, \& {Vietri}}]{Harrison2018}
{Harrison}, C.~M., {Costa}, T., {Tadhunter}, C.~N., {et~al.} 2018, Nature
  Astronomy, 2, 198

\bibitem[{{Hatcher} {et~al.}(2021){Hatcher}, {Kirkpatrick}, {Fornasini},
  {Civano}, {Lambrides}, {Kocesvski}, {Carroll}, {Giavalisco}, {Hickox}, \&
  {Ji}}]{Hatcher2021}
{Hatcher}, C., {Kirkpatrick}, A., {Fornasini}, F., {et~al.} 2021, \aj, 162, 65

\bibitem[{{Hern{\'a}n-Caballero} {et~al.}(2014){Hern{\'a}n-Caballero},
  {Alonso-Herrero}, {P{\'e}rez-Gonz{\'a}lez}, {Barro}, {Aird}, {Ferreras},
  {Cava}, {Cardiel}, {Esquej}, {Gallego}, {Nandra}, \&
  {Rodr{\'\i}guez-Zaur{\'\i}n}}]{Hernan2014}
{Hern{\'a}n-Caballero}, A., {Alonso-Herrero}, A., {P{\'e}rez-Gonz{\'a}lez},
  P.~G., {et~al.} 2014, \mnras, 443, 3538

\bibitem[{{Hickox} {et~al.}(2009){Hickox}, {Jones}, {Forman}, {Murray},
  {Kochanek}, {Eisenstein}, {Jannuzi}, {Dey}, {Brown}, {Stern}, {Eisenhardt},
  {Gorjian}, {Brodwin}, {Narayan}, {Cool}, {Kenter}, {Caldwell}, \&
  {Anderson}}]{Hickox2009}
{Hickox}, R.~C., {Jones}, C., {Forman}, W.~R., {et~al.} 2009, \apj, 696, 891

\bibitem[{Hickox {et~al.}(2014)Hickox, Mullaney, Alexander, Chen, Civano, \&
  Goulding}]{Hickox2014}
Hickox, R.~C., Mullaney, J.~R., Alexander, D.~M., {et~al.} 2014, ApJ, 782, 11

\bibitem[{{Hopkins} \& {Hernquist}(2006)}]{Hopkins2006b}
{Hopkins}, P.~F. \& {Hernquist}, L. 2006, \apjs, 166, 1

\bibitem[{{Hopkins} {et~al.}(2008){Hopkins}, {Hernquist}, {Cox}, \&
  {Keres}}]{Hopkins2008a}
{Hopkins}, P.~F., {Hernquist}, L., {Cox}, T.~J., \& {Keres}, D. 2008, ApJS,
  175, 356

\bibitem[{{Hopkins} {et~al.}(2006){Hopkins}, {Hernquist}, {Cox}, {Robertson},
  {Di Matteo}, \& {Springel}}]{Hopkins2006a}
{Hopkins}, P.~F., {Hernquist}, L., {Cox}, T.~J., {et~al.} 2006, ApJ, 639, 700

\bibitem[{{Hopkins} {et~al.}(2010)}]{Hopkins2010}
{Hopkins}, P.~F. {et~al.} 2010, ApJ, 724, 915

\bibitem[{{Just} {et~al.}(2007){Just}, {Brandt}, {Shemmer}, {Steffen},
  {Schneider}, {Chartas}, \& {Garmire}}]{Just2007}
{Just}, D.~W., {Brandt}, W.~N., {Shemmer}, O., {et~al.} 2007, \apj, 685, 1004

\bibitem[{{Kauffmann} {et~al.}(2003){Kauffmann}, {Heckman}, {Tremonti},
  {Brinchmann}, {Charlot}, {White}, {Ridgway}, {Brinkmann}, {Fukugita}, {Hall},
  {Ivezi{\'c}}, {Richards}, \& {Schneider}}]{Kauffmann2003b}
{Kauffmann}, G., {Heckman}, T.~M., {Tremonti}, C., {et~al.} 2003, \mnras, 346,
  1055

\bibitem[{{Kelly} \& {Shen}(2013)}]{Kelly_Shen2013}
{Kelly}, B.~C. \& {Shen}, Y. 2013, \apj, 764, 45

\bibitem[{{King} \& {Pounds}(2015)}]{KingPounds2015}
{King}, A. \& {Pounds}, K. 2015, \araa, 53, 115

\bibitem[{{Kocevski} {et~al.}(2015){Kocevski}, {Brightman}, {Nandra},
  {Koekemoer}, {Salvato}, {Aird}, {Bell}, {Hsu}, {Kartaltepe}, {Koo}, {Lotz},
  {McIntosh}, {Mozena}, {Rosario}, \& {Trump}}]{Kocevski2015}
{Kocevski}, D.~D., {Brightman}, M., {Nandra}, K., {et~al.} 2015, \apj, 814, 104

\bibitem[{{Kocevski} {et~al.}(2012){Kocevski}, {Faber}, {Mozena}, {Koekemoer},
  {Nandra}, {Rangel}, {Laird}, {Brusa}, {Wuyts}, {Trump}, {Koo}, {Somerville},
  {Bell}, {Lotz}, {Alexander}, {Bournaud}, {Conselice}, {Dahlen}, {Dekel},
  {Donley}, {Dunlop}, {Finoguenov}, {Georgakakis}, {Giavalisco}, {Guo},
  {Grogin}, {Hathi}, {Juneau}, {Kartaltepe}, {Lucas}, {McGrath}, {McIntosh},
  {Mobasher}, {Robaina}, {Rosario}, {Straughn}, {van der Wel}, \&
  {Villforth}}]{Kocevski2012}
{Kocevski}, D.~D., {Faber}, S.~M., {Mozena}, M., {et~al.} 2012, \apj, 744, 148

\bibitem[{Kormendy \& Ho(2013)}]{KormendyHo2013}
Kormendy, J. \& Ho, L.~C. 2013, ARAA, 51, 511

\bibitem[{{Kormendy} \& {Kennicutt}(2004)}]{Kormendy2004}
{Kormendy}, J. \& {Kennicutt}, Robert~C., J. 2004, \araa, 42, 603

\bibitem[{{Koss} {et~al.}(2010){Koss}, {Mushotzky}, {Veilleux}, \&
  {Winter}}]{Koss2010}
{Koss}, M., {Mushotzky}, R., {Veilleux}, S., \& {Winter}, L. 2010, \apjl, 716,
  L125

\bibitem[{{Koss} {et~al.}(2018){Koss}, {Blecha}, {Bernhard}, {Hung}, {Lu},
  {Trakhtenbrot}, {Treister}, {Weigel}, {Sartori}, {Mushotzky}, {Schawinski},
  {Ricci}, {Veilleux}, \& {Sanders}}]{Koss2018}
{Koss}, M.~J., {Blecha}, L., {Bernhard}, P., {et~al.} 2018, \nat, 563, 214

\bibitem[{{Koutoulidis} {et~al.}(2022){Koutoulidis}, {Mountrichas},
  {Georgantopoulos}, {Pouliasis}, \& {Plionis}}]{Koutoulidis2022}
{Koutoulidis}, L., {Mountrichas}, G., {Georgantopoulos}, I., {Pouliasis}, E.,
  \& {Plionis}, M. 2022, \aap, 658, A35

\bibitem[{Laigle {et~al.}(2016)Laigle, McCracken, Ilbert, Hsieh, Davidzon,
  Capak, Hasinger, Silverman, Pichon, Coupon, Aussel, Borgne, Caputi, Cassata,
  Chang, Civano, Dunlop, Fynbo, Kartaltepe, Koekemoer, F{\`{e}}vre, Floc'h,
  Leauthaud, Lilly, Lin, Marchesi, Milvang-Jensen, Salvato, Sanders, Scoville,
  Smolcic, Stockmann, Taniguchi, Tasca, Toft, Vaccari, \& Zabl}]{Laigle2016}
Laigle, C., McCracken, H.~J., Ilbert, O., {et~al.} 2016, ApJS, 224, 24

\bibitem[{{Lanzuisi} {et~al.}(2018){Lanzuisi}, {Civano}, {Marchesi},
  {Comastri}, {Brusa}, {Gilli}, {Vignali}, {Zamorani}, {Brightman},
  {Griffiths}, \& {Koekemoer}}]{Lanzuisi2018}
{Lanzuisi}, G., {Civano}, F., {Marchesi}, S., {et~al.} 2018, \mnras, 480, 2578

\bibitem[{{Lanzuisi} {et~al.}(2017)}]{Lanzuisi2017}
{Lanzuisi}, G. {et~al.} 2017, \aap, 602, 13

\bibitem[{{Lusso} {et~al.}(2012)}]{Lusso2012}
{Lusso}, E. {et~al.} 2012, MNRAS, 425, 623

\bibitem[{{Lynden-Bell}(1969)}]{Lynden1969}
{Lynden-Bell}, D. 1969, \nat, 223, 690

\bibitem[{Magorrian {et~al.}(1998)}]{Magorrian1998}
Magorrian, J. {et~al.} 1998, AJ, 115, 2285

\bibitem[{Maiolino \& Rieke(1995)}]{Maiolino1995}
Maiolino, R. \& Rieke, G.~H. 1995, \apj, 454, 95

\bibitem[{{Maiolino} {et~al.}(1997){Maiolino}, {Ruiz}, {Rieke}, \&
  {Papadopoulos}}]{Maiolino1997}
{Maiolino}, R., {Ruiz}, M., {Rieke}, G.~H., \& {Papadopoulos}, P. 1997, \apj,
  485, 552

\bibitem[{Malizia {et~al.}(2020)Malizia, Bassani, Stephen, Bazzano, \&
  Ubertini}]{Malizia2020}
Malizia, A., Bassani, L., Stephen, J.~B., Bazzano, A., \& Ubertini, P. 2020,
  Astronomy {\&} Astrophysics, 639, A5

\bibitem[{{Malkan} {et~al.}(1998){Malkan}, {Gorjian}, \& {Tam}}]{Malkan1998}
{Malkan}, M.~A., {Gorjian}, V., \& {Tam}, R. 1998, \apjs, 117, 25

\bibitem[{{Marchesi} {et~al.}(2016{\natexlab{a}}){Marchesi}, {Civano}, {Elvis},
  {Salvato}, {Brusa}, {Comastri}, {Gilli}, {Hasinger}, {Lanzuisi}, {Miyaji},
  {Treister}, {Urry}, {Vignali}, {Zamorani}, {Allevato}, {Cappelluti},
  {Cardamone}, {Finoguenov}, {Griffiths}, {Karim}, {Laigle}, {LaMassa},
  {Jahnke}, {Ranalli}, {Schawinski}, {Schinnerer}, {Silverman}, {Smolcic},
  {Suh}, \& {Trakhtenbrot}}]{Marchesi2016a}
{Marchesi}, S., {Civano}, F., {Elvis}, M., {et~al.} 2016{\natexlab{a}}, \apj,
  817, 34

\bibitem[{{Marchesi} {et~al.}(2016{\natexlab{b}}){Marchesi}, {Lanzuisi},
  {Civano}, {Iwasawa}, {Suh}, {Comastri}, {Zamorani}, {Allevato}, {Griffiths},
  {Miyaji}, {Ranalli}, {Salvato}, {Schawinski}, {Silverman}, {Treister},
  {Urry}, \& {Vignali}}]{Marchesi2016b}
{Marchesi}, S., {Lanzuisi}, G., {Civano}, F., {et~al.} 2016{\natexlab{b}},
  \apj, 830, 100

\bibitem[{{Marconi} {et~al.}(2004){Marconi}, {Risaliti}, {Gilli}, {Hunt},
  {Maiolino}, \& {Salvati}}]{Marconi2004}
{Marconi}, A., {Risaliti}, G., {Gilli}, R., {et~al.} 2004, \mnras, 351, 169

\bibitem[{Masoura {et~al.}(2021)Masoura, Mountrichas, Georgantopoulos, \&
  Plionis}]{Masoura2021}
Masoura, V.~A., Mountrichas, G., Georgantopoulos, I., \& Plionis, M. 2021,
  \aap, 646, A167

\bibitem[{Masoura {et~al.}(2018)Masoura, Mountrichas, Georgantopoulos, Ruiz,
  Magdis, \& Plionis}]{Masoura2018}
Masoura, V.~A., Mountrichas, G., Georgantopoulos, I., {et~al.} 2018, \aap, 618,
  31

\bibitem[{Merloni {et~al.}(2014)Merloni, Bongiorno, Brusa, Iwasawa, Mainieri,
  Magnelli, Salvato, Berta, Cappelluti, Comastri, Fiore, Gilli, \&
  Koekemoer}]{Merloni2014}
Merloni, A., Bongiorno, A., Brusa, M., {et~al.} 2014, \mnras, 437, 3550

\bibitem[{{Merritt} \& {Ferrarese}(2001)}]{Merritt_Ferrarese2001}
{Merritt}, D. \& {Ferrarese}, L. 2001, \mnras, 320, L30

\bibitem[{{Mineo} {et~al.}(2012){Mineo}, {Gilfanov}, \& {Sunyaev}}]{Mineo2012}
{Mineo}, S., {Gilfanov}, M., \& {Sunyaev}, R. 2012, \mnras, 426, 1870

\bibitem[{{Mountrichas} {et~al.}(2021b){Mountrichas}, {Buat},
  {Georgantopoulos}, {Yang}, {Masoura}, {Boquien}, \&
  {Burgarella}}]{Mountrichas2021b}
{Mountrichas}, G., {Buat}, V., {Georgantopoulos}, I., {et~al.} 2021b, \aap,
  653, A70

\bibitem[{Mountrichas {et~al.}(2021a)Mountrichas, Buat, Yang, Boquien,
  Burgarella, \& Ciesla}]{Mountrichas2021a}
Mountrichas, G., Buat, V., Yang, G., {et~al.} 2021a, \aap, 646, A29

\bibitem[{{Mountrichas} {et~al.}(2021c){Mountrichas}, {Buat}, {Yang},
  {Boquien}, {Burgarella}, {Ciesla}, {Malek}, \& {Shirley}}]{Mountrichas2021c}
{Mountrichas}, G., {Buat}, V., {Yang}, G., {et~al.} 2021c, \aap, 653, A74

\bibitem[{{Mountrichas} {et~al.}(2022{\natexlab{a}}){Mountrichas}, {Buat},
  {Yang}, {Boquien}, {Burgarella}, {Ciesla}, {Malek}, \&
  {Shirley}}]{Mountrichas2022b}
{Mountrichas}, G., {Buat}, V., {Yang}, G., {et~al.} 2022{\natexlab{a}}, \aap,
  663, A130

\bibitem[{{Mountrichas} {et~al.}(2022{\natexlab{b}}){Mountrichas}, {Buat},
  {Yang}, {Boquien}, {Ni}, {Pouliasis}, {Burgarella}, {Theule}, \&
  {Georgantopoulos}}]{Mountrichas2022c}
{Mountrichas}, G., {Buat}, V., {Yang}, G., {et~al.} 2022{\natexlab{b}}, \aap,
  667, A145

\bibitem[{{Mountrichas} {et~al.}(2022{\natexlab{c}}){Mountrichas}, {Masoura},
  {Xilouris}, {Georgantopoulos}, {Buat}, \& {Paspaliaris}}]{Mountrichas2022a}
{Mountrichas}, G., {Masoura}, V.~A., {Xilouris}, E.~M., {et~al.}
  2022{\natexlab{c}}, \aap, 661, A108

\bibitem[{{Mountrichas} \& {Shankar}(2023)}]{Mountrichas2023}
{Mountrichas}, G. \& {Shankar}, F. 2023, \mnras, 518, 2088

\bibitem[{{Mullaney} {et~al.}(2015){Mullaney}, {Del-Moro}, {Aird}, {Alexander},
  {Civano}, {Hickox}, {Lansbury}, {Ajello}, {Assef}, {Ballantyne}, {Balokovi},
  {Bauer}, {Brandt}, {Boggs}, {Brightman}, {Christensen}, {Comastri}, {Craig},
  {Elvis}, {Forster}, {Gandhi}, {Grefenstette}, {Hailey}, {Harrison}, {Koss},
  {LaMassa}, {Luo}, {Madsen}, {Puccetti}, {Saez}, {Stern}, {Treister}, {Urry},
  {Wik}, {Zappacosta}, \& {Zhang}}]{Mullaney2015}
{Mullaney}, J.~R., {Del-Moro}, A., {Aird}, J., {et~al.} 2015, \apj, 808, 185

\bibitem[{{Muzzin} {et~al.}(2013){Muzzin}, {Marchesini}, {Stefanon}, {Franx},
  {Milvang-Jensen}, {Dunlop}, {Fynbo}, {Brammer}, {Labb{\'e}}, \& {van
  Dokkum}}]{Muzzin2013}
{Muzzin}, A., {Marchesini}, D., {Stefanon}, M., {et~al.} 2013, \apjs, 206, 8

\bibitem[{Park {et~al.}(2006)Park, Kashyap, Siemiginowska, van Dyk, Zezas,
  Heinke, \& Wargelin}]{Park2006}
Park, T., Kashyap, V.~L., Siemiginowska, A., {et~al.} 2006, \apj, 652, 610

\bibitem[{{Perna} {et~al.}(2018){Perna}, {Sargent}, {Brusa}, {Daddi},
  {Feruglio}, {Cresci}, {Lanzuisi}, {Lusso}, {Comastri}, {Coogan}, {D'Amato},
  {Gilli}, {Piconcelli}, \& {Vignali}}]{Perna2018}
{Perna}, M., {Sargent}, M.~T., {Brusa}, M., {et~al.} 2018, \aap, 619, A90

\bibitem[{{Pouliasis} {et~al.}(2022){Pouliasis}, {Mountrichas},
  {Georgantopoulos}, {Ruiz}, {Gilli}, {Koulouridis}, {Akiyama}, {Ueda},
  {Garrel}, {Nagao}, {Paltani}, {Pierre}, {Toba}, \& {Vignali}}]{Pouliasis2022}
{Pouliasis}, E., {Mountrichas}, G., {Georgantopoulos}, I., {et~al.} 2022, \aap,
  667, A56

\bibitem[{{Ranalli} {et~al.}(2003){Ranalli}, {Comastri}, \&
  {Setti}}]{Ranalli2003}
{Ranalli}, P., {Comastri}, A., \& {Setti}, G. 2003, \aap, 399, 39

\bibitem[{{Ricci} {et~al.}(2017){Ricci}, {Trakhtenbrot}, {Koss}, {Ueda},
  {Schawinski}, {Oh}, {Lamperti}, {Mushotzky}, {Treister}, {Ho}, {Weigel},
  {Bauer}, {Paltani}, {Fabian}, {Xie}, \& {Gehrels}}]{Ricci2017}
{Ricci}, C., {Trakhtenbrot}, B., {Koss}, M.~J., {et~al.} 2017, \nat, 549, 488

\bibitem[{{Rodighiero} {et~al.}(2015){Rodighiero}, {Brusa}, {Daddi},
  {Negrello}, {Mullaney}, {Delvecchio}, {Lutz}, {Renzini}, {Franceschini},
  {Baronchelli}, {Pozzi}, {Gruppioni}, {Strazzullo}, {Cimatti}, \&
  {Silverman}}]{Rodighiero2015}
{Rodighiero}, G., {Brusa}, M., {Daddi}, E., {et~al.} 2015, \apjl, 800, L10

\bibitem[{{Rosario} {et~al.}(2012){Rosario}, {Santini}, {Lutz}, {Shao},
  {Maiolino}, {Alexander}, {Altieri}, {Andreani}, {Aussel}, {Bauer}, {Berta},
  {Bongiovanni}, {Brandt}, {Brusa}, {Cepa}, {Cimatti}, {Cox}, {Daddi}, {Elbaz},
  {Fontana}, {F{\"o}rster Schreiber}, {Genzel}, {Grazian}, {Le Floch},
  {Magnelli}, {Mainieri}, {Netzer}, {Nordon}, {P{\'e}rez Garcia}, {Poglitsch},
  {Popesso}, {Pozzi}, {Riguccini}, {Rodighiero}, {Salvato}, {Sanchez-Portal},
  {Sturm}, {Tacconi}, {Valtchanov}, \& {Wuyts}}]{Rosario2012}
{Rosario}, D.~J., {Santini}, P., {Lutz}, D., {et~al.} 2012, \aap, 545, A45

\bibitem[{{Rovilos} {et~al.}(2012){Rovilos}, {Comastri}, {Gilli},
  {Georgantopoulos}, {Ranalli}, {Vignali}, {Lusso}, {Cappelluti}, {Zamorani},
  {Elbaz}, {Dickinson}, {Hwang}, {Charmandaris}, {Ivison}, {Merloni}, {Daddi},
  {Carrera}, {Brandt}, {Mullaney}, {Scott}, {Alexander}, {Del Moro},
  {Morrison}, {Murphy}, {Altieri}, {Aussel}, {Dannerbauer}, {Kartaltepe},
  {Leiton}, {Magdis}, {Magnelli}, {Popesso}, \& {Valtchanov}}]{rovilos2012}
{Rovilos}, E., {Comastri}, A., {Gilli}, R., {et~al.} 2012, \aap, 546, A58

\bibitem[{{Sanders} {et~al.}(1988){Sanders}, {Soifer}, {Elias}, {Neugebauer},
  \& {Matthews}}]{Sanders1988}
{Sanders}, D.~B., {Soifer}, B.~T., {Elias}, J.~H., {Neugebauer}, G., \&
  {Matthews}, K. 1988, \apjl, 328, L35

\bibitem[{{Schawinski} {et~al.}(2011){Schawinski}, {Treister}, {Urry},
  {Cardamone}, {Simmons}, \& {Yi}}]{Schawinski2011}
{Schawinski}, K., {Treister}, E., {Urry}, C.~M., {et~al.} 2011, \apjl, 727, L31

\bibitem[{{Schreiber} {et~al.}(2015){Schreiber}, {Pannella}, {Elbaz},
  {B{\'e}thermin}, {Inami}, {Dickinson}, {Magnelli}, {Wang}, {Aussel}, {Daddi},
  {Juneau}, {Shu}, {Sargent}, {Buat}, {Faber}, {Ferguson}, {Giavalisco},
  {Koekemoer}, {Magdis}, {Morrison}, {Papovich}, {Santini}, \&
  {Scott}}]{Schreiber2015}
{Schreiber}, C., {Pannella}, M., {Elbaz}, D., {et~al.} 2015, \aap, 575, A74

\bibitem[{{Schulze} {et~al.}(2015){Schulze}, {Bongiorno}, {Gavignaud},
  {Schramm}, {Silverman}, {Merloni}, {Zamorani}, {Hirschmann}, {Mainieri},
  {Wisotzki}, {Shankar}, {Fiore}, {Koekemoer}, \& {Temporin}}]{Schulze2015}
{Schulze}, A., {Bongiorno}, A., {Gavignaud}, I., {et~al.} 2015, \mnras, 447,
  2085

\bibitem[{{Schulze} \& {Wisotzki}(2010)}]{Schulze_Wisotzki2010}
{Schulze}, A. \& {Wisotzki}, L. 2010, \aap, 516, A87

\bibitem[{{Scoville} {et~al.}(2007){Scoville}, {Aussel}, {Benson}, {Blain},
  {Calzetti}, {Capak}, {Ellis}, {El-Zant}, {Finoguenov}, {Giavalisco}, {Guzzo},
  {Hasinger}, {Koda}, {Le F{\`e}vre}, {Massey}, {McCracken}, {Mobasher},
  {Renzini}, {Rhodes}, {Salvato}, {Sanders}, {Sasaki}, {Schinnerer}, {Sheth},
  {Shopbell}, {Taniguchi}, {Taylor}, \& {Thompson}}]{Scoville2007}
{Scoville}, N., {Aussel}, H., {Benson}, A., {et~al.} 2007, \apjs, 172, 150

\bibitem[{{Scoville} {et~al.}(2017){Scoville}, {Lee}, {Vanden Bout},
  {Diaz-Santos}, {Sanders}, {Darvish}, {Bongiorno}, {Casey}, {Murchikova},
  {Koda}, {Capak}, {Vlahakis}, {Ilbert}, {Sheth}, {Morokuma-Matsui}, {Ivison},
  {Aussel}, {Laigle}, {McCracken}, {Armus}, {Pope}, {Toft}, \&
  {Masters}}]{Scoville2017}
{Scoville}, N., {Lee}, N., {Vanden Bout}, P., {et~al.} 2017, \apj, 837, 150

\bibitem[{Shimizu {et~al.}(2015)Shimizu, Mushotzky, Mel{\'{e}}ndez, Koss, \&
  Rosario}]{Shimizu2015}
Shimizu, T.~T., Mushotzky, R.~F., Mel{\'{e}}ndez, M., Koss, M., \& Rosario,
  D.~J. 2015, \mnras, 452, 1841

\bibitem[{Shimizu {et~al.}(2017)Shimizu, Mushotzky, Mel{\'{e}}ndez, Koss,
  Barger, \& Cowie}]{Shimizu2017}
Shimizu, T.~T., Mushotzky, R.~F., Mel{\'{e}}ndez, M., {et~al.} 2017, \mnras,
  466, 3161

\bibitem[{Shirley {et~al.}(2019)Shirley, Roehlly, Hurley, Buat, del Carmen
  Campos~Varillas, Duivenvoorden, Duncan, Efstathiou, Farrah, Solares, Malek,
  Marchetti, McCheyne, Papadopoulos, Pons, Scipioni, Vaccari, \&
  Oliver}]{Shirley2019}
Shirley, R., Roehlly, Y., Hurley, P.~D., {et~al.} 2019, \mnras, 490, 634

\bibitem[{{Silk} \& {Rees}(1998)}]{Silk1998}
{Silk}, J. \& {Rees}, M.~J. 1998, \aap, 331, L1

\bibitem[{{Silverman} {et~al.}(2009){Silverman}, {Lamareille}, {Maier},
  {Lilly}, {Mainieri}, {Brusa}, {Cappelluti}, {Hasinger}, {Zamorani},
  {Scodeggio}, {Bolzonella}, {Contini}, {Carollo}, {Jahnke}, {Kneib}, {Le
  F{\`e}vre}, {Merloni}, {Bardelli}, {Bongiorno}, {Brunner}, {Caputi},
  {Civano}, {Comastri}, {Coppa}, {Cucciati}, {de la Torre}, {de Ravel},
  {Elvis}, {Finoguenov}, {Fiore}, {Franzetti}, {Garilli}, {Gilli}, {Iovino},
  {Kampczyk}, {Knobel}, {Kova{\v{c}}}, {Le Borgne}, {Le Brun}, {Mignoli},
  {Pello}, {Peng}, {Perez Montero}, {Ricciardelli}, {Tanaka}, {Tasca},
  {Tresse}, {Vergani}, {Vignali}, {Zucca}, {Bottini}, {Cappi}, {Cassata},
  {Fumana}, {Griffiths}, {Kartaltepe}, {Koekemoer}, {Marinoni}, {McCracken},
  {Memeo}, {Meneux}, {Oesch}, {Porciani}, \& {Salvato}}]{Silverman2009}
{Silverman}, J.~D., {Lamareille}, F., {Maier}, C., {et~al.} 2009, \apj, 696,
  396

\bibitem[{{Somerville} {et~al.}(2008){Somerville}, {Hopkins}, J., {Robertson},
  \& {Hernquist}}]{Somerville2008}
{Somerville}, R.~S., {Hopkins}, P.~F., J., C.~T., {Robertson}, B.~E., \&
  {Hernquist}, L. 2008, MNRAS, 391, 481

\bibitem[{{Speagle} {et~al.}(2014){Speagle}, {Steinhardt}, {Capak}, \&
  {Silverman}}]{Speagle2014}
{Speagle}, J.~S., {Steinhardt}, C.~L., {Capak}, P.~L., \& {Silverman}, J.~D.
  2014, \apjs, 214, 15

\bibitem[{Stalevski {et~al.}(2012)Stalevski, Fritz, Baes, Nakos, \&
  Popovi{\'{c}}}]{Stalevski2012}
Stalevski, M., Fritz, J., Baes, M., Nakos, T., \& Popovi{\'{c}}, L.~{\v{C}}.
  2012, \mnras, 420, 2756

\bibitem[{Stalevski {et~al.}(2016)Stalevski, Ricci, Ueda, Lira, Fritz, \&
  Baes}]{Stalevski2016}
Stalevski, M., Ricci, C., Ueda, Y., {et~al.} 2016, \mnras, 458, 2288

\bibitem[{{Stanley} {et~al.}(2017){Stanley}, {Alexander}, {Harrison},
  {Rosario}, {Wang}, {Aird}, {Bourne}, {Dunne}, {Dye}, {Eales}, {Knudsen},
  {Micha{\l}owski}, {Valiante}, {De Zotti}, {Furlanetto}, {Ivison}, {Maddox},
  \& {Smith}}]{Stanley2017}
{Stanley}, F., {Alexander}, D.~M., {Harrison}, C.~M., {et~al.} 2017, \mnras,
  472, 2221

\bibitem[{Stanley {et~al.}(2015)Stanley, Harrison, Alexander, Swinbank, Aird,
  Moro, Hickox, \& Mullaney}]{Stanley2015}
Stanley, F., Harrison, C.~M., Alexander, D.~M., {et~al.} 2015, \mnras, 453, 591

\bibitem[{{Straatman} {et~al.}(2018){Straatman}, {van der Wel}, {Bezanson},
  {Pacifici}, {Gallazzi}, {Wu}, {Noeske}, {Bari{\v{s}}i{\'c}}, {Bell},
  {Brammer}, {Calhau}, {Chauke}, {Franx}, {van Houdt}, {Labb{\'e}}, {Maseda},
  {Mu{\~n}oz-Mateos}, {Muzzin}, {van de Sande}, {Sobral}, \&
  {Spilker}}]{Straatman2018}
{Straatman}, C. M.~S., {van der Wel}, A., {Bezanson}, R., {et~al.} 2018, \apjs,
  239, 27

\bibitem[{{Suh} {et~al.}(2019){Suh}, {Civano}, {Hasinger}, {Lusso}, {Marchesi},
  {Schulze}, {Onodera}, {Rosario}, \& {Sanders}}]{Suh2019}
{Suh}, H., {Civano}, F., {Hasinger}, G., {et~al.} 2019, \apj, 872, 168

\bibitem[{{Sutherland} \& {Saunders}(1992)}]{Sutherland_and_Saunders1992}
{Sutherland}, W. \& {Saunders}, W. 1992, MNRAS, 259, 413

\bibitem[{{Tacconi} {et~al.}(2018){Tacconi}, {Genzel}, {Saintonge}, {Combes},
  {Garc{\'\i}a-Burillo}, {Neri}, {Bolatto}, {Contini}, {F{\"o}rster Schreiber},
  {Lilly}, {Lutz}, {Wuyts}, {Accurso}, {Boissier}, {Boone}, {Bouch{\'e}},
  {Bournaud}, {Burkert}, {Carollo}, {Cooper}, {Cox}, {Feruglio}, {Freundlich},
  {Herrera-Camus}, {Juneau}, {Lippa}, {Naab}, {Renzini}, {Salome}, {Sternberg},
  {Tadaki}, {{\"U}bler}, {Walter}, {Weiner}, \& {Weiss}}]{Tacconi2018}
{Tacconi}, L.~J., {Genzel}, R., {Saintonge}, A., {et~al.} 2018, \apj, 853, 179

\bibitem[{{Torbaniuk} {et~al.}(2021){Torbaniuk}, {Paolillo}, {Carrera},
  {Cavuoti}, {Vignali}, {Longo}, \& {Aird}}]{Torbaniuk2021}
{Torbaniuk}, O., {Paolillo}, M., {Carrera}, F., {et~al.} 2021, \mnras, 506,
  2619

\bibitem[{{van der Wel} {et~al.}(2021){van der Wel}, {Bezanson}, {D'Eugenio},
  {Straatman}, {Franx}, {van Houdt}, {Maseda}, {Gallazzi}, {Wu}, {Pacifici},
  {Barisic}, {Brammer}, {Munoz-Mateos}, {Vervalcke}, {Zibetti}, {Sobral}, {de
  Graaff}, {Calhau}, {Kaushal}, {Muzzin}, {Bell}, \& {van
  Dokkum}}]{VanderWel2021}
{van der Wel}, A., {Bezanson}, R., {D'Eugenio}, F., {et~al.} 2021, \apjs, 256,
  44

\bibitem[{{van der Wel} {et~al.}(2016){van der Wel}, {Noeske}, {Bezanson},
  {Pacifici}, {Gallazzi}, {Franx}, {Mu{\~n}oz-Mateos}, {Bell}, {Brammer},
  {Charlot}, {Chauk{\'e}}, {Labb{\'e}}, {Maseda}, {Muzzin}, {Rix}, {Sobral},
  {van de Sande}, {van Dokkum}, {Wild}, \& {Wolf}}]{VanderWel2016}
{van der Wel}, A., {Noeske}, K., {Bezanson}, R., {et~al.} 2016, \apjs, 223, 29

\bibitem[{{Vasudevan} {et~al.}(2013){Vasudevan}, {Mushotzky}, \&
  {Gandhi}}]{Vasudevan2013}
{Vasudevan}, R.~V., {Mushotzky}, R.~F., \& {Gandhi}, P. 2013, \apjl, 770, L37

\bibitem[{{Wilman} {et~al.}(2008){Wilman}, {Pierini}, {Tyler}, {McGee},
  {Oemler}, {Morris}, {Balogh}, {Bower}, \& {Mulchaey}}]{Wilman2008}
{Wilman}, D.~J., {Pierini}, D., {Tyler}, K., {et~al.} 2008, \apj, 680, 1009

\bibitem[{{Worthey} \& {Ottaviani}(1997)}]{Worthey1997}
{Worthey}, G. \& {Ottaviani}, D.~L. 1997, \apjs, 111, 377

\bibitem[{{Wu} {et~al.}(2018){Wu}, {van der Wel}, {Gallazzi}, {Bezanson},
  {Pacifici}, {Straatman}, {Franx}, {Bari{\v{s}}i{\'c}}, {Bell}, {Brammer},
  {Calhau}, {Chauke}, {van Houdt}, {Maseda}, {Muzzin}, {Rix}, {Sobral},
  {Spilker}, {van de Sande}, {van Dokkum}, \& {Wild}}]{Wu2018}
{Wu}, P.-F., {van der Wel}, A., {Gallazzi}, A., {et~al.} 2018, \apj, 855, 85

\bibitem[{{Yang} {et~al.}(2022){Yang}, {Boquien}, {Brandt}, {Buat},
  {Burgarella}, {Ciesla}, {Lehmer}, {Ma{\l}ek}, {Mountrichas}, {Papovich},
  {Pons}, {Stalevski}, {Theul{\'e}}, \& {Zhu}}]{Yang2022}
{Yang}, G., {Boquien}, M., {Brandt}, W.~N., {et~al.} 2022, \apj, 927, 192

\bibitem[{Yang {et~al.}(2020)Yang, Boquien, Buat, Burgarella, Ciesla, Duras,
  Stalevski, Brandt, \& Papovich}]{Yang2020}
Yang, G., Boquien, M., Buat, V., {et~al.} 2020, \mnras, 491, 740

\bibitem[{Zou {et~al.}(2019)Zou, Yang, Brandt, \& Xue}]{Zou2019}
Zou, F., Yang, G., Brandt, W.~N., \& Xue, Y. 2019, \apj, 878, 11

\bibitem[{Zubovas {et~al.}(2013)Zubovas, Nayakshin, King, \&
  Wilkinson}]{Zubovas2013}
Zubovas, K., Nayakshin, S., King, A., \& Wilkinson, M. 2013, \mnras, 433, 3079

\end{thebibliography}
\bibliographystyle{aa}
\appendix

\onecolumn

\begin{longtable}{cccccrcccrrrcc}
\caption{\label{tableMain} Catalogue of the 73 sources used in our analysis and their derived properties.}\\
\hline\hline
ID  & RA & DEC & z &  $\rm L_{X}$ & $\rm N_H$ & $\rm D_n$ & $\rm H\delta$ & $\rm M_\star$ & SFR & sSFR & $\rm SFR_{NORM}$ & $\rm \lambda_{EDD}$ & $\rm M_{BH}$\\
(1) & (2) & (3) & (4) & (5) & (6) & (7) & (8) & (9) & (10) & (11) & (12) & (13)  & (14) \\
\hline& \\[-1.5ex]
\endfirsthead
\caption{continued.}\\
\hline\hline
ID  & RA & DEC & z &  $\rm L_{X}$ & $\rm N_H$ & $\rm D_n$ & $\rm H\delta$ & $\rm M_\star$ & SFR & sSFR & $\rm SFR_{NORM}$ & $\rm \lambda_{EDD}$ & $\rm M_{BH}$\\
(1) & (2) & (3) & (4) & (5) & (6) & (7) & (8) & (9) & (10) & (11) & (12) &  (13)  & (14) \\
\hline& \\[-1.5ex]
\endhead
\hline
\endfoot
17 & 150.39723 & 1.83487 & 0.96 & 43.04 & 23.18 & 1.63 & 3.73 & 3.57 & 0.21 & -3.22 & -1.02 &  -2.05  & 8.20\\
39 & 150.41814 & 1.97671 & 0.86 & 43.71 & 23.04 & 1.31 & 3.94 & 2.23 & 49.9 & -0.65 & 1.34 &  -1.50  & 8.35\\
77 & 150.24255 & 1.7664 & 0.62 & 44 & 21.69 & 1.41 & 0.82 & 2.51 & 2.1 & -2.08 & 0.55 &  -0.97  & 8.14\\
106 & 150.2998 & 1.7838 & 0.71 & 43.38 & 20 & 1.63 & -0.28 & 1.68 & 0.44 & -2.58 & -0.42 &  -1.57  & 8.07\\
130 & 149.95947 & 1.80147 & 0.68 & 42.66 &  >23.15  & 1.73 & 1.01 & 1.18 & 1.36 & -1.94 & 0.05 &  -2.80  & 8.56\\
137 & 149.94092 & 1.80623 & 0.83 & 42.61 &  <20.00  & 1.38 & 5.17 & 0.91 & 3.12 & -1.46 & 0.19 &  -1.70  & 7.40\\
147 & 150.36452 & 1.81068 & 0.97 & 43.39 & 23.27 & 1.42 & 0.93 & 1.4 & 5.09 & -1.44 & 0.18 &  -2.13  & 8.63\\
209 & 149.99289 & 1.85787 & 0.83 & 43.09 &  <21.74  & 1.49 & 1.07 & 1.85 & 0.15 & -3.09 & -1.23 &  -1.31  & 7.51\\
332 & 150.24869 & 1.97235 & 0.67 & 43.04 & 23.73 & 1.34 & 3.55 & 2.18 & 35.41 & -0.79 & 1.7 &  -2.11  & 8.26\\
442 & 149.82056 & 1.81172 & 0.75 & 43.08 &  <21.55  & 1.27 & 4.6 & 1.7 & 28.71 & -0.77 & 1.15 &  -1.12  & 7.31\\
448 & 149.52786 & 1.81717 & 0.79 & 42.84 &  <20.00  & 1.5 & 2.15 & 2.07 & 2.42 & -1.93 & 0.18 &  -1.86  & 7.80\\
451 & 149.69179 & 1.81819 & 0.74 & 43.22 & 20 & 1.52 & 2.87 & 3.5 & 1.71 & -2.31 & -0.21 &  -2.19  & 8.52\\
774 & 150.42722 & 2.08308 & 0.96 & 43.52 &  <21.70  & 1.38 & 2.85 & 0.67 & 1.81 & -1.57 & -0.44 &  --  & --\\
815 & 150.49304 & 2.11559 & 0.85 & 43.01 & 20 & 1.66 & 2.27 & 1.84 & 0.96 & -2.28 & -0.41 &  -2.14  & 8.26\\
879 & 150.57307 & 2.20349 & 0.82 & 42.85 &  <21.77  & 1.37 & 3.64 & 2.94 & 30.91 & -0.98 & 1.18 &  -1.96  & 7.91\\
919 & 150.6729 & 2.2313 & 0.9 & 43.01 & 22.68 & 1.41 & 5.48 & 0.51 & 5.53 & -0.96 & 0.03 &  -1.48  & 7.59\\
953 & 150.67397 & 2.26266 & 0.97 & 43.01 & 23.06 & 1.63 & 3.14 & 2.14 & 0.41 & -2.72 & -0.88 &  -2.42  & 8.53\\
1004 & 150.53792 & 2.3105 & 0.84 & 43.75 & 22.5 & 1.29 & 3.82 & 0.51 & 7.12 & -0.85 & 0.03 &  -0.50  & 7.39\\
1049 & 150.6673 & 2.34656 & 0.98 & 42.89 & 20 & 1.32 & 6.99 & 2.49 & 4.71 & -1.72 & 0.22 &  -2.23  & 8.21\\
1055 & 150.37733 & 2.35152 & 0.93 & 42.62 & 20 & 1.63 & 2.58 & 2.22 & 1.78 & -2.1 & -0.19 &  -2.24  & 7.95\\
1163 & 150.27464 & 1.98883 & 0.92 & 43.35 & 22.87 & 1.5 & 3.52 & 1.71 & 1.44 & -2.07 & -0.28 &  -1.47  & 7.94\\
1183 & 150.32401 & 2.00395 & 0.96 & 43.19 & 20 & 1.35 & 3.35 & 1.78 & 3.54 & -1.7 & 0.03 &  --  & --\\
1199 & 150.25063 & 2.01504 & 0.67 & 43.76 & 23.67 & 1.36 & 7.31 & 0.77 & 0.93 & -1.92 & -0.23 &  -0.78  & 7.68\\
1228 & 150.29488 & 2.03449 & 0.95 & 43.13 &  >23.34  & 1.71 & 0.85 & 3.04 & 2.6 & -2.07 & 0.06 &  -2.53  & 8.76\\
1252 & 150.26643 & 2.04985 & 0.96 & 42.93 & 22.51 & 1.7 & 1.99 & 6.45 & 1.67 & -2.59 & -0.09 &  -2.32  & 8.35\\
1267 & 150.14873 & 2.0606 & 0.72 & 43.43 & 23.18 & 1.35 & 4.21 & 7.39 & 0.66 & -3.05 & 2.77 &  -1.40  & 7.95\\
1388 & 150.32422 & 2.1765 & 0.91 & 43.13 &  >23.42  & 1.58 & 2.83 & 0.88 & 1.08 & -1.91 & -0.19 &  --  & --\\
1485 & 150.05592 & 2.23346 & 0.94 & 43.55 & 22.33 & 1.4 & 3.51 & 4.14 & 75.23 & -0.74 & 1.6 &  --  & --\\
1496 & 150.04961 & 2.24024 & 0.6 & 42.69 & 23.02 & 1.38 & 2.49 & 0.51 & 1.32 & -1.59 & -0.38 &  -1.91  & 7.69\\
1507 & 149.92113 & 2.25036 & 0.88 & 43.15 &  >23.19  & 1.32 & 4.72 & 2.21 & 40.77 & -0.73 & 1.23 &  -1.69  & 7.94\\
1677 & 149.97012 & 2.33537 & 0.93 & 43.45 & 23.31 & 1.34 & 2.57 & 5.42 & 0.45 & -3.08 & -0.67 &  -2.12  & 8.69\\
1815 & 150.0556 & 2.44917 & 0.73 & 43.27 & 23.27 & 1.72 & 1.17 & 3.04 & 2.32 & -2.12 & 0.34 &  -1.97  & 8.35\\
1856 & 149.84212 & 2.00165 & 0.74 & 43.32 &  >23.00  & 1.59 & 0.31 & 0.79 & 0.48 & -2.22 & -0.47 &  -1.43  & 7.87\\
1971 & 149.73769 & 2.06521 & 0.68 & 43.15 &  <20.91  & 1.38 & 3.43 & 3.68 & 43.07 & -0.93 & 1.28 &  -1.54  & 7.80\\
1994 & 149.52129 & 2.07939 & 0.67 & 43.09 & 20 & 1.47 & 2.11 & 1.52 & 5.34 & -1.45 & 0.63 &  -1.68  & 7.88\\
2061 & 149.70183 & 2.12183 & 0.89 & 43.4 & 20 & 1.64 & 1.02 & 1.84 & 1.65 & -2.05 & -0.21 &  -1.39  & 7.90\\
2105 & 149.88545 & 2.17717 & 0.66 & 43.52 & 23.03 & 1.44 & 2.5 & 3.62 & 8.62 & -1.62 & 0.49 &  -1.46  & 8.11\\
2108 & 149.53911 & 2.1826 & 0.99 & 43.4 &  <21.47  & 1.3 & 4.7 & 2.36 & 16.15 & -1.16 & 0.75 &  --  & --\\
2153 & 149.64964 & 2.20925 & 0.95 & 42.97 & 20 & 1.81 & -0.52 & 3.59 & 0.27 & -3.12 & -0.86 &  -2.42  & 8.49\\
2157 & 149.61758 & 2.21556 & 0.68 & 42.81 &  <21.14  & 1.27 & 2.91 & 0.94 & 11.96 & -0.9 & 1.11 &  -0.95  & 6.86\\
2162 & 149.86823 & 2.21866 & 0.79 & 43.07 & 20 & 1.74 & 0.37 & 1.45 & 0.17 & -2.92 & -1.01 &  -1.84  & 8.02\\
2226 & 149.82198 & 2.25467 & 0.93 & 43.12 &  <21.75  & 1.26 & 3.53 & 2.01 & 27.11 & -0.87 & 0.95 &  --  & --\\
2257 & 149.85678 & 2.27314 & 0.76 & 42.72 & 23.08 & 1.4 & 4.02 & 2.45 & 5.41 & -1.66 & 0.55 &  -2.48  & 8.30\\
2333 & 149.48697 & 2.32644 & 0.84 & 43.1 & 22.71 & 1.43 & 5.84 & 1.45 & 0.64 & -2.35 & -0.49 &  -0.48  & 6.68\\
2339 & 149.79379 & 2.32721 & 0.89 & 43.58 & 20 & 1.53 & 2.2 & 2.72 & 0.2 & -3.13 & -1.08 &  -1.82  & 8.53\\
2377 & 149.67668 & 2.35335 & 0.95 & 42.83 &  <20.00  & 1.28 & 5.28 & 1.06 & 47.19 & -0.35 & 1.14 &  -0.79  & 6.72\\
2385 & 149.64417 & 2.35919 & 0.95 & 43.08 &  <20.00  & 1.89 & 1.78 & 2.57 & 2.49 & -2.01 & -0.04 &  -2.50  & 8.68\\
2396 & 149.84178 & 2.36706 & 0.85 & 42.99 &  <20.00  & 1.45 & 6.59 & 0.44 & 0.47 & -1.97 & -1.04 &  -1.66  & 7.75\\
2416 & 149.78671 & 2.37833 & 0.76 & 42.67 & 20 & 1.76 & 0.43 & 0.77 & 0.38 & -2.31 & -0.73 &  -2.60  & 8.37\\
2531 & 150.63396 & 2.49045 & 0.8 & 43.01 &  <21.57  & 1.32 & 2.87 & 1.05 & 14.65 & -0.85 & 1.03 &  -1.14  & 7.25\\
2537 & 150.47704 & 2.49411 & 0.68 & 42.71 &  >22.96  & 1.26 & 7.88 & 0.71 & 0.46 & -2.19 & -0.61 &  -1.99  & 7.79\\
2747 & 150.50719 & 2.65357 & 0.81 & 42.79 &  <20.00  & 1.33 & 4.53 & 2.11 & 28.85 & -0.86 & 1.26 &  -0.70  & 6.59\\
2793 & 150.44531 & 2.71916 & 0.82 & 44.09 & 23.99 & 1.43 & 3.42 & 1.89 & 1.93 & -1.99 & 0.01 &  -1.17  & 8.43\\
2819 & 150.45914 & 2.7334 & 0.79 & 42.82 &  >23.27  & 1.47 & 3.1 & 4.92 & 4.23 & -2.07 & 0.4 &  -2.02  & 7.94\\
2929 & 149.99065 & 2.4643 & 0.89 & 43.05 & 22.59 & 1.25 & 4.86 & 2.09 & 88.45 & -0.37 & 1.57 &  -0.11  & 6.26\\
2964 & 150.15025 & 2.47516 & 0.69 & 43.95 & 22.96 & 1.34 & 8.09 & 3.2 & 118.31 & -0.43 & 1.86 &  -0.70  & 7.81\\
3059 & 149.92026 & 2.51424 & 0.7 & 42.95 & 21.57 & 1.47 & 4.49 & 0.48 & 1.25 & -1.58 & -0.34 &  -1.53  & 7.58\\
3085 & 149.91573 & 2.52133 & 0.73 & 42.81 & 21.21 & 1.92 & -1.48 & 4.41 & 0.12 & -3.57 & -1.13 &  -3.05  & 8.96\\
3097 & 150.1385 & 2.52937 & 0.9 & 43.24 &  >23.44  & 1.56 & 2.76 & 1.79 & 0.12 & -3.17 & -1.35 &  -1.45  & 7.80\\
3167 & 150.02681 & 2.56205 & 0.75 & 43.45 & 20 & 1.53 & 2.25 & 1.57 & 0.55 & -2.46 & -0.55 &  -1.35  & 7.92\\
3238 & 150.1991 & 2.59763 & 0.9 & 43.83 & 22.85 & 1.73 & 0.76 & 2.24 & 0.5 & -2.65 & -0.7 &  -1.70  & 8.68\\
3250 & 150.3288 & 2.60545 & 0.92 & 42.77 &  <20.00  & 1.33 & 5.73 & 1.49 & 29.42 & -0.7 & 1.04 &  -1.40  & 7.27\\
3281 & 150.12984 & 2.62471 & 0.9 & 42.94 & 21.71 & 1.69 & -0.5 & 1.74 & 0.48 & -2.56 & -0.71 &  -1.87  & 7.91\\
3555 & 149.85017 & 2.45224 & 0.71 & 42.91 & 22.64 & 1.31 & 5 & 1.48 & 6.15 & -1.38 & 0.64 &  -1.80  & 7.81\\
3565 & 149.53326 & 2.45826 & 0.84 & 42.96 & 23.14 & 1.53 & 2.45 & 2.39 & 1.54 & -2.19 & -0.09 &  -2.03  & 8.09\\
3568 & 149.70566 & 2.46027 & 0.97 & 43.06 & 20 & 1.27 & 5.14 & 1.19 & 25.93 & -0.66 & 0.78 &  -1.64  & 7.80\\
3581 & 149.65997 & 2.4651 & 0.96 & 43.78 & 23.51 & 1.85 & -0.86 & 4.78 & 0.3 & -3.21 & -0.84 &  --  & --\\
3671 & 149.87674 & 2.51872 & 0.73 & 43.14 &  <20.00  & 1.51 & 4.21 & 1.75 & 1.23 & -2.15 & -0.16 &  -1.65  & 7.90\\
3757 & 149.8474 & 2.56896 & 0.91 & 42.79 &  <20.00  & 1.44 & 4.98 & 4.8 & 55.98 & -0.93 & 1.44 &  -2.06  & 7.94\\
3760 & 149.80627 & 2.57244 & 0.96 & 42.99 & 22.75 & 1.49 & 5.26 & 2.16 & 2.6 & -1.92 & -0.06 &  -1.74  & 7.83\\
3826 & 149.88414 & 2.60991 & 0.76 & 43.35 & 22.84 & 1.28 & 6.19 & 1.92 & 47.75 & -0.6 & 1.48 &  -1.69  & 8.16\\
3833 & 149.62067 & 2.61433 & 0.95 & 43.5 &  >23.51  & 1.54 & 3.87 & 1.9 & 5.22 & -1.56 & 0.24 &  -1.09  & 7.71\\
3907 & 149.41666 & 2.69366 & 0.97 & 43.31 &  <21.83  & 1.31 & 4.18 & 1.52 & 12.3 & -1.09 & 0.61 &  --  & --\\
\end{longtable}
\tablefoot{(1): Unique identifier in the LEGA-C catalogue. (2, 3): LEGA-C optical right ascension and declination [degrees]. (4): Spectroscopic redshift. (5): X-ray luminosity [log($\rm erg~s^{-1}$)].  (6): Absorbing column density [$\rm \log (cm^{-2})$]. The $<$ and $>$ symbols indicate lower and upper limits, respectively. (7): $\rm D_n(4000)$ index indicating the amplitude of the calcium break at 4000 $\rm \AA$. (8): $\rm H_\delta$ Balmer line [$\rm \AA$]. (9): Stellar mass in units of $\rm 10^{11}~M_\odot$ (10): Star formation rate [$\rm M_\odot~yr^{-1}$]. (11): Specific star formation rate [log$\rm (Gyr^{-1}$)]. (12): Normalised star formation rate in logarithmic scale. \textbf{(13): Eddington ratio in logarithmic scale (Eq.~\ref{eq_edd}). (14): Black hole mass estimates using Eq.~\ref{eq_bhmass} in units of [log$\rm (M_\odot$)].}}

\end{document}